\documentclass[aps,prd,superscriptaddress,floatfix,nofootinbib,notitlepage]{revtex4}

\usepackage{graphicx}
\usepackage{amsmath}
\usepackage{longtable}
\usepackage{aas_macros}
\usepackage{xcolor}

\graphicspath{{./figures/}}

\usepackage[sort&compress]{natbib}
\usepackage{hyperref}
\newcommand{\irf}[1]{\texttt{#1}}

\newcommand{\Eqref}[1]{Eq.~\eqref{#1}}

\newcommand{\Fermi}{{\textit{Fermi}}}
\newcommand{\fermi}{\Fermi}

\usepackage{color}

\ifdefined\bwfigures
\else
\fi

\newcommand\fdg{\mbox{$.\!\!^\circ$}}

\begin{document}

\title{Search for Gamma-ray Emission from Dark Matter Annihilation in the Large Magellanic Cloud with the \fermi\ Large Area Telescope}

\date{\today}

\author{Matthew R.~Buckley}
\affiliation{Department of Physics and Astronomy, Rutgers University, Piscataway, NJ 08854, USA}
\author{Eric Charles}
\affiliation{W. W. Hansen Experimental Physics Laboratory, Kavli Institute for Particle Astrophysics and Cosmology, Department of Physics and SLAC National Accelerator Laboratory, Stanford University, Stanford, CA 94305, USA}
\author{Jennifer M.~Gaskins}
\affiliation{California Institute of Technology, Pasadena, CA 91125, USA}
\affiliation{GRAPPA, University of Amsterdam, 1098 XH Amsterdam, Netherlands}
\author{Alyson M.~Brooks}
\affiliation{Department of Physics and Astronomy, Rutgers University, Piscataway, NJ 08854, USA}
\author{Alex Drlica-Wagner}
\affiliation{Center for Particle Astrophysics, Fermi National Accelerator Laboratory, Batavia, IL 60510}
\author{Pierrick Martin}
\affiliation{Institut de Recherche en Astrophysique et Plan\'{e}tologie, UPS/CNRS, UMR5277, 31028 Toulouse cedex 4, France}
\author{Geng Zhao}
\affiliation{W. W. Hansen Experimental Physics Laboratory, Kavli Institute for Particle Astrophysics and Cosmology, Department of Physics and SLAC National Accelerator Laboratory, Stanford University, Stanford, CA 94305, USA}

\begin{abstract}
At a distance of 50~kpc and with a dark matter mass of $\sim10^{10}$~M$_{\odot}$, the Large Magellanic Cloud (LMC) is a natural target for indirect dark matter searches.  
We use five years of data from the {\it Fermi} Large Area Telescope (LAT) 
and updated models of the gamma-ray emission from standard astrophysical
components to search for a dark matter annihilation
signal from the LMC\@.  We perform a rotation curve analysis to
determine the dark matter distribution, setting a robust minimum on the amount 
of dark matter in the LMC\@, which we use to set conservative bounds on the annihilation
cross section. The LMC emission is generally very well described by the standard astrophysical
sources, with at most a $1-2\sigma$ excess identified
near the kinematic center of the LMC once systematic uncertainties
are taken into account.   We place competitive bounds on the dark
matter annihilation cross section as a function of dark matter particle mass and annihilation channel. 

\end{abstract}
\maketitle

\section{INTRODUCTION}\label{sec:intro}

The structure of the visible Universe cannot be explained only by the known physics of the Standard Model. Measurements of galactic rotation curves \cite{Rubin:1980zd} and galaxy cluster dynamics \cite{Zwicky:1933gu}, precision measurements of the Cosmic Microwave Background \cite{Ade:2013zuv}, observations of the primordial abundances of heavy isotopes produced by Big Bang Nucleosynethesis \cite{Olive:2003iq}, and other lines of evidence provide orthogonal sets of data that all point to a significant component of the Universe's energy density being made up of a new form of matter without significant interaction with the Standard Model. Further evidence comes from the excellent concordance between observation and computer simulations of large-scale structure when cold dark matter is included.   Observationally, we know dark matter interacts gravitationally, is non-relativistic during the formation of large-scale structure, and does not have large scattering cross sections with either itself \cite{Markevitch:2003at} or the Standard Model \cite{Akerib:2013tjd}. No particle in the Standard Model meets the necessary requirements to make up the dark matter energy density. Other than these pieces of information, we have no solid experimental or theoretical understanding of the fundamental nature of dark matter.

While it is not a necessary condition for a successful model of dark matter, it is theoretically well motivated to expect the dark matter to be composed of heavy ($m_\chi \gtrsim 1$~GeV) particles that have a significant annihilation cross section into Standard Model particles. The canonical example of such dark matter is a non-relativistic thermal relic that froze out of equilibrium with the Standard Model particle bath in the early Universe. A dark matter particle with the thermally averaged annihilation cross section $\langle \sigma v\rangle\sim 3\times 10^{-26}$~cm$^3$/s can yield the measured dark matter energy density today, $\Omega h^2 = 0.1199\pm 0.0027$ \cite{Ade:2013zuv}. This can be realized in models with $SU(2)_L$ weak interactions, though other models can also work \cite{Feng:2008ya}. 

While significant annihilation would cease during freeze-out, if the dark matter pair annihilation is due to an $s$-wave process and therefore velocity independent, low levels of annihilation would continue to the present day. The end products of this annihilation can be searched for as excesses relative to products from Standard Model astrophysical processes.  (We will refer such background processes as ``baryonic,'' to distinguish them from the sought-after dark matter signals.)  As we do not know the nature of the dark matter itself, we cannot know with any certainty which Standard Model channels are the most likely to contain evidence of this annihilation. Furthermore, relatively simple modifications of the canonical thermal relic theory can result in present-day annihilation cross sections that differ by many orders of magnitude from the standard assumption of $\langle \sigma v\rangle \sim 3\times 10^{-26}$~cm$^3$/s \cite{ArkaniHamed:2008qn}.
Thus, such indirect searches must be performed in as many channels as possible, including photons, neutrinos, positrons, antiprotons, and heavier antinuclei. We must also remain open to annihilation rates far from that expected of a simple thermal relic.

Of particular interest is the indirect search for dark matter annihilating into gamma rays. Such signatures are the result of many possible annihilation channels, and so are a generic expectation of dark matter annihilation. In addition to annihilation into pairs of gamma rays, which have a characteristic line spectrum with $E_\gamma = m_\chi$, dark matter may convert into pairs (or a larger multiplicity) of quarks, leptons, gluons, or $SU(2)_L$ gauge bosons, all of which will result in a continuum spectrum of gamma rays, as unstable particles decay, light quarks hadronize, and the showered mesons themselves decay into states that include gamma rays. It is this continuum emission that we search for in this paper. As gamma rays travel relatively unimpeded through the Universe compared to charged cosmic rays (CRs), the dependence on propagation models is reduced compared to charged-particle final states, though not completely eliminated.

The Large Area Telescope (LAT) on board the {\it Fermi Gamma-ray Space Telescope} ({\it Fermi} LAT) is currently the most sensitive instrument for indirect searches via gamma rays in the energy range from $\sim 100$~MeV to over 300 GeV \cite{Atwood:2009ez}. At present, the {\it Fermi} LAT is the only instrument sensitive to gamma-ray signals of dark matter annihilation in the ${\cal O}(10-100~\mbox{GeV})$ mass range with cross sections that are on the order of a thermal relic.

Gamma rays from dark matter annihilation would be preferentially detected from nearby overdense regions of dark matter. Prior to this work, searches for indirect signals using {\it Fermi} LAT data have been performed targeting dwarf spheroidal galaxies orbiting the Milky Way \cite{Ackermann:2011wa,Ackermann:2013yva,GeringerSameth:2011iw,Geringer-Sameth:2014qqa}, unresolved halo substructure \cite{Ackermann:2012nb,Belikov:2011pu,Berlin:2013dva,Buckley:2010vg}, galaxy clusters \cite{Ackermann:2010rg,Dugger:2010ys}, the isotropic gamma-ray background \cite{Abdo:2010dk,Ackermann:2012uf,2015arXiv150105316D,2015arXiv150105464T}, and the Milky Way Galactic Center \cite{Abazajian:2012pn,Abazajian:2014fta,Boyarsky:2010dr,Daylan:2014rsa,Goodenough:2009gk,Gordon:2013vta,Hooper:2010mq,Hooper:2011ti,Hooper:2012sr,Hooper:2013rwa,Huang:2013pda}. In a number of these analyses of the Galactic Center, a spatially extended anomalous excess has been reported in the {\it Fermi} LAT data. This excess has not been positively identified with any previously known astrophysical source, though some possibilities have been considered as the source of these gamma rays. For example, a previously unknown population of several hundred millisecond pulsars in the Galactic Center \cite{Abazajian:2010zy,Wharton:2011dv,Hooper:2013nhl,Petrovic:2014xra,2013MNRAS.436.2461M},  a larger-than-expected CR proton flux \cite{Carlson:2014cwa}, or CR electrons injected in a past burst event \cite{Petrovic:2014uda} could be non-exotic astrophysical explanations for the observed signal. Alternatively, this excess can be well fit by dark matter with a standard halo density profile, annihilating into Standard Model quarks or leptons with an approximately thermal cross section. The origin of these gamma rays remains a topic of much debate and the source of a  great deal of model-building interest \cite{Abdullah:2014lla,Agrawal:2014una,Arina:2014yna,Basak:2014sza,Berlin:2014pya,Boehm:2014bia,Boehm:2014hva,Buckley:2010ve,Cerdeno:2014cda,Cheung:2014lqa,Cline:2014dwa,Hooper:2010im,Huang:2014cla,Ipek:2014gua,Ko:2014gha,Marshall:2011mm,Martin:2014sxa,Wang:2014elb,Wharton:2011dv,Zhu:2011dz,Cerdeno:2015ega}.

Given the uncertainties and observational limitations in the Galactic Center, resolving the origin of this signal will likely require input from observations of other targets.  The most sensitive indirect constraints have come from combined observations of dwarf spheroidal galaxies.
Unfortunately, the sensitivity of that search is currently too weak to resolve the controversy \cite{Ackermann:2011wa,Geringer-Sameth:2014qqa}. New sky surveys \cite{Abbott:2005bi,Abell:2009aa} are likely to identify additional dwarf galaxies in the near future, which would improve the sensitivity of a combined satellite search. However, the current bound is driven by a small number of dwarfs with high expected fluxes and low backgrounds; and it is by no means assured that the upcoming surveys will identify another such ``good'' dwarf, which would be needed for large improvements \cite{He:2013jza}.

With that motivation in mind, it is desirable to identify a new target for indirect detection with the potential for sensitivity competitive with the dwarf spheroidal galaxy search. In this paper, we present for the first time the indirect detection constraints derived from {\it Fermi} LAT observations of the Large Magellanic Cloud (LMC), the largest satellite galaxy of the Milky Way. Galaxies in the mass range of the LMC are expected to be dark matter-rich, and evidence suggests that the LMC is on its first infall to the Milky Way and has not been tidally stripped \cite{Besla2007,Kallivayalil:2013xb}. Due to its large dark matter mass and relative ($\sim 50$~kpc) proximity to Earth \cite{Matsunaga:2009hu,Pietrzynski:2009ij}, the LMC would be the second brightest source of gamma rays from dark matter annihilations in the sky, after the Galactic Center, and a promising target \cite{Tasitsiomi:2003ue,Tasitsiomi:2003vw}. Unlike the dwarf spheroidal galaxies, the LMC has significant baryonic backgrounds. Despite this, we can place robust and conservative upper bounds on the dark matter annihilation signal that are competitive with those extracted from the dwarf spheroidal observations. 

The paper is organized as follows. Section~\ref{sec:dm} covers the theory of indirect detection of dark matter annihilation, our derivation of the dark matter profile of the LMC, and implications for the expected indirect detection signal. In Section~\ref{sec:baryons}, we discuss the baryonic backgrounds present in the LMC, and our methods of separating them from dark matter indirect detection signals. The {\it Fermi} LAT instrument, data selection, and data preparation are described in Section~\ref{sec:data}.  Our statistical techniques and data analysis are presented in Section~\ref{sec:stats}, and we show the resulting bounds in Section~\ref{sec:constraints}. We place our results into the larger context and propose directions for future work in the concluding Section~\ref{sec:conclusion}.

\section{Dark Matter Annihilation in the LMC}\label{sec:dm}

The gamma-ray flux from dark matter annihilation depends on the the product of factors related to the particle physics and the spatial distribution of dark matter.  Gamma-ray observatories viewing a solid angle $\Delta \Omega$ will see a differential flux of photons from dark matter annihilation given by
\begin{equation}
\frac{d\phi}{dE_\gamma} = \left( \frac{x \langle \sigma v\rangle}{8\pi}\frac{dN_\gamma}{dE_\gamma}\frac{1}{m_\chi^2}\right) \left( \int_{\Delta\Omega} d\Omega \int_\text{l.o.s.} d\ell ~\rho_\chi^2(\vec{\ell}\,)\right), \label{eq:generalflux}
\end{equation}
where $x=1$ if dark matter is its own antiparticle, $x=1/2$ if it is not, and $dN_\gamma/dE_\gamma$ is the differential spectrum of gamma rays from annihilation of a pair of dark matter particles \cite{Ullio:2002pj}. In this paper we make the standard assumption that $x=1$. 

The elements inside the first set of parentheses of Eq.~\eqref{eq:generalflux} depend on the particle physics of dark matter, and are the same for all targets of indirect detection. These are completely unknown experimentally, though we may have theoretical reasons to assume certain ranges of masses and final states.  For our search we scan over these assumed ranges, testing for a signal at each combination of mass $m_\chi$ and annihilation channel.  These choices, along with the astrophysical factors discussed next, are sufficient to determine the differential flux of gamma rays up to an overall normalization, allowing us to place bounds on the total thermally averaged annihilation cross section $\langle\sigma v\rangle$. We will return to the choices of differential spectrum $dN_\gamma/dE_\gamma$ in Section~\ref{sec:dm}C.

The factors in the second set of parentheses in Eq.~\eqref{eq:generalflux} are the astrophysical quantities that are target-dependent. Finding an astronomical object that maximizes this quantity then is a key step in designing a sensitive search for indirect signals of dark matter. This integral depends on the dark matter density profile $\rho$ as a function of position $\vec{\ell}$ in the direction of the line-of-sight (l.o.s.). The integral of the density squared over a solid angle $\Delta\Omega$ is known as the $J$-factor:
\begin{equation}
J(\Delta\Omega) \equiv \int_{\Delta\Omega} d\Omega \int_\text{l.o.s.} d\ell ~\rho_\chi^2(\vec{\ell}\,).
\end{equation}
Note that the definition of the $J$-factor depends implicitly on the distance to the dark matter target.   The density profiles of dark matter halos as a function of position must be determined from a combination of observation and simulation. In this work, we adopt the six-parameter generalized dark matter density profile as a function of the distance $r$ from the profile center \cite{1990ApJ...356..359H,1996MNRAS.278..488Z,1998ApJ...502...48K}:
\begin{equation}
\rho(r) = \frac{\rho_0}{\left(\frac{r}{r_S}\right)^\gamma \left[ 1+\left(\frac{r}{r_S}\right)^\alpha\right]^{\frac{\beta-\gamma}{\alpha}}} \Theta(r_\text{max}-r), \label{eq:general_nfw}
\end{equation}
where $\Theta(x)$ is the Heaviside step function. Here, the characteristic density $\rho_0$, the scale radius $r_S$, and the coefficients $\alpha$, $\beta$, and $\gamma$ are all free and must be fit to a particular dark matter halo. We terminate the profile at some distance $r_\text{max} \sim 100$~kpc. Setting $(\alpha,\beta,\gamma) = (1,3,1)$ yields the classic NFW profile~\cite{REF:1996ApJ...462..563N}, transitioning from an inner slope of $-1$ to $-3$ at large radii. An isothermal profile 
has a core rather than an NFW-like cusp, and can be obtained from Eq.~\eqref{eq:general_nfw} with $(\alpha,\beta,\gamma) = (2,2,0)$. \medskip

As can be seen from the definition of $J$, huge gains in the sensitivity to the annihilation cross section can be made by targeting those objects that are both dark-matter-dense and nearby. Prior to this work, the most likely targets for indirect searches have been the center of the Milky Way and the dwarf galaxies orbiting the Milky Way, as these have the largest $J$-factors relative to their baryonic backgrounds. However, the LMC is both very massive and relatively nearby. Though there is uncertainty in the dark matter profile of the LMC, we will show that, even under conservative assumptions, our largest Galactic satellite is the second-brightest target for dark matter annihilation searches, after the Galactic Center itself.

\subsection{The LMC Dark Matter Profile }

Proper motion data for the LMC indicate that it may be on its first infall into the Milky Way's virial halo \cite{Besla2007}.  If true, then little dark matter may have been lost from the LMC through tidal stripping with the Milky Way \cite{Penarrubia2010,BZ14}, which gives our search for dark matter annihilation an added advantage.  The LMC has a prominent stellar bar, suggesting that it may have been a barred spiral before capture by the Milky Way, but now generally has a more irregular morphology.  Unlike the Galactic Center, which is viewed edge on, we view the LMC closer to face on, at low inclination.  This orientation makes it difficult to measure the inclination angle precisely, hence uncertainty in the inclination is the largest source of error in determining the LMC dark matter density profile from rotation curve data.   

In addition, the gravitational center of the LMC is uncertain to within $\sim 1\fdg 5$.  The observed stellar kinematics favor rotation about a center located near the eastern end of the stellar bar \cite{vdM02} (denoted in this paper as the {\tt stellar} center), while the kinematics of the H\,{\sc i} gas favor rotation about a center located at the western end \cite{Kim98} (the {\tt HI} center).  These two locations are $1\fdg 41 \pm 0\fdg 43$ apart.  A recent determination of the center of the LMC based on proper motion data favors a position in agreement with the H\,{\sc i} center to within errors \cite{vdMK14}.  For our study, we adopt three centers as benchmarks: the previously mentioned {\tt stellar} and {\tt HI} centers derived from the stellar and H\,{\sc i}  rotation curves, and an {\tt outer} center defined as the center of the outer lines of equal surface brightness (corrected for viewing angle).
The {\tt HI} and {\tt stellar} centers are roughly at the edges of the LMC bar, and therefore define the extremes of our profile center uncertainties. The coordinates of these centers are listed in Table~\ref{tab:centers}. In addition to these center locations motivated by astronomical observations, we will perform scans of center locations over the entire LMC, as the dark matter center is not necessarily exactly co-located with any of the rotation centers of the visible LMC (see {\it e.g.}, Ref.~\cite{2013ApJ...765...10K})

\begin{table}[ht]
\begin{tabular}{ccccc}
\hline \hline
Center & $\ell$ ($^{\circ}$) & $b$ ($^{\circ}$)\\ \hline
{\tt stellar} & $280.54$ & $-32.51$ \\ 
{\tt HI} & $279.78$ & $-33.77$ \\ 
{\tt outer} & $280.07$ & $-32.46$ \\ \hline \hline
\end{tabular}

\caption{Coordinates of our three benchmark LMC centers, in both right ascension/declination and Galactic coordinates. \label{tab:centers}}
\end{table}

Given these uncertainties, as well as others ({\it e.g.}, how to convert the light of the stars into stellar mass), we choose not to determine the ``best'' fit to the dark matter distribution of the LMC, but rather to find the range of allowed distributions.  Below, we use the observed rotation curve data to place upper and lower limits on the dark matter density profile in the LMC, from which we derive a range of potential $J$-factors for this target. As we will show, the observational data place a robust floor of $\sim10^{20}$ GeV$^2$/cm$^5$ on the integrated $J$-factor of the LMC, though the stellar rotation curves are also consistent with much larger $J$-factors. 
Future observational work might reduce this uncertainty, but we again emphasize that even under the most conservative assumption, the LMC is a viable source of dark matter annihilation products. \medskip

Under the assumption of circular orbits, a measurement of the rotational velocity of a galaxy is a direct measurement of the 
mass enclosed as a function of radius, $v_\text{rot}^2 = GM(<r)/r$.  For the inner 3 kpc of the LMC, we adopted the H\,{\sc i} rotation curve of Ref.~\cite{Kim98}.\footnote{Ref.~\cite{Kim98} used Gaussian fits to the H\,{\sc i} data to determine the velocities as a function of radius.  To better fit non-circular motions in the H\,{\sc i} data, Hermite polynomials are a better choice \cite{deblok2008}.  The fact that we have neglected non-circular motions means that the rotation curve could rise more quickly in the center than Ref.~\cite{Kim98} determined.  Hence, all of our fits will be lower limits on the contribution from dark matter to the rotation curve.  Likewise, Ref.~\cite{Kim98} adopted a high transverse motion of the LMC on the sky that has since been updated with new proper motion measurements.  We make no correction here, but again note that this makes our dark matter fits conservative underestimations.}  
The distribution of H\,{\sc i}  velocities was binned in 100 pc radial bins, and the 1$\sigma$ variation within those bins were adopted as the 
errors in the H\,{\sc i} velocities.  Beyond 3~kpc, we adopted the flat rotation curve observed in stellar kinematics \cite{Kunkel1997}. For these large radii, we adopted the value $v_\text{flat} = 97.7 \pm 18.8$ km/s determined by Ref.~\cite{vdMK14}, but we corrected it to the same inclination as the data from Ref.~\cite{Kim98} (we discuss the inclination angle in greater detail below).  To determine the dark matter contribution to the rotation curve, the contributions from the H\,{\sc i} gas and stars were subtracted in quadrature (as the enclosed mass is proportional to velocity squared). We adopted the H\,{\sc i}+He mass as a function of radius from Ref.~\cite{LuksRohlfs92}.  For stars, we assumed an exponential stellar disk (neglecting the obvious bar) with total stellar mass of $2.7\times10^9~M_{\odot}$ within 8.9~kpc \cite{vdM02} and scale length of 1.5~kpc \cite{vdM01}.  We allowed the stellar mass contribution to vary (see below), equivalent to allowing a range of mass-to-light ratios. This procedure adopted the same position for the kinematic center of the LMC as the Ref.~\cite{Kim98} data (which is our {\tt HI} center, see Table~\ref{tab:centers}).  

\begin{figure}[ht]
\includegraphics[width=0.5\columnwidth]{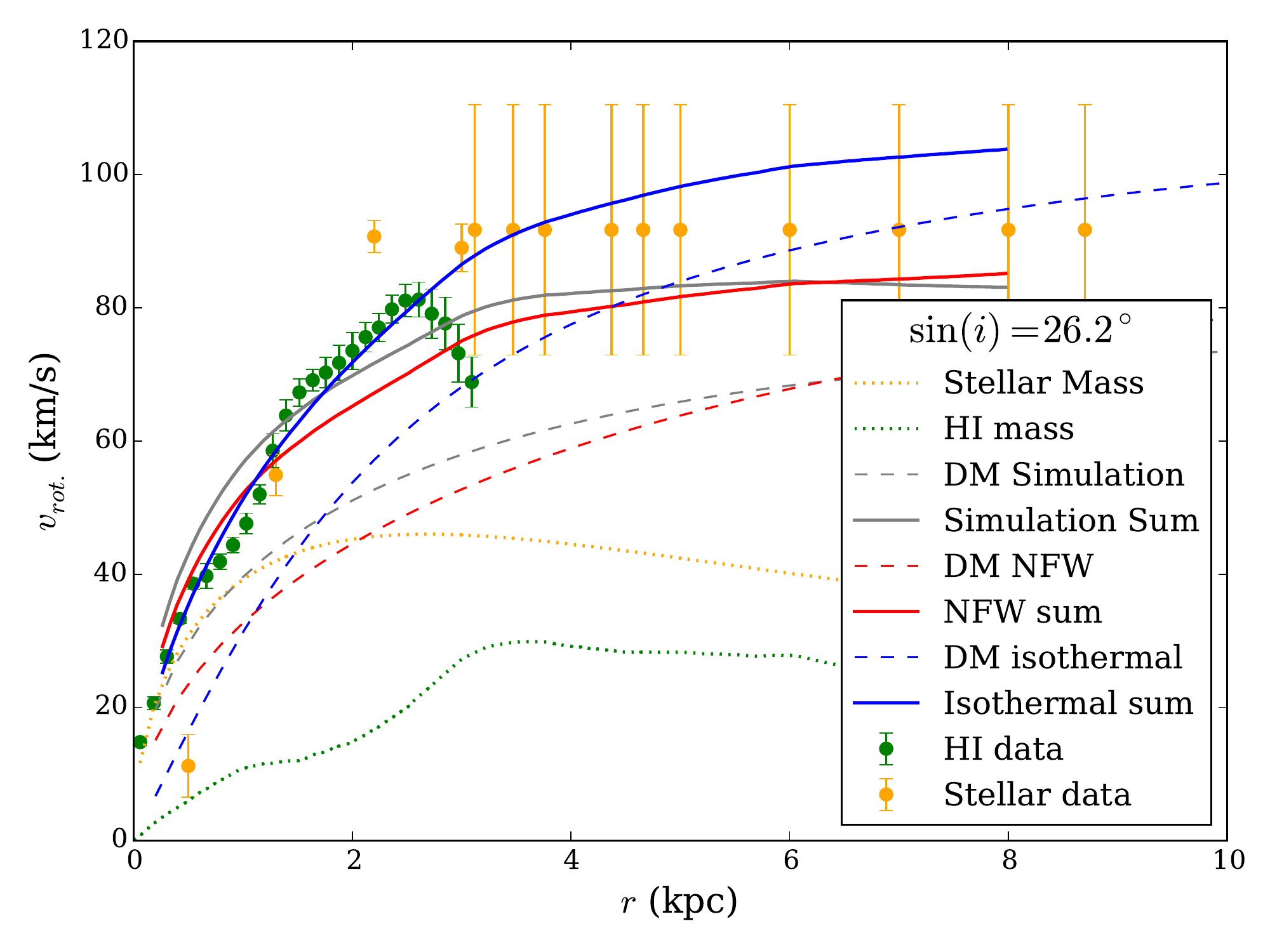}\includegraphics[width=0.5\columnwidth]{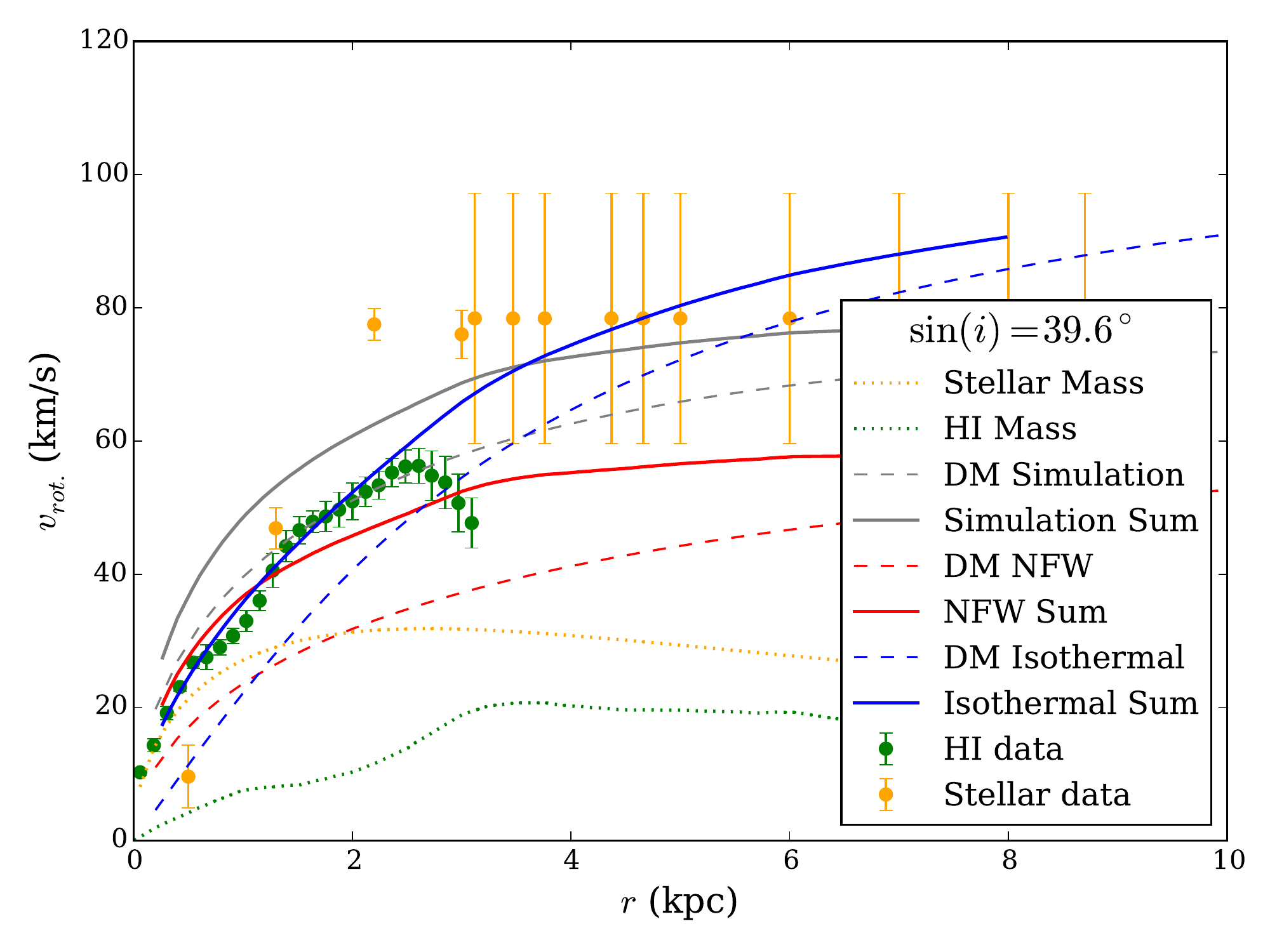}
\caption{LMC rotation curve data, assuming an inclination $i$ that maximizes (left) and minimizes (right) the dark matter density. Stellar $v_\text{rot}$ data are shown with orange points \cite{vdMK14}, and H\,{\sc i} $v_\text{rot}$ data \cite{Kim98} in green. The orange dotted line denotes the contribution to $v_\text{rot}$ from the stellar mass, and the contribution from the H\,{\sc i}+He gas is shown in dotted green \cite{LuksRohlfs92}. The $v_\text{rot}$ values predicted by NFW and isothermal profiles fit to data are shown by red and  blue dashed lines, respectively. Solid lines show $v_\text{rot}$ of the dark matter profiles plus contribution from the stars and gas, with the maximum values in the left plot and the minimum on the right. Grey lines show the mean profile of dark matter fit from simulations of LMC-like galaxies (dashed is dark matter-only, solid is dark matter plus stars and gas), and are not fit to the stellar and H\,{\sc i} data points. The simulated dark matter rotation curve is independent of inclination angle, and the flat rotation curve beyond 3~kpc is based on the results of Ref.~\cite{vdMK14}.
\label{fig:LMC_vrot}}
\end{figure}

The inclination angle $i$ is the largest source of uncertainty in interpreting the LMC's rotation curve.  Hence, we fit for dark matter contributions at the extremes of what is allowed by the inclination and velocity errors.  The H\,{\sc i} data favor an inclination of 33$^{\circ}$ \cite{Kim98}, but the kinematics of young stars favors a lower inclination of $26\fdg 2\pm 5\fdg 9$.
The proper motion data alone favor $39\fdg 6 \pm 4\fdg 5$ \cite{vdMK14}.  Taking the central values of these two extremes and neglecting the errors on the individual measurements, the uncertainty of the inclination angle spans 14$^{\circ}$.  Adopting a lower inclination raises the normalization of the rotation curve, while higher inclination values lower it.  Hence, we find a minimum contribution from the dark matter by adopting $i = 39.6^{\circ}$ and rescaling the rotation velocities accordingly, and a maximum by adopting $i = 26\fdg 2$.  At each inclination extremum, we perform a Levenberg-Marquardt least-squares fit to both a purely isothermal density profile $(\alpha,\beta,\gamma) = (2,2,0)$, and an Navarro-Frenk-White (NFW) density profile $(\alpha,\beta,\gamma) = (1,3,1)$.  As mentioned above, the stellar contribution was allowed to vary so as to contribute the largest possible mass to the inner rotation curve in each case. By maximizing the stellar contribution consistent with the rotation curve, we ensure that our dark matter contributions are always lower limits. In practice, the stellar mass varied between $1.2 \times 10^9~{\rm M}_\odot$ at maximum inclination and $2.4\times 10^9~{\rm M}_\odot$ at minimum inclination.  In Figure~\ref{fig:LMC_vrot}, we plot the rotation curve data for the LMC, at the maximum and minimum inclination angles, along with the best-fit profiles.  The data points beyond 3 kpc represent a flat rotation curve, as found in Ref.~\cite{vdMK14} based on data from Ref.~\cite{Kunkel1997}.  We will use {\tt nfw-max} and {\tt iso-max} to denote the NFW and isothermal profiles fit to the data at $i = 26\fdg 2$, and {\tt nfw-min} and {\tt iso-min} the results of the fit with an inclination angle of $i = 39\fdg 6$.

The assumptions of pure NFW or isothermal profiles are simplifications that we do not expect to be realized in the actual LMC. Thus, we have taken a separate approach to determine what the ``typical'' dark matter density profile of an LMC--mass 
galaxy might be.  Recent cosmological simulation results have demonstrated that energetic feedback from stars and supernovae can transform an initially steep inner density profile into a shallower profile \cite{Governato2012, Teyssier2013, diCintio2013}.  The degree of transformation is sensitive to the mass of stars formed \cite{Penarrubia:2012bb,diCintio2013}, and the stellar mass is dependent on halo mass \cite{Behroozi2013, Moster2012}.  Ref.~\cite{diCintio2014b} has provided a general relation for the generalized NFW parameters ($\alpha,\beta,\gamma$) as a function of stellar-to-halo mass ratio. Therefore, we can extract a range of generalized NFW profiles appropriate for the LMC from simulations, provided we know the stellar and halo masses of the galaxy.  

\begin{figure}[ht]
\includegraphics[width=0.7\columnwidth]{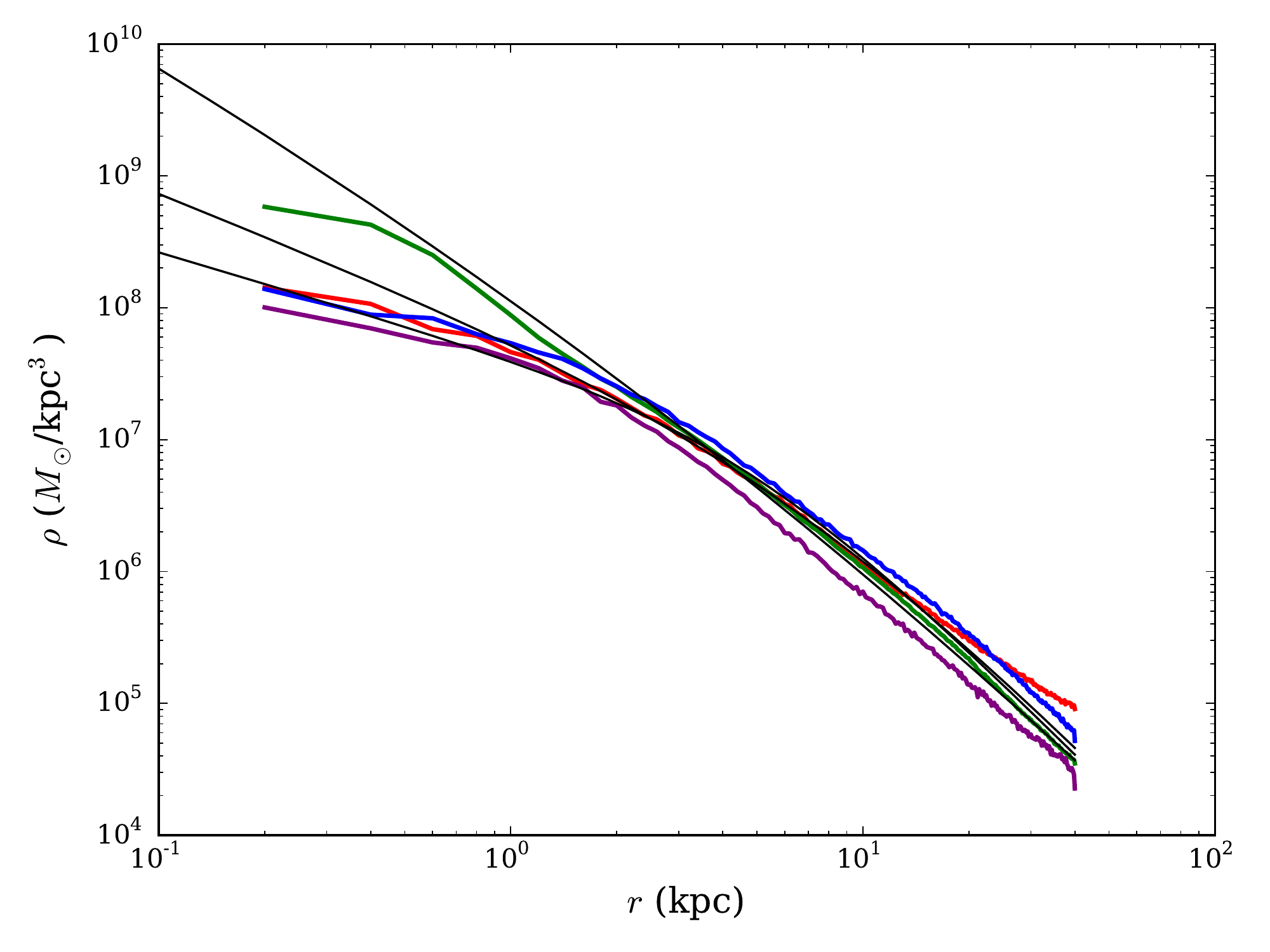}
\caption{Density profiles of the four LMC-mass cosmological simulations (red, blue, green, purple lines), and maximum, minimum, and average of the  fitted generalized NFW profiles with $(\alpha,\beta,\gamma)$ values derived from Ref.~\cite{diCintio2014b}, which extend to down to $r=0$ (solid black lines). \label{fig:envelope_simulations}}
\end{figure}

\begin{figure}[ht]
\includegraphics[width=0.7\columnwidth]{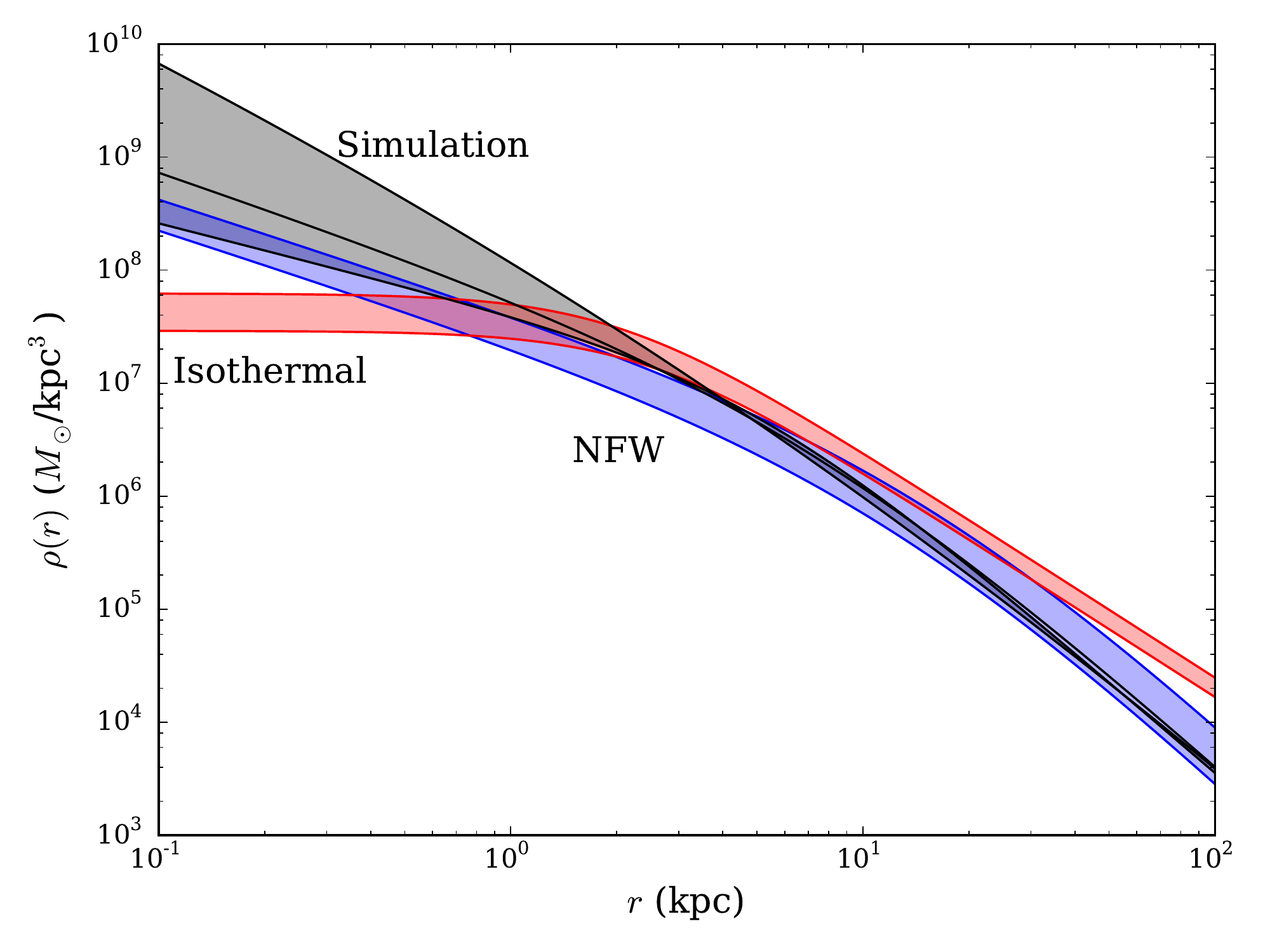}
\caption{Density profiles as a function of radius $r$ from the LMC center for the benchmark models (listed in Table~\ref{tab:profiles}). Maximum and minimum NFW (blue), isothermal (red), and range of simulated (black) profiles constitute the upper, lower edges of the shaded regions. The average simulated profile is shown as a line in the shaded black region.  \label{fig:rho}}
\end{figure}

We adopt a stellar mass of $2.7\times10^9~M_{\odot}$ from Ref.~\cite{vdM02}.  The allowed dark matter halo mass range of the LMC is uncertain by an order of magnitude, {\it e.g.}, $(3$ -- $25)\times 10^{10}~M_{\odot}$ \cite{K13}, 
and allows for the whole range of density profiles between isothermal and NFW.  To better constrain the stellar-to-halo mass ratio, we use a sample of cosmologically simulated galaxies from Ref.~\cite{Munshi2013} that has been shown to match the observed 
stellar-to-halo mass relation.  This sample was chosen to have halo masses in the range $(3$ -- $25)\times10^{10}~M_{\odot}$, stellar masses $\geq 10^9~M_{\odot}$, and logarithmic stellar-to-halo mass ratios ranging from $-1.2$ to $-1.7$.  We have adopted the ($\alpha,\beta,\gamma$) values for the extrema of these halos from Ref.~\cite{diCintio2014b}, which provide an ``envelope'' of typical  dark matter density profiles in an LMC--mass galaxy predicted by state-of-the-art cosmological simulations.
We take the average values of ($\alpha,\beta,\gamma$), defining the mean simulated profile. Figure~\ref{fig:envelope_simulations} shows the density profiles of the simulated galaxies, and the overlaid best-fit profiles. The resulting generalized NFW parameters of these three simulated profiles are shown in Table~\ref{tab:profiles}. In Figure~\ref{fig:rho}, we plot the density profiles $\rho(r)$ of our benchmark models: the two NFW and isothermal models, and our three generalized NFW profiles forming the range of results from simulation. 

In Figure~\ref{fig:LMC_vrot}, showing the rotation curve data to which the NFW and isothermal profile parameters were fit, we overlay the simulated profiles. Note that dark matter distributions drawn from simulations are not directly fit to the LMC data and are not corrected for inclination angle.

\subsection{\texorpdfstring{$J$}{J}-factors of the LMC}

We now have three different classes of dark matter profiles that span the measured rotation curves for the LMC. From our fits to the rotation curve data, we have profiles that maximize and minimize the LMC dark matter density, assuming both an NFW profile ({\tt nfw-max} and {\tt nfw-min}) and an isothermal profile ({\tt iso-max} and {\tt iso-min}). Fitting the measured stellar-to-halo mass ratio to simulation, we also have a range of profiles fit to a generalized NFW. In addition to the profile parameters consistent with the mass ratio that maximize and minimize the dark matter density consistent with simulation ({\tt sim-max} and {\tt sim-min}), we also include a profile that has the average values of the ($\alpha,\beta,\gamma$) parameters from the simulated galaxies ({\tt sim-mean}). As the $J$-factor depends on the integrated density profile squared, the maximum and minimum profiles within a specific class of profiles ({\it i.e.}, NFW, isothermal, or simulated) will also have the maximum or minimum $J$-factor within their class of halo profiles. Recall that, in order to be maximally conservative in our NFW and isothermal dark matter profiles, we at every opportunity maximized the baryonic contributions to the observed rotation curves, minimizing the assumed dark matter density.

\begin{figure}[ht]
\includegraphics[width=0.5\columnwidth]{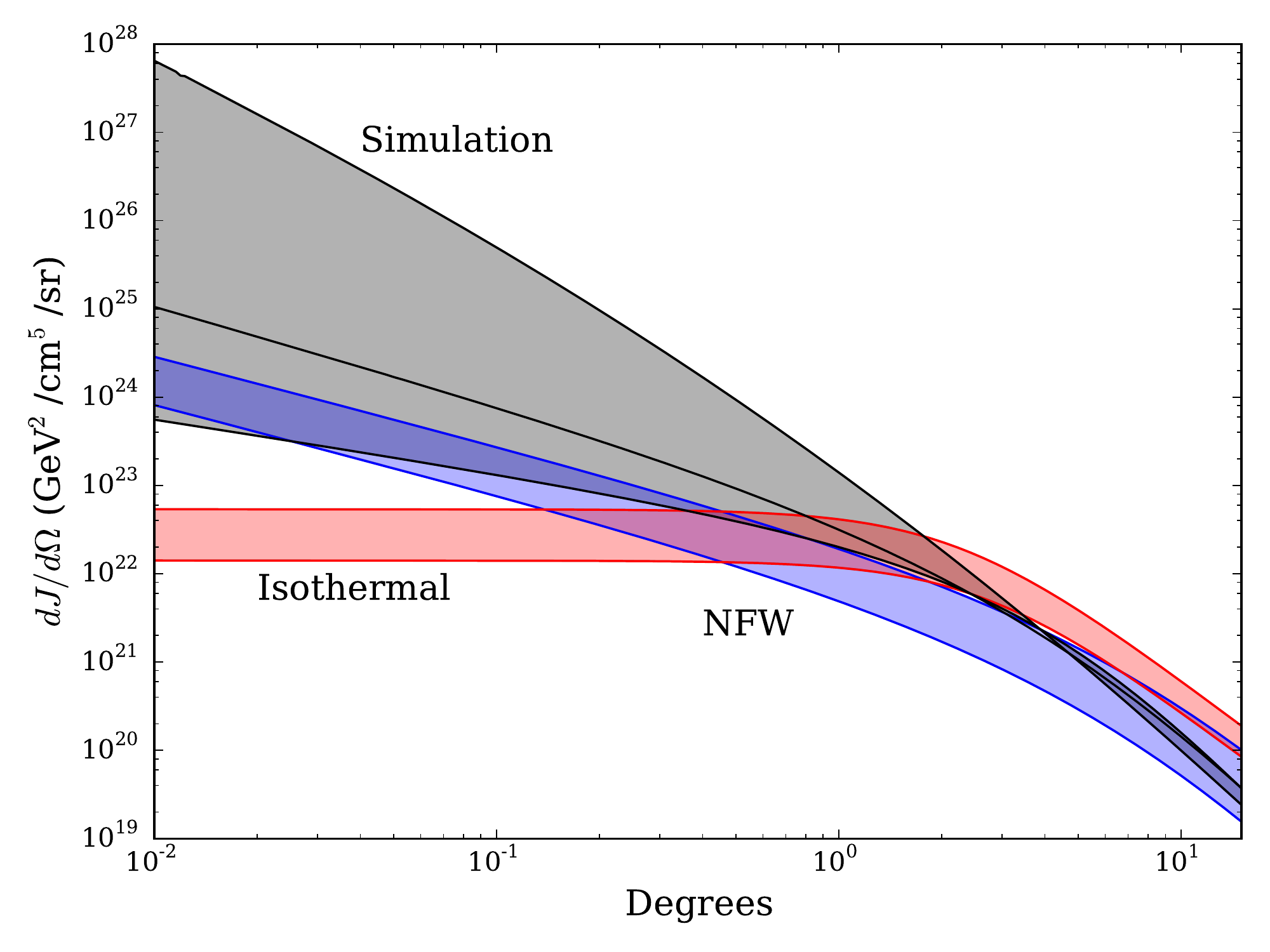}\includegraphics[width=0.5\columnwidth]{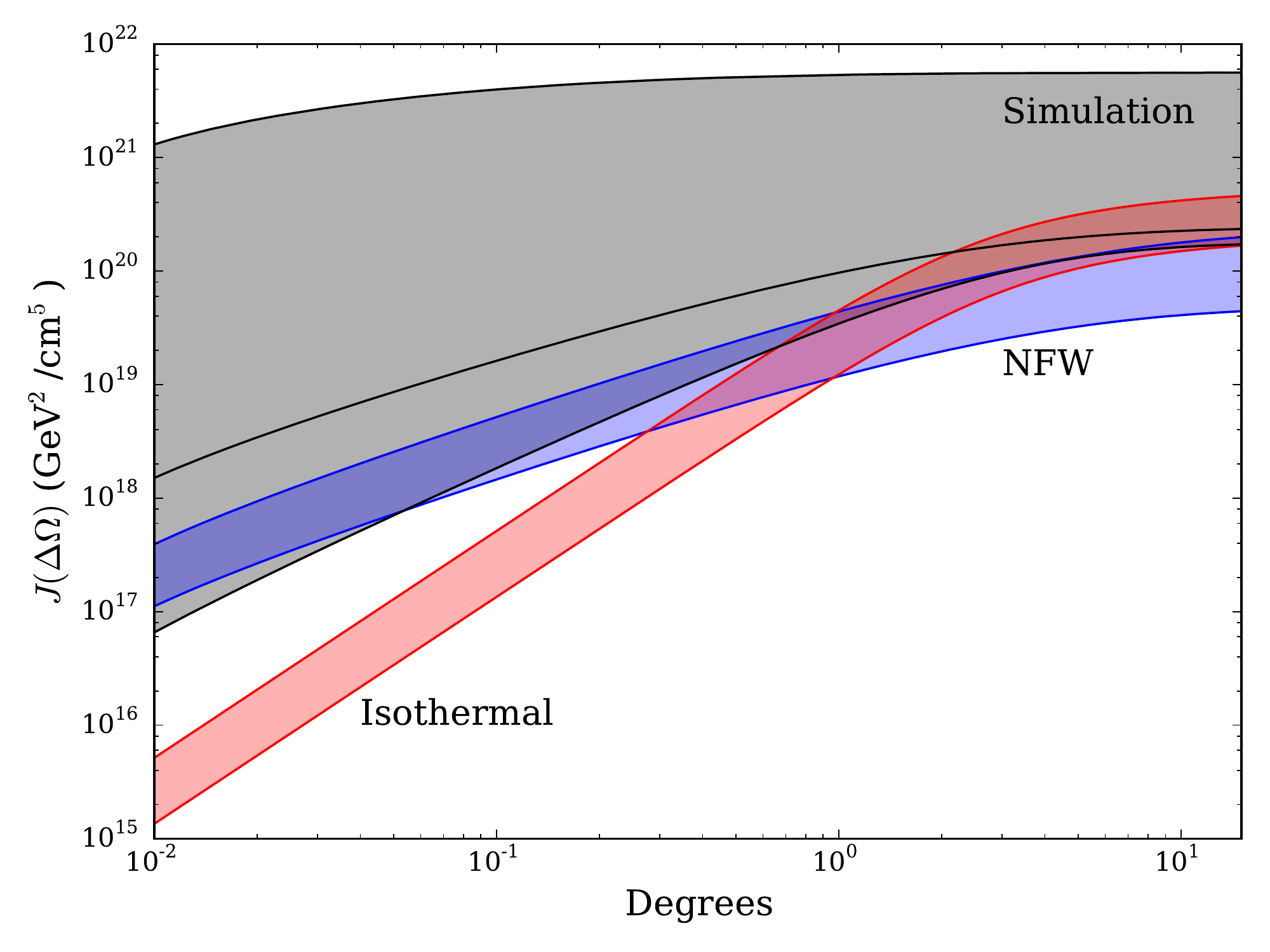}
\caption{Differential (left) and integrated (right) $J$-factors as a function of angle from the LMC center for the benchmark models (listed in Table~\ref{tab:profiles}). Labeling and color coding is as in Fig.~\ref{fig:rho}.}
\label{fig:J}
\end{figure}

We summarize these benchmark models in Table~\ref{tab:profiles}, including the integrated $J$-factor out to 15$^\circ$ (though we note that the majority of the contribution to the $J$-factor comes from the inner few degrees of the LMC), and the dark matter mass within 8.7~kpc.  The range of dark matter masses inferred from these fits is consistent with the observed total (dark matter plus baryon) mass of the LMC inside this radius $M(8.7~\mbox{kpc}) = (1.7\pm0.7)\times 10^{10}~M_\odot$ \cite{vanderMarel:2013jza}. In Figure~\ref{fig:J}, we plot the differential $J$-factor $dJ/d\Omega$ as a function of observation angle from the profile center, as well as the integrated $J$-factor. As can be seen, despite the range of profile choices available, the total $J$-factor of the LMC is remarkably consistent for six of our seven benchmarks, with $\log_{10} J/(\text{GeV}^2/\text{cm}^5) \sim 19.5-20.5$. For comparison, the most promising dwarf spheroidal galaxies have $\log_{10} J/(\text{GeV}^2/\text{cm}^5) \sim 19-19.5$ \cite{Martinez2013,Geringer-Sameth:2014yza}, while the Galactic Center within $1^\circ$ has $\log_{10} J/(\text{GeV}^2/\text{cm}^5) \gtrsim 21-24$ (depending on assumptions for the inner slope of the dark matter profile, see {\it e.g.}~\cite{Daylan:2014rsa}).

When setting bounds on dark matter annihilation, we will take the average for each of these classes of dark matter profiles. For the NFW and isothermal profiles, we take the geometric mean of the maximum and minimum profiles, and use the logarithmic difference of the maximum and minimum $J$-factors as an estimate of the $1\sigma$ uncertainty on the $J$-factor. We will refer to these two profiles as {\tt nfw-mean} and {\tt iso-mean}. For our generalized profiles taken from simulation, we will use the {\tt sim-mean} profile. 

Recall that the mean profile is obtained from the average generalized NFW parameters $(\alpha,\beta,\gamma)$ fit from simulation, rather than averaging the $J$-factors of the simulated profiles.  This distinction is important as the extreme generalized NFW profile {\tt sim-max} has a much higher $J$-factor than any other profile we consider, despite having a total dark matter mass that is consistent with the other benchmarks. This is because this profile has a very steep inner slope, and the annihilation is proportional to density squared. 

It is possible that future observations of the LMC can be used to reduce the uncertainties in our derivation of the LMC dark matter distribution. If the resulting profile is in the upper range of the generalized NFW envelope obtained from simulation, the LMC would set the best bounds on dark matter annihilation by far, compared to other targets. However, this extreme profile is an outlier that, while consistent with the total mass within 8.7~kpc, seems inconsistent with the rotation curve data.  Instead, we focus here on the conservative profiles profiles using the averaged {\tt nfw-mean}, {\tt iso-mean}, and {\tt sim-mean}, which still have $J$-factors that are larger than those of the ``best'' individual dwarf spheroidal galaxies. This gain is tempered by the higher baryonic backgrounds (discussed in detail in Section~\ref{sec:baryons}).     

\begin{table}[ht]
\begin{tabular}{cccccccc}
\hline\hline
Profile & $\alpha$ & $\beta$ & $\gamma$ & $r_S$~(kpc) & $\rho_0$~($M_\odot/\text{kpc}^3$) & $J$~(GeV$^2$/cm$^5$) & $M(8.7~\mbox{kpc})$~($M_\odot$) \\ \hline 
\tt{nfw-max} & 1 & 3 & 1 & 17.0 & $2.5 \times 10^6$ & $2.0\times 10^{20}$ & $1.1\times 10^{10}$ \\ 
\tt{nfw-mean} & 1 & 3 & 1 & 12.6 & $2.6\times 10^6$ & $9.4 \times 10^{19}$  & $7.7\times 10^9$ \\ 
\tt{nfw-min} & 1 & 3 & 1 & 12.6 & $1.8 \times 10^6$ & $4.4\times 10^{19}$  & $5.3 \times 10^9$ \\ 
\tt{iso-max} & 2 & 2 & 0 & 2.0 & $6.2 \times 10^7$ & $4.6\times 10^{20}$ & $2.0 \times 10^{10}$\\ 
\tt{iso-mean} & 2 & 2 & 0 & 2.4 & $3.7 \times 10^7$ & $2.8 \times 10^{20}$ & $1.5 \times 10^{10}$ \\ 
\tt{iso-min} & 2 & 2 & 0 & 2.4 & $2.9\times 10^7$ & $1.7\times 10^{20}$ & $1.2 \times 10^{10}$\\ 
\tt{sim-max} & 0.35 & 3 & 1.3 & 5.4 & $1.1\times 10^8$ & $5.6\times 10^{21}$ & $1.6\times 10^{10}$\\ 
{\tt sim-mean} & 0.96 & 2.85 & 1.05 & 7.2 & $8.4\times 10^6$ & $2.3\times 10^{20}$ & $1.4 \times 10^{10}$ \\ 
\tt{sim-min} & 1.56 & 2.69 & 0.79 & 4.9 & $1.2\times 10^7$ & $1.7\times 10^{20}$ & $1.3 \times 10^{10}$ \\ \hline \hline
\end{tabular}
\caption{Parameters of LMC benchmark profiles, along with derived quantities $J$ and the mass $M$ enclosed up to 8.7~kpc. $J$ is calculated out to $15^\circ$ (12.8~kpc). Average values for the isothermal and NFW $J$-factors are obtained from the geometric mean of the maximum and minimum profiles. \label{tab:profiles}}
\end{table}

\subsection{Gamma-Ray Spectrum}

As the particle physics of dark matter is as yet unknown, we do not know the mass or the final state products of the annihilation of dark matter. However, if dark matter annihilates into a pair of Standard Model particles other than neutrinos, be it $W/Z$ gauge bosons, gluons, quarks, or charged leptons, then (with the exception of the stable $e^\pm$), those particles must decay or hadronize. This leads to a cascade of Standard Model particles, decaying down to electrons, protons, their antipartners, and a large multiplicity of photons with gamma-ray energies. Photons are also emitted as final state radiation from the charged particles, including $e^+e^-$ pairs.

As a result of this cascade, the gamma rays from dark matter annihilation do not feature a sharp line at $E_\gamma = m_\chi$, but rather a continuous spectrum with characteristic energies significantly lower than the dark matter mass. Indeed, in this analysis we do not perform a line-search for dark matter annihilating directly into photons. The annihilation channels we consider in this work are
\begin{equation}
\chi\chi \to s\bar{s},~b\bar{b},~t\bar{t},~gg,~W^-W^+,~e^+e^-,~\mu^+\mu^-,~\mbox{and}~\tau^-\tau^+. \label{eq:channels}
\end{equation}
Annihilation into pairs of $u$ or $d$ quarks produces a similar spectrum as annihilation into gluon pairs, $c\bar{c}$ is similar to $s\bar{s}$, as are $ZZ$ and $W^-W^+$, so bounds on such channels can be roughly extrapolated from the subset of channels we analyze in detail. We scan over all dark matter masses between 5~GeV and 10~TeV. 
Channels of dark matter annihilating to massive particles are only open above the mass threshold, when the dark matter mass is equal to that of the heavy Standard Model particle in the final state.

\begin{figure}[ht]
\includegraphics[width=0.5\columnwidth]{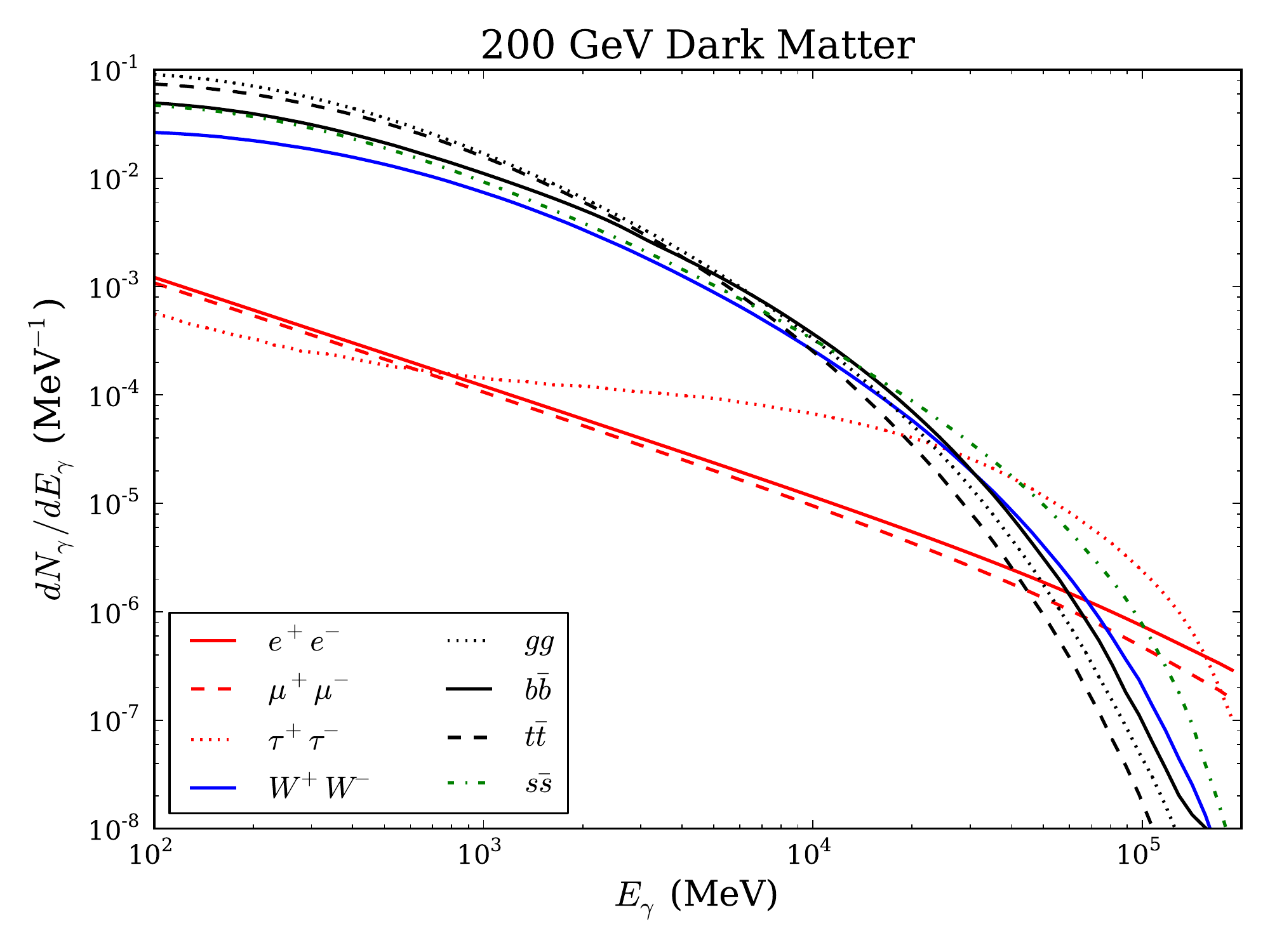}\includegraphics[width=0.5\columnwidth]{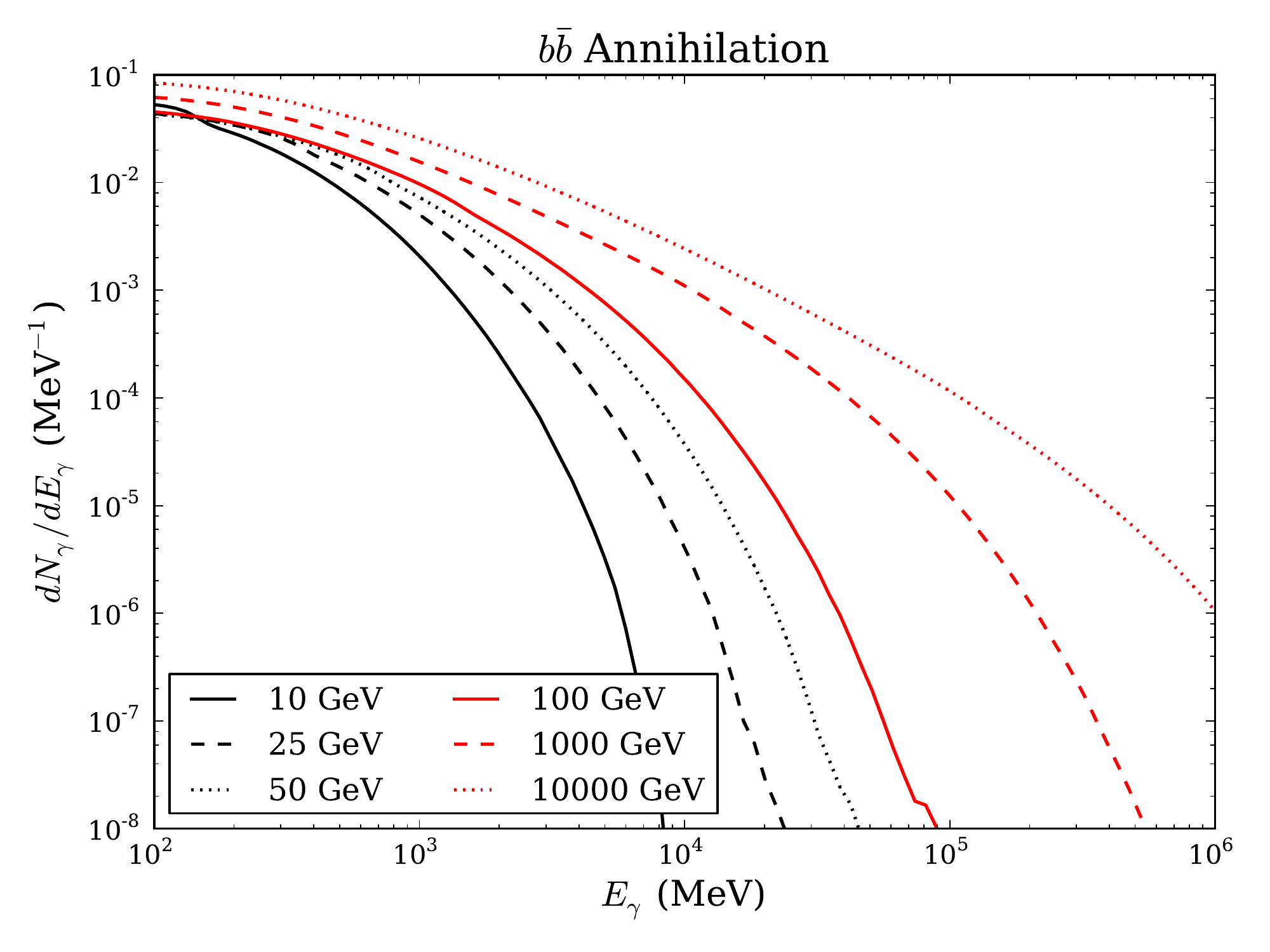}
\includegraphics[width=0.5\columnwidth]{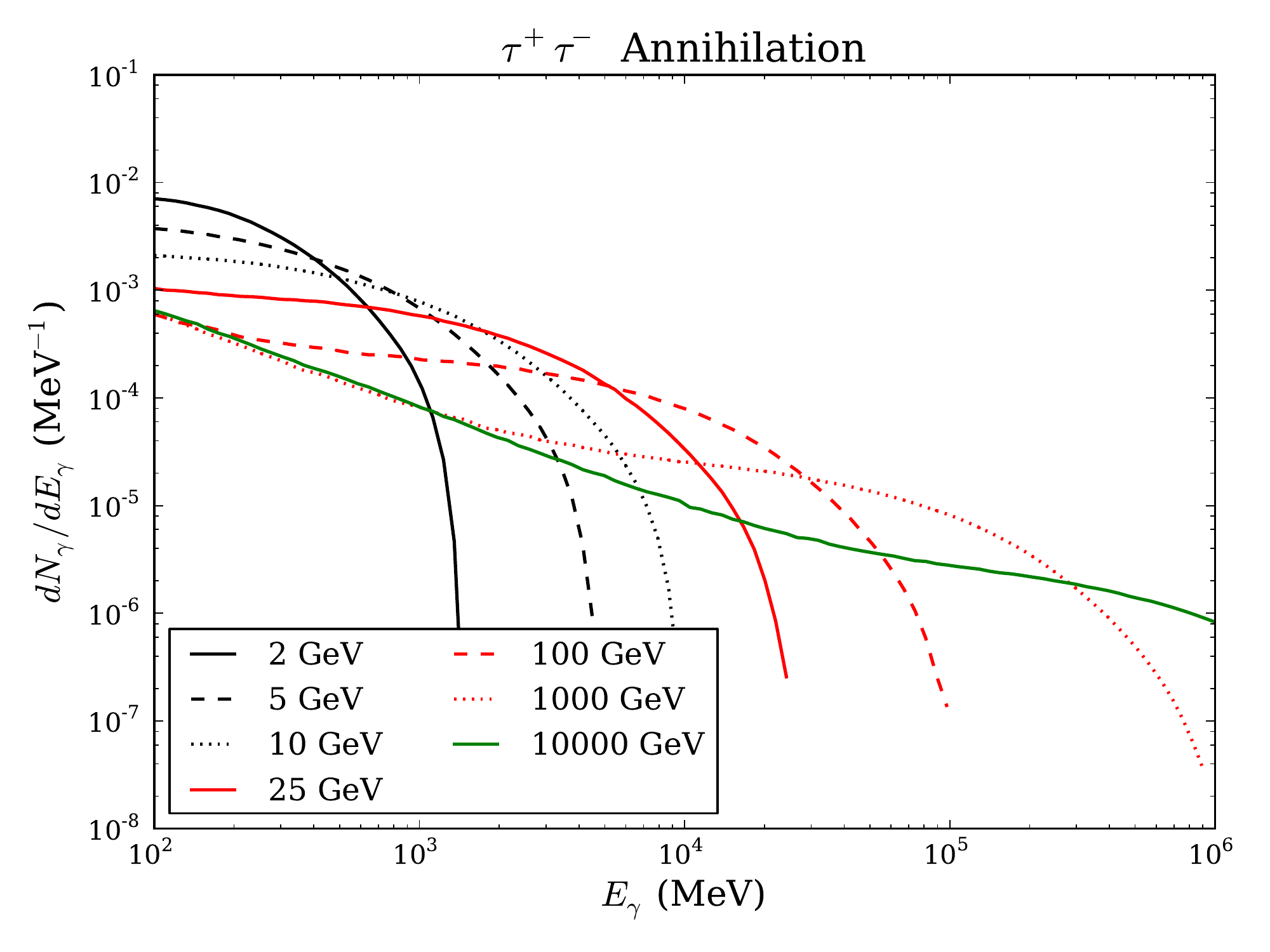}\includegraphics[width=0.5\columnwidth]{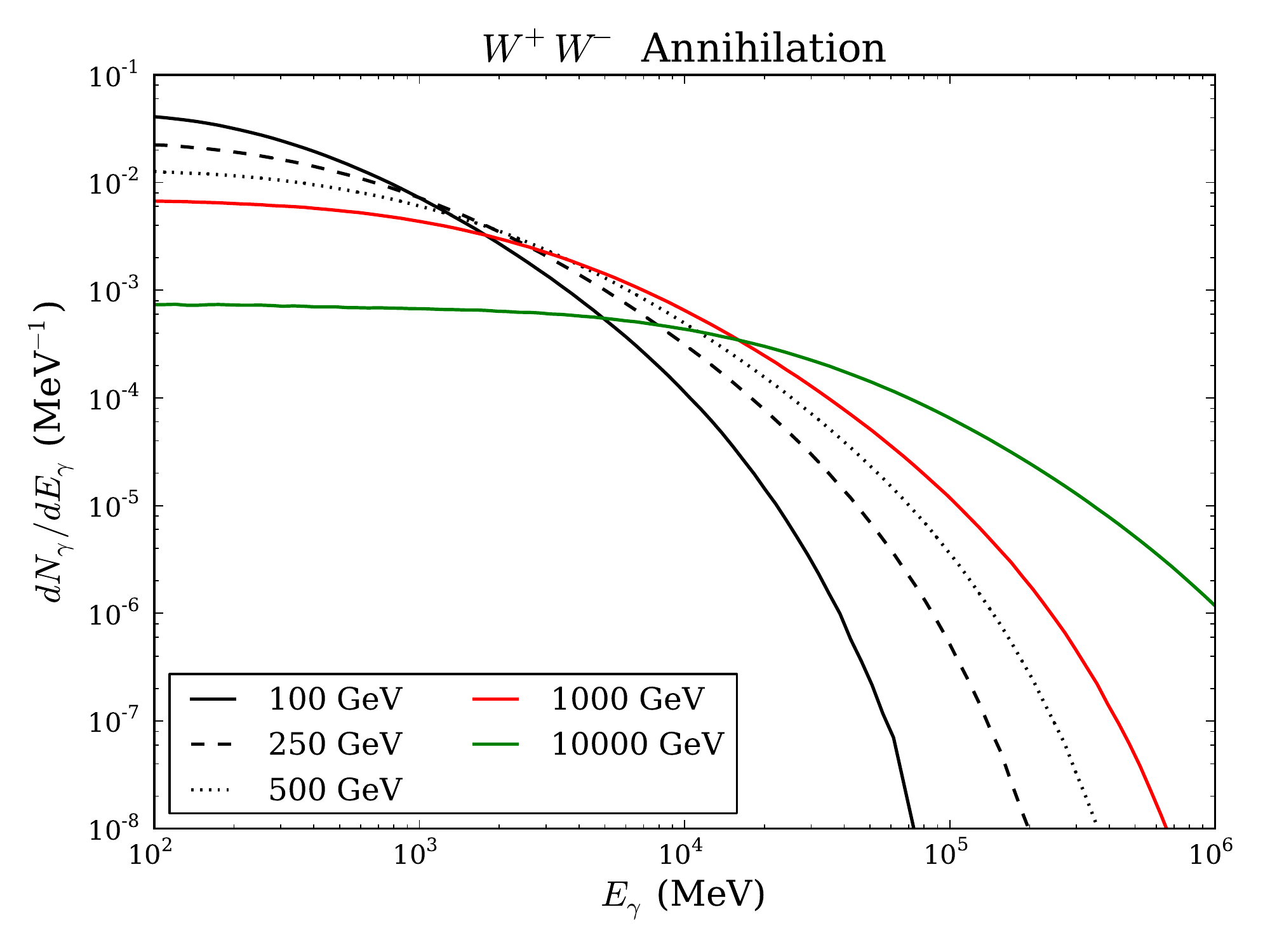}

\caption{Gamma-ray spectra $dN_\gamma/dE_\gamma$ of dark matter pair annihilation. Upper left: Annihilation spectra of 200~GeV dark matter into each of the channels we consider in this work. Upper right: Annihilation spectrum into $b\bar{b}$ for a range of dark matter masses. Lower left: Annihilation spectrum into $\tau^+\tau^-$ for a range of dark matter masses. Lower right: Annihilation spectrum into $W^+W^-$ for a range of dark matter masses.  \label{fig:annihilation_spectra}}
\end{figure}

For each final state, we calculate the resulting spectrum of gamma rays as a function of dark matter mass using code available as part of the {\it Fermi} LAT {\em ScienceTools}.\footnote{The {\tt DMFitFuction} spectral model described at: \url{http://fermi.gsfc.nasa.gov/ssc/data/analysis/documentation/Cicerone/Cicerone_Likelihood/Model_Selection.html}, see also Ref.~\cite{Jeltema:2008hf}.  We note that this formulation does not include electroweak corrections \cite{2007PhRvD..76f3516K,2002PhRvL..89q1802B,2009PhRvD..80l3533K,2010PhRvD..82d3512C,Ciafaloni:2010ti}.   The electroweak corrections are expected to be important
(assuming s-wave annihilation) when the dark matter mass is much heavier than 1 TeV, and would alter the spectra substantially for the $W^+W^-$, $e^+e^-$, $\mu^+\mu^-$ and $\tau^+\tau^-$ channels, increasing the number of expected $\gamma$ 
rays per dark matter annihilation below $\sim 10$~GeV~\cite{Ciafaloni:2010ti,Cirelli:2010xx}.  
However, the bounds in the high mass regime come primarily from the highest energy bins.  
Even for 10~TeV dark matter masses in the most affected channels, 
including the electroweak corrections improves the limits on $\langle \sigma v
\rangle$ by $\lesssim 20$\%.}
In Figure~\ref{fig:annihilation_spectra}, we show representative spectra $dN/dE_\gamma$ per pair annihilation for a range of channels and dark matter masses. 

\section{Baryonic Backgrounds}\label{sec:baryons}

The gamma-ray emission from the LMC was first detected by the EGRET instrument aboard the {\it Compton Gamma Ray Observatory} \cite{1993ApJS...86..629T,1999ApJS..123..203E}, operating from 1991 to 2000.~\cite{1992ApJ...400L..67S}. The LMC was established as an extended source, but the limited angular resolution of EGRET prevented a deep investigation of the origin and composition of the high-energy emission. With 
more than an order of magnitude improvement in sensitivity, better angular resolution, and extended energy coverage compared to its predecessor, the {\it Fermi} LAT instrument enabled a strong detection of the LMC early in the mission. From 11 months of continuous all sky-survey observations,~\cite{2010A&A...512A...7A} reported a detection of the LMC with formal significance $\sim 33\sigma$ in $\sim$100\,MeV--10\,GeV gamma rays and confirmed the extended nature of the source. The emission is relatively strong in the direction of the 30 Doradus star-forming region, but more generally the emission seems spatially correlated to classical tracers of star formation activity (such as the H$\alpha$ emission).
The extension and spectrum of the source suggest that the observed gamma rays originate from CRs interacting with the interstellar medium through inverse-Compton scattering, bremsstrahlung, and hadronic interactions. Yet, contributions from discrete objects such as pulsars could not be (and were not) ruled out at that time.

Compared to this early work, we now utilize five years of LAT data. These data are of better quality than the initial data set, thanks to improvements in the instrument calibration, event reconstruction, and background rejection ({\it i.e.},~Pass 7 reprocessed data instead of Pass 6 data).\footnote{\url{http://fermi.gsfc.nasa.gov/ssc/data/analysis/documentation/Pass7REP_usage.html}} Recently, a new analysis of the high-energy gamma-ray emission of the LMC was performed using 5.5 years of Pass 7 reprocessed LAT data, which resulted in a more accurate description of the source. This new effort will be presented in detail elsewhere.\footnote{{\it Fermi} LAT Collaboration, in preparation, see also \url{http://fermi.gsfc.nasa.gov/science/mtgs/symposia/2014/program/05_Martin.pdf}}  The present work is based on an intermediate version of the diffuse emission model from that work, with only very minor differences compared to the final model described in the upcoming paper.  We briefly summarize here the main features of the emission model and the approach followed to derive it. This is of prime importance to understand the possible limitations and systematic effects that may affect the search for dark matter signals on top of this astrophysical background. 

A region of interest (ROI) specific to the LMC was defined as a $10^\circ \times 10^\circ$ square centered on 
$(\mbox{RA},\mbox{DEC})=(80\fdg 894,-69\fdg 756)$ 
and aligned on equatorial coordinates (J2000.0 epoch). The energy range considered in that analysis was 200~MeV--50~GeV and counts were binned in six logarithmic bins per decade. The lower energy bound was dictated by the poor angular resolution at the lowest energies, while the upper bound was imposed by the limited statistics at the highest energies.   The data-set used to build the background model largely overlaps with (but is not identical to) the data-set we use in the remainder of the paper to perform our search for dark matter.

The emission model is built from a fitting procedure using a maximum likelihood approach for binned data and Poisson statistics. A given model is composed of several components, accounting for different sources in the field. Each component has a spatial description, a spectral description, and a certain number of free parameters.   The expected distribution of counts in energy and across the ROI is obtained by convolution of the model with the point-spread function (PSF), taking into account the exposure achieved for the data set. The free parameters of the model are then adjusted until the distribution of expected counts provides the highest likelihood given the actual binned spatial-energy cube of observed counts.

As a first step in the process of modeling the emission over the ROI, and before developing a model for the LMC, we have to account for known background and foreground emission, in the form of diffuse and/or isolated sources. The base model is composed of the isotropic contribution (extragalactic emission and residual charged-particle background misclassified as gamma rays) the Galactic diffuse model (from CRs interacting with the interstellar medium in our Galaxy)\footnote{The diffuse background models are available at \url{http://fermi.gsfc.nasa.gov/ssc/data/access/lat/BackgroundModels.html} as {\tt iso\_clean\_v05.txt} and {\tt gll\_iem\_v05.fits}.}, and all objects listed in the second {\it Fermi} LAT source catalog \citep{2012ApJS..199...31N} within the ROI but outside the LMC boundaries (including sources as far as 2$^\circ$ away from the edges of the ROI to account for spill-over effects due to the poor angular resolution at low energies). 

Starting from this base model, we aim to describe the remaining emission with a combination of point-like sources and extended spatial intensity distributions, adding new components successively. Point-like sources can easily be found if they have hard spectra, because the angular resolution is relatively good at high energies, or if they exhibit a variability pattern reminiscent of an already-known object. In the case of the LMC, three new point sources were recognized in this way.\footnote{In the recently released third {\it Fermi} source catalog (3FGL) produced with four years of data these sources were not individually detected but rather absorbed into the extended LMC source, 3FGL J0526.6$-$6825e~\cite{REF:2015.3FGL}.}  For the rest of the emission, an iterative procedure is required to develop the model.

At each step, a scan over position in the LMC and size of the source is performed to identify the new component that provides the best fit to the data. For each trial position and size, a fit is performed assuming a power-law spectral shape for the new component (which is a  good approximation for most components). If the improvement in the likelihood is significant -- that is, has a log-likelihood test statistic (TS, see Section~\ref{sec:stats}) greater than 25 -- then the component is added to the model and a new iteration starts. The process stops when adding a new component yields a TS lower than 25. At the end of this process, a nearly complete model is obtained. Next, we again optimize the positions and sizes of the extended components within this nearly complete model, from the brightest to the faintest in turn. The final stage consists of deriving bin-by-bin spectra for all components to check that the initially adopted power-law spectral shape is appropriate. If not, it is replaced by a power law with exponential cutoff or a log-parabola shape, depending on which provides the best fit and a significant improvement relative to the power law.

In the case of the LMC, the best model was obtained under the assumption that extended emission arises from large-scale populations of CRs interacting with the interstellar medium. In the $\sim$100~MeV to 100~GeV range, the interstellar radiation is dominated by gas-related processes, especially hadronic interactions in which CR nuclei interact with interstellar gas to produce mesons that decay into gamma rays. The corresponding gamma-ray emission follows the gas distribution (see, for instance, Ref.~\cite{2012ApJ...750....3A}). For the LMC, we therefore modeled each extended emission component as the product of the gas column density distribution with a two-dimensional Gaussian emissivity distribution whose position and size were iteratively optimized.  One advantage of this assumption is that the model retains the small-scale structure that the gamma-ray emission may have.

\begin{figure}[ht]
\includegraphics[width=0.5\columnwidth]{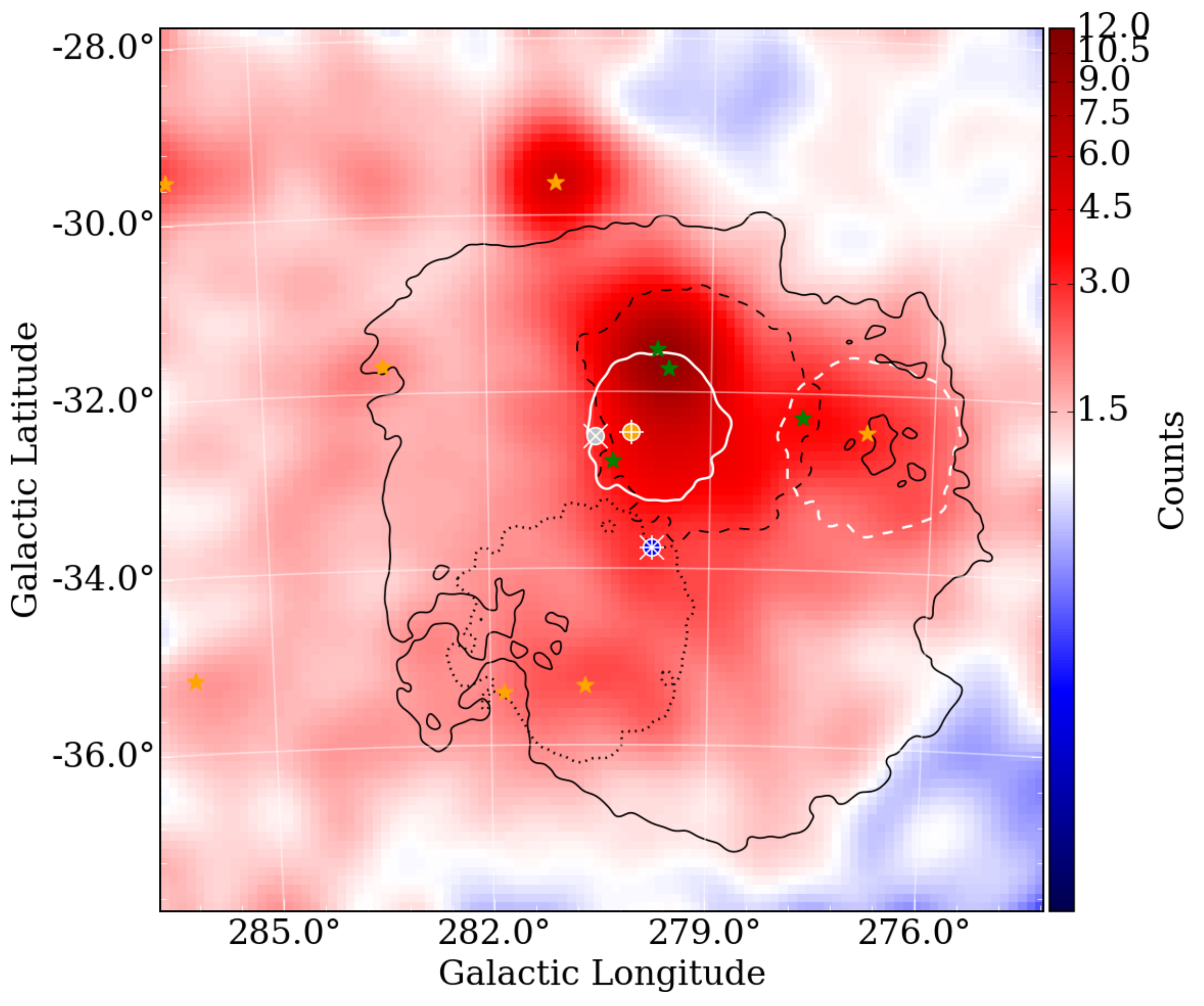}\includegraphics[width=0.5\columnwidth]{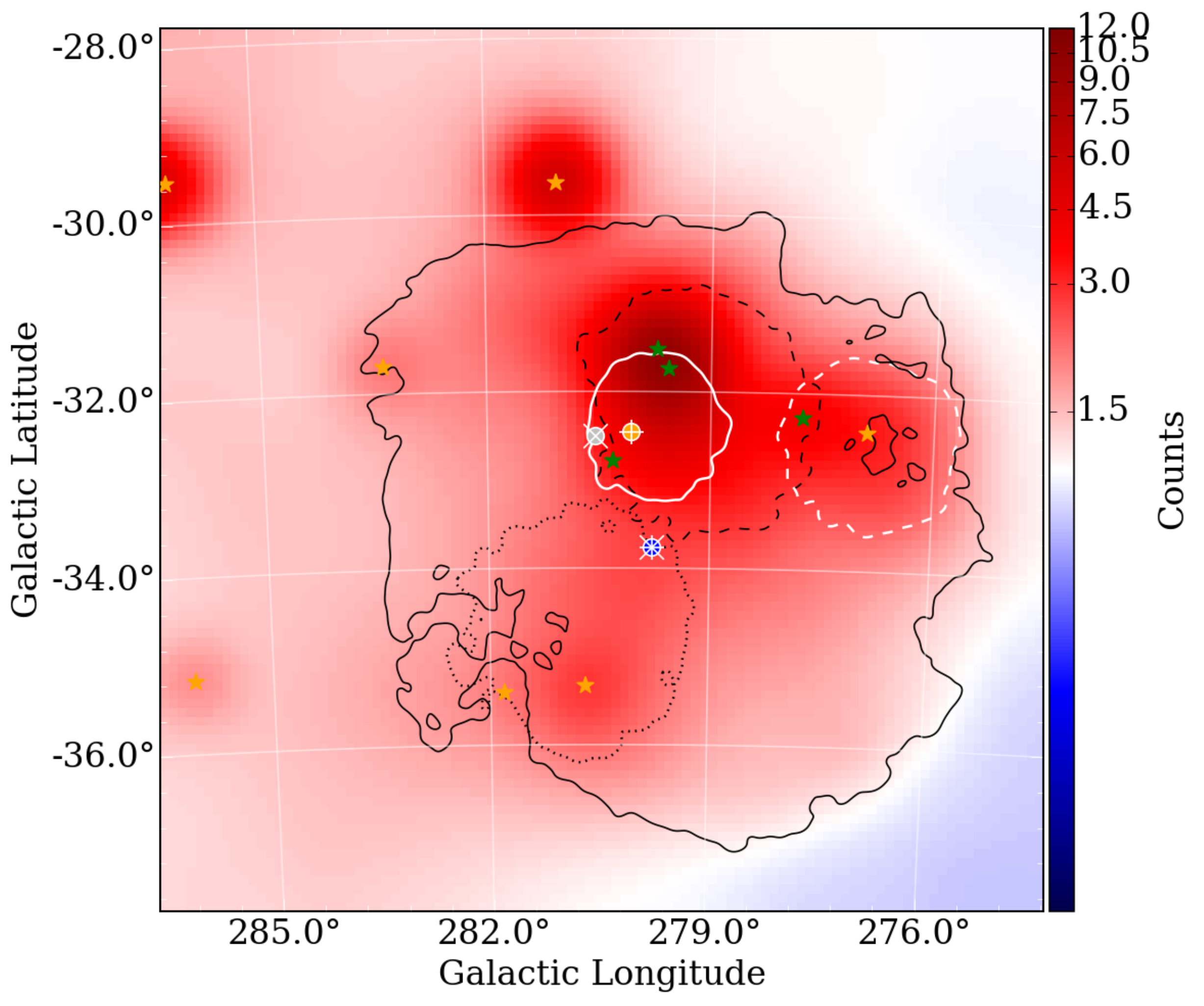}
\caption{Left: Counts map of the LMC region, in the energy range from 792\,MeV to 12.6\,GeV. Right: Model map of the same region and for the same energy range created from the emission model (see text for details).   Both maps are binned in $0\fdg 1 \times 0\fdg 1$ pixels and smoothed with a $\sigma=0\fdg 3$ Gaussian kernel.  The possible locations of the LMC center (Tab.~\ref{tab:centers}) are shown: {\tt stellar} (white circle with $\times$ cross), {\tt outer} (orange circle with $+$ cross), and {\tt HI} (blue circle with $\rlap{+}{\times}$ cross). Smoothed contours of extended components of the background emission model are also shown: {\tt E0} (solid black lines), {\tt E1} (dashed black), {\tt E2} (white dashed), {\tt E3} (white solid), and {\tt E4} (black dotted);  the contours are drawn at 2\% of the peak level for each of the extended sources.  Green stars mark the point-like objects PS1 to PS4 in our background emission model, orange stars are point sources in the 2$^{\rm nd}$ {\it Fermi}-LAT point source catalog.  Recall that the extended emission sources are correlated with the gas column density, resulting in the irregular shapes.  The effective angular resolution can be inferred from the distribution of counts around the point-like sources.   Galactic diffuse emission is visible outside of the LMC region.  
\label{fig:background_map}}
\end{figure}

This model-building procedure resulted in an emission model with nine components: four point-like objects and five extended components. The former are denoted {\tt PS1}, {\tt PS2}, {\tt PS3}, and {\tt PS4}, while we call the latter {\tt E0}, {\tt E1}, {\tt E2}, {\tt E3}, and {\tt E4}. The corresponding full model map is compared to the counts map in Figure~\ref{fig:background_map}, where the layout of the various emission components is overlaid.

One point should be emphasized. By design, this iterative building of a model for the LMC aims to account for any emission component, point-like or extended. Therefore, should any dark matter signal be present in the data, part or all of it may be absorbed in one or more of the above mentioned (extended) components. A large part of our efforts in our treatment of the statistical and systematic errors (Section~\ref{sec:stats}) will focus on placing conservative bounds in just this case. Fortunately, the expected dark matter distributions presented in the previous section seem to differ notably from the standard astrophysical background presented above. Additionally, the specific dark matter signal spectra differ from the typical spectra we inferred for the various emission components. Nevertheless, this possible bias should be kept in mind and will be discussed in detail.

\section{LAT INSTRUMENT AND DATA SELECTION}\label{sec:data}

The {\it Fermi} LAT is a pair-conversion telescope: incoming gamma rays convert to $e^{+}e^{-}$ 
pairs that are tracked in the instrument.   The data analysis is
event based; the energies and directions of the incoming gamma rays
are estimated from the tracks and energy depositions of the pair in
the LAT.    Detailed descriptions of the LAT and of its performance can
be found elsewhere~\cite{Atwood:2009ez,REF:2009.OnOrbitCalib,REF:2012.P7Perf}.

For the analysis of a complicated region such as the LMC, the
PSF is crucial for resolving the 
contributions from different spatial components.  The 68\% containment radius of the PSF
($R_{\rm 68}$) averaged over the LAT field-of-view is $\sim 1^{\circ}$
($\sim 1\fdg 8$) at 500~MeV for
events that convert in the front (back) of the LAT tracking volume.

For our data sets we use the \irf{P7REP\_CLEAN}\ event selection (``Pass
7 Reprocessed'' data) on data taken between 
2008 August 4, and 2013 August 4 by the {\it Fermi} LAT.  We chose to use the 
stringent \irf{P7REP\_CLEAN}\ event selection since it has low residual 
CR contamination compared to the gamma-ray flux.  We used the 
\texttt{P7REP\_CLEAN\_V15} version of the instrument response functions (IRFs).
The data reduction and exposure calculations were performed using the
{\it Fermi} LAT \texttt{ScienceTools} version 09-34-00.\footnote{\url{http://fermi.gsfc.nasa.gov/ssc/data/analysis/software/}} 

We used events with reconstructed energies from 500~MeV to 500~GeV. 
Depending on the dark matter annihilation channel, this gives the 
analysis reasonable sensitivity for dark matter particles masses down to $\sim 2$~GeV.    
Extending the analysis to lower energies would introduce significant
complications because of the increasing width of the PSF.
We only use events with a measured zenith angle less than 100$^{\circ}$ to 
remove the emission from the Earth's limb ({\it i.e.}, gamma rays from CR 
interactions in the upper atmosphere).  We also apply the standard
selection criteria for good time intervals\footnote{To date the only time intervals marked as having poor quality
  data are during bright Solar flares, when extremely high X-ray
  fluxes saturated the LAT anticoincidence detector;  see Appendix
  A of \cite{REF:2012.P7Perf} for more details.} and to remove data taken in non-standard 
operating and observing modes.  Note that the adopted rocking angle cut is only
applicable to data taken prior to 2013 December 6 when the LAT
observing strategy
changed,\footnote{\url{http://fermi.gsfc.nasa.gov/ssc/proposals/alt_obs/obs_modes.html}}
which is the case for our data set.  

The details about the data selection criteria are summarized in
Table~\ref{tab:data}.   This data selection very similar, but not
identical, to the selection used to build the background model
described in Section~\ref{sec:baryons}.   Both selections include the 
entire LMC.
 In Figure~\ref{fig:background_map}, we show a
map of the gamma rays collected in the LMC ROI, along with the
identified baryonic backgrounds of Section~\ref{sec:baryons} and the three positions considered to be potentially the kinematic center of the LMC (Table~\ref{tab:centers}). This counts map shows gamma rays in the energy range from 792~MeV to 12.6~GeV, which covers 13 of the 30 logarithmically spaced energy bins used in our analysis.

\begin{table}[ht]
  \centering
  \begin{tabular}{lc}
    \hline\hline
  Selection & Criteria \\
  \hline
  Observation Period &  2008 Aug. 4 to 2013 Aug.\ 4 \\
  Mission Elapsed Time (s)\footnote{\Fermi\ Mission Elapsed Time is defined as seconds since 2001 January 1, 00:00:00 UTC.} & 239557414 to 397345414 \\
  Energy range (GeV) & 0.5 to 500 \\
  Fit Region &  $10^{\circ}\times10^{\circ}$ centered on $(\ell,b) =
  (277\fdg 86,-32\fdg 41)$ \\
  Zenith range (deg) & $\theta_{\rm z} < 100$ \\
  Rocking angle range (deg)\footnote{Applied by selecting on 
    \texorpdfstring{\texttt{ROCK\_ANGLE}}{ROCK_ANGLE} 
    with the {\emph{gtmktime ScienceTool}}} & $|\theta_{\rm r}| < 52$ \\
  Data quality cut\footnote{Standard data quality selection:
    \texorpdfstring{\texttt{DATA\_QUAL == 1 \&\& LAT\_CONFIG ==
        1}}{DATA_QUAL == 1 \&\& LAT_CONFIG == 1}  
    with the {\emph{gtmktime ScienceTool}}.} & yes \\
  \hline\hline
  \end{tabular}
  \caption{\label{tab:data}Summary table of {\it Fermi} LAT data selection criteria used for this paper's analysis.}
\end{table}

\section{Statistical Techniques}\label{sec:stats}

The parameters of the dark matter model space over which we must search are the coordinates of the
center of the dark matter distribution ($l_{\rm DM},b_{\rm DM}$), the
parameters of the dark matter radial profile ({\it i.e.}, $\alpha$, $\beta$, $\gamma$, $r_S$, $\rho_0$, and $r_{\rm max}$), the final states for the dark matter annihilation, and the mass of the dark matter particle
$m_{\chi}$. For each channel, our goal is to set an upper limit on the cross section $\langle \sigma v\rangle$, or, if a statistically significant excess is seen over the background model, determine the maximally likely dark matter mass and annihilation channel that fits the observation. As stated in Section~\ref{sec:dm}, we parametrized the six-dimensional dark matter profile fit to the LMC via three different classes of profiles: NFW, isothermal, and generalized NFW profiles fit to simulation.   As we have argued, these profiles cover the realistic (though conservative) range of possible dark matter distributions in the LMC, though reducing astrophysical uncertainties would help to further constrain the dark matter annihilation profile. 
From our fits to the LMC rotation curves, as well as extracted results from $N$-body simulation (Section~\ref{sec:dm}), the three benchmark profiles we will use to extract constraints on annihilation in this work are the averaged results of fits to these three types of dark matter profiles: the NFW {\tt nfw-mean}, the isothermal {\tt iso-mean}, and generalized NFW fit to simulation {\tt sim-mean}.

While the baryonic backgrounds are well fit by our gamma-ray models, these models are empirically fit to a LAT data set that overlaps the set from which our bounds will be drawn. Additionally, despite the good fit of the models, we cannot expect that they perfectly predict the observed gamma rays in our ROI. As a result we must control for systematic uncertainties in addition to the statistical errors as we fit our dark matter models to data, and account for the possibility that the baryonic backgrounds are hiding a potential dark matter signal. We will address these systematic issues later in this section. We first discuss the statistical methods we use to constrain the dark matter annihilation rate. 

\subsection{Fitting method}\label{subsec:fitting_method}

We use a multi-step likelihood fitting procedure that has previously been applied to LAT searches for dark matter signals in dwarf spheroidal galaxies~\cite{Ackermann:2013yva} and the Smith high-velocity cloud~\cite{Drlica-Wagner:2014yca}. This approach requires us to assume a spatial distribution of the dark matter component.  In principle, one could perform a search for an assumed spectrum while remaining agnostic regarding the spatial distribution, but given the comparative theoretical uncertainties, it is more sensible to restrict the dark matter profile to our limited range of possibilities, and allow the spectrum of gamma-ray annihilation for each spatial distribution to be fit using the procedure described below.   Thus, our approach is to assume a dark matter profile and location, and use the procedure described below to fit for the cross section in each dark matter mass and annihilation channel.  We then scan over possible values for the profile parameters and center locations.  

\subsubsection{Broadband Fitting}\label{subsec:broadband_fitting}

For the first step of the fitting procedure we use the standard LAT binned Poisson likelihood, defined as
\begin{equation}\label{eq:binned_likelihood}
  \mathcal{L}({\boldsymbol \mu},{\boldsymbol \theta} | \mathcal{D}) = \prod_{k}
  \frac{\lambda_{k}^{n_{k}}e^{-\lambda_{k}}}{n_{k}!},
\end{equation}
which depends on the gamma-ray data $\mathcal{D}$, signal
parameters ${\boldsymbol \mu}$, and nuisance ({\it i.e.}, background) parameters ${\boldsymbol \theta}$. The number of observed counts in each energy and spatial bin, indexed by $k$, depends on the data $n_{k}(\mathcal{D})$, while the model-predicted counts depend on the input parameters
$\lambda_{k}({\boldsymbol \mu},{\boldsymbol \theta})$.   The likelihood function includes information about the observed counts, 
instrument response, exposure and model components.  The nuisance parameters are the scaling coefficients and spectral 
indices of the identified baryonic backgrounds of Section~\ref{sec:baryons}.  

In principle the signal parameters ${\boldsymbol \mu}$ are, as previously stated, the dark matter profile parameters ($\rho_0$, $r_s$, $\alpha$, $\beta$, and $\gamma$), the coordinates of the dark matter profile center, the annihilation channel, the dark matter mass, and the total annihilation cross section $\langle \sigma v\rangle$.   However, as discussed, the dark matter profile parameters are reduced to those of our three benchmark models. This likelihood function is evaluated by fitting source spectra across all energy
bins simultaneously, and is thus necessarily dependent on the spectral model assumed for the source of interest.  Specifically, each choice of dark matter mass and channel results in a different spectrum of gamma rays.  However, performing a likelihood fit for each of these dark matter parameters would be inefficient.  Instead, in the first ``broadband'' fitting step of the analysis, we model the 
spectral form of the dark matter component as a power law with index $\Gamma = 2$ and fit only for the normalization of the dark matter and background components.   The purpose of this broadband fit is to establish baselines for
the background components, not to derive estimates for the dark matter contribution.   The dark matter component is included in the broadband
fit to reduce the potential bias on the background components, in the case a signal is present. For the analysis of dwarf spheroidals 
omitting the dark matter component entirely from the broadband fitting resulted in a change of $< 1$\% in the background parameters \cite{Ackermann:2013yva}.   However, 
because of the stronger coupling between the dark matter spatial models and the background components in the LMC, omitting the dark matter component from 
the broadband fits in the case of the LMC biases the background model parameters and reduces the coverage of the upper limits in simulated realizations of the analysis 
with injected signals from 95\% to $\sim 80$\%.   Modeling the dark matter component as a power law with $\Gamma =2$ at this stage of the analysis 
is sufficient to reduce that potential bias to negligible levels and to produce the correct coverage in simulated realizations, as we will demonstrate in Sec.~\ref{sec:systematic_errors}.
We then take into account the spectral shape of each annihilation channel in the second step of the analysis.

\subsubsection{Bin-by-bin Fitting}\label{subsec:bin_by_bin_fitting}

In the second step of the procedure, rather than refitting for each dark matter spectrum ({\it i.e.}, 
for each choice of mass and annihilation channel), 
we mitigate the spectral dependence by independently fitting a spectral model in each energy bin
$j$, to create a spectral energy distribution for a source of an assumed spatial morphology.    
This expands the global parameters ${\boldsymbol \mu}$
and ${\boldsymbol \theta}$ into sets of independent parameters for each energy bin $j$: ${\boldsymbol
\mu_{j}}$ and ${\boldsymbol \theta_{j}}$.   

Likewise, the likelihood function in \Eqref{eq:binned_likelihood} can be expressed as a product of likelihood functions for the individual energy bins,
\begin{equation}\label{eq:bin_by_bin_likelihood}
  \mathcal{L}(\{{\boldsymbol \mu_{j}}\},\{{\boldsymbol \theta_{j}}\} | \mathcal{D}) =
  \prod_{j} \mathcal{L}_{j}({\boldsymbol \mu_{j}},{\boldsymbol \theta_{j}} | \mathcal{D}_{j}).
\end{equation}
In \Eqref{eq:bin_by_bin_likelihood} the terms in the product are independent binned Poisson likelihood functions, 
akin to \Eqref{eq:binned_likelihood}, but running over spatial bins only. The end result is a likelihood function in each spectral bin for the dark matter flux component, assuming only a specific spatial morphology. As the choice of binning will affect the likelihood, the end result retains an explicit dependence on the binning, $\{{\boldsymbol \mu}_j\}$ and $\{{\boldsymbol \theta}_j\}$.
 
We use the results of the broadband fits from the first step of the procedure to constrain 
the nuisance parameters in these ``bin-by-bin'' fits.
Previous works using this methodology performed the bin-by-bin
likelihood fitting with the nuisance parameters fixed to their
global maximum likelihood estimates,
\begin{equation}
{\boldsymbol {\hat{\theta} } } = {\rm arg}_{{\boldsymbol \theta}} {\rm max} \left[ \mathcal{L} ({\boldsymbol \mu},{\boldsymbol \theta} | \mathcal{D}) \right].
\end{equation}
Fixing the parameters of the background sources at their globally fit
values avoids numerical instabilities resulting from the fine binning
in energy and the degeneracy of the diffuse background components at
high Galactic latitude.   However, for this analysis, since the background
model is an empirical description of the LMC region, we must consider
the possibility that some of dark matter signal could be absorbed into the
background model.

We must therefore quantify and incorporate the degeneracies between
the dark matter models we are testing and the components of the
background model.   We do so by first identifying the background model
components that are degenerate with the various dark matter models, and then
allowing the normalizations of those components to vary within the
statistical uncertainties of the ``global'' fit when performing the
``bin-by-bin'' fitting. 

To quantify the energy-dependent degeneracy between a dark matter
spatial profile and the components of the background model, and to identify the background 
model components with the largest degeneracy with the dark matter signal we study
a few representative dark matter spatial profiles.   For each profile 
we fit for the normalizations of all of the background components as
well as the dark matter component in each energy bin independently and then
extract the correlation factors between the dark matter component and the
various background components as a function of energy.   The dark matter spectrum 
is taken as a power law with index $\Gamma = 2$.

The correlation factor at a given energy between the dark matter component and the $i^\text{th}$
background component in energy bin $j$ can be obtained from the covariance matrices for the parameters 
once the likelihood function has be maximized:
\begin{equation}
  \rho_{i,{\rm DM}}(j) = \frac{ {\rm cov}_{i,{\rm DM}}(j)}{\sigma_{i}(j)\sigma_{\rm DM}(j)},
\end{equation}
where $\sigma_{k}(j) = \sqrt{ {\rm cov}_{k,k}(j) }$ are the variances on the
normalizations of $k^{\rm th}$ model component in the $j^{\rm th}$ energy bin. 

\begin{figure}[th]
\includegraphics[width=0.7\columnwidth]{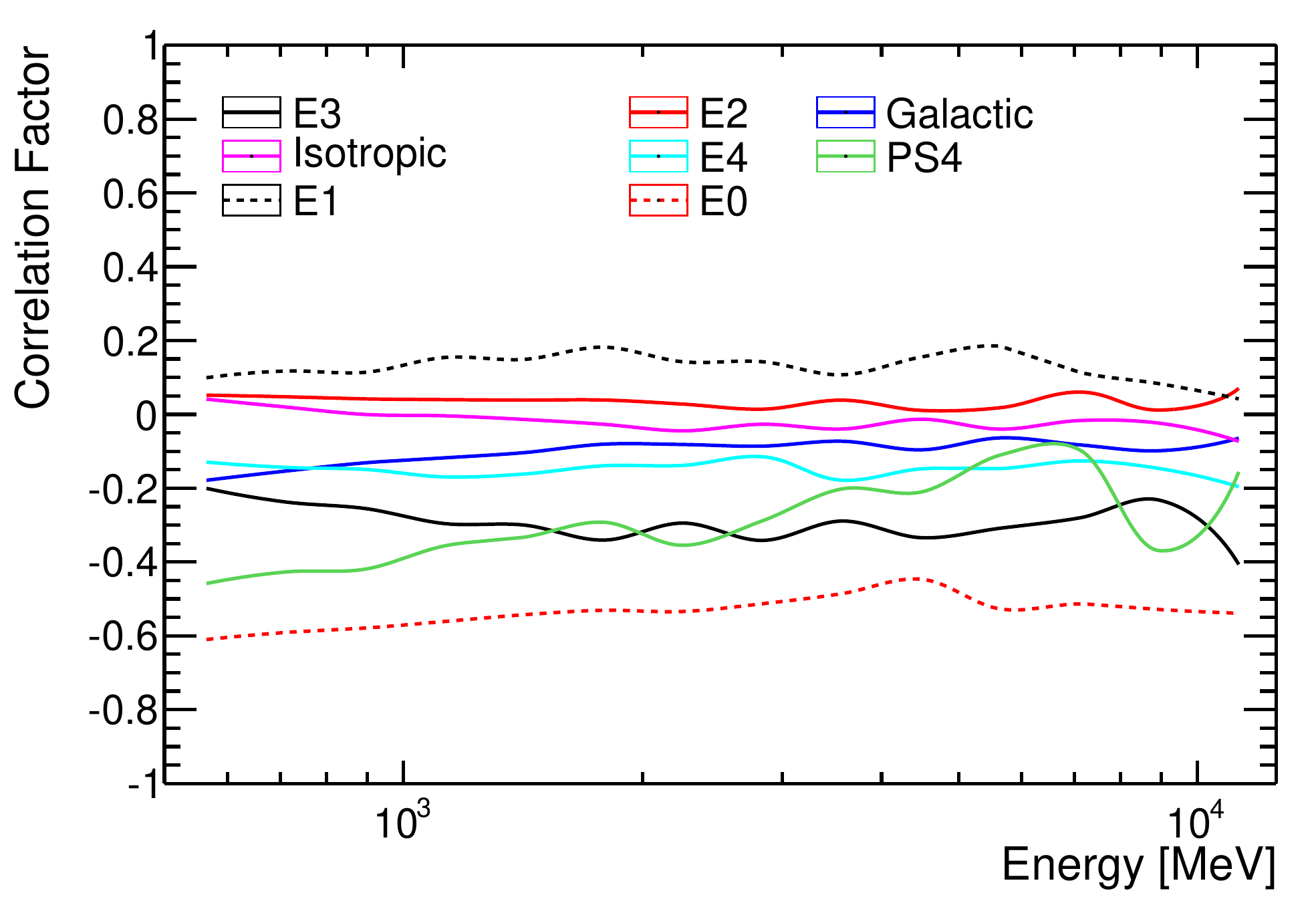}
\caption{Average spatial correlation factor between the major baryonic background components and dark matter signal as a function of photon energy.  The averaging was done over all three kinematic centers and the full range of dark matter spatial profiles we considered. The dark matter annihilation spectrum was taken as a power law with index $\Gamma = 2$. \label{fig:correlation}}
\end{figure}

Averages of the energy-dependent correlation factors between several of
the background model components and the various dark matter signal models are shown in Figure~\ref{fig:correlation}. 
Specifically, we show the correlation factor between the dark matter and the background component as a function of photon energy, averaged over each dark matter profile (envelope, NFW, and isothermal) and for all three of our possible center locations.   As can be seen, several of the extended LMC baryonic backgrounds have a non-negligible correlation between their spectra and that of the dark matter model.  
We identified the components of the background model with the largest
correlation factors with the dark matter spatial templates as the {\tt E0} and {\tt E3}
extended components and the point source {\tt PS4} (tentatively identified as the supernova remnant N132D).
\footnote{We also found large correlation factors between some of the fits in the control regions described in 
Section~\ref{sec:systematic_errors} and the {\tt E2} and {\tt E4} extended components.} 

For these components that have significant correlations with dark matter, if we were to fix the nuisance parameters ${\boldsymbol \theta}_j$ in the bin-by-bin analysis to the values derived from the global maximum likelihood estimates ${\boldsymbol {\hat{\theta} } }$, it is quite conceivable that in the maximization of the likelihood function, any potential dark matter signal could be assigned to one (or more) of the known baryonic backgrounds. That is, were dark matter annihilation to exist in the data at a level significant enough to be detectable, the standard methodology could result in assigning too much of that annihilation signal into the baryonic backgrounds. In general this will result in overly optimistic bounds set on dark matter annihilation signals that have large correlations with the LMC baryonic backgrounds.

To address this possibility, for these five ({\tt E0},{\tt E2},{\tt E3},{\tt E4} and {\tt PS4}) components we wish to allow the ${\boldsymbol \theta}_{jk}$, where the index $k$ runs over the five components, to vary within the uncertainties estimated from the global fit rather than fixing them to the global maximum likelihood estimates $\hat{\theta_{k}}$.   Specifically, in the bin-by-bin fits we allow the $\theta_{jk}$ to vary subject to a Gaussian prior:
\begin{equation}
  \mathcal{L}(\{{\boldsymbol \mu_{j}}\},\{{\boldsymbol \theta_{j}}\} | \mathcal{D}) =
  \prod_{j} \mathcal{L}_{j}({\boldsymbol \mu_{j}},{\boldsymbol \theta_{j}} | \mathcal{D}_{j}) \times \prod_{k} \frac{e^{-(\theta_{jk}-\hat{\theta_{k}})^{2}/(2 \sigma_{k}^{2})}}{\sqrt{2\pi}\sigma_{k}}.\label{eq:binbybinprior}
\end{equation}

The error associated with each background component must be derived from the results of the broadband fit. To account for the reduced statistics in the bin-by-bin fits as compared to the broadband fits, we assign the width of the Gaussian prior on the nuisance parameters as ten times the uncertainties on the parameters in the broadband fits: $\sigma_{k} = 10 \delta \hat\theta_k$. We arrived at this factor of 10 empirically, {\it i.e.}, we performed tests with simulated data varying the width of the Gaussian prior by the same factor and found that using a factor of 10 allowed the $\theta_{jk}$ to vary within the uncertainty bounds $\hat{\theta_k} \pm \delta \hat{\theta_k}$ and also resulted in the correct coverage properties for upper limits on simulated data.   We note that the factor depends on the number of energy bins used in the data analysis.  With this modification to the calculation of the bin-by-bin likelihood, we can, for each set of dark matter halo parameter values chosen, estimate the significance of any observed excess and calculate the upper limit on the cross section into a specific final state, as a function of dark matter mass. For any given fit, the only free parameter describing the dark matter component is its overall normalization ({\it i.e.}, either the power-law prefactor or $\langle \sigma v \rangle$);  we explore the variation in the other parameters by scanning discrete mass values for each choice of annihilation channels and dark matter spatial profiles.

To estimate the significance of an excess we define the TS in terms of the likelihood ratio with respect to the null ({\it i.e.}, background-only) hypothesis:
\begin{equation}
\mbox{TS} =  2\ln\frac{\mathcal{L}({\boldsymbol \mu},{\boldsymbol \theta} | \mathcal{D})}{\mathcal{L}_{\rm null}({\boldsymbol \theta} | \mathcal{D})}.\label{eq:TS}
\end{equation}
For the energy bins up to about 10\,GeV the statistics are large enough that Chernoff's theorem applies, and we expect the TS-distribution to follow a $\chi^{2}$ distribution~\cite{REF:Chernoff}.  At higher energies, the counts per bin are in the Poisson regime and the $\chi^{2}$ distribution moderately over-predicts the number of high TS trials observed in simulated data.

Similarly, we evaluate the one-sided 95\% confidence level (CL) exclusion limit on the flux as the point at which the $p$-value for a $\chi^{2}$ distribution with 1 degree of freedom is 0.05 when we take the maximum likelihood estimate as the null hypothesis.  That is, the 95\% CL upper limit on the flux assigned to dark matter is the value at which the log-likelihood decreases by 1.35 with respect to its maximum value.

\begin{figure}[th]
\includegraphics[width=0.7\columnwidth]{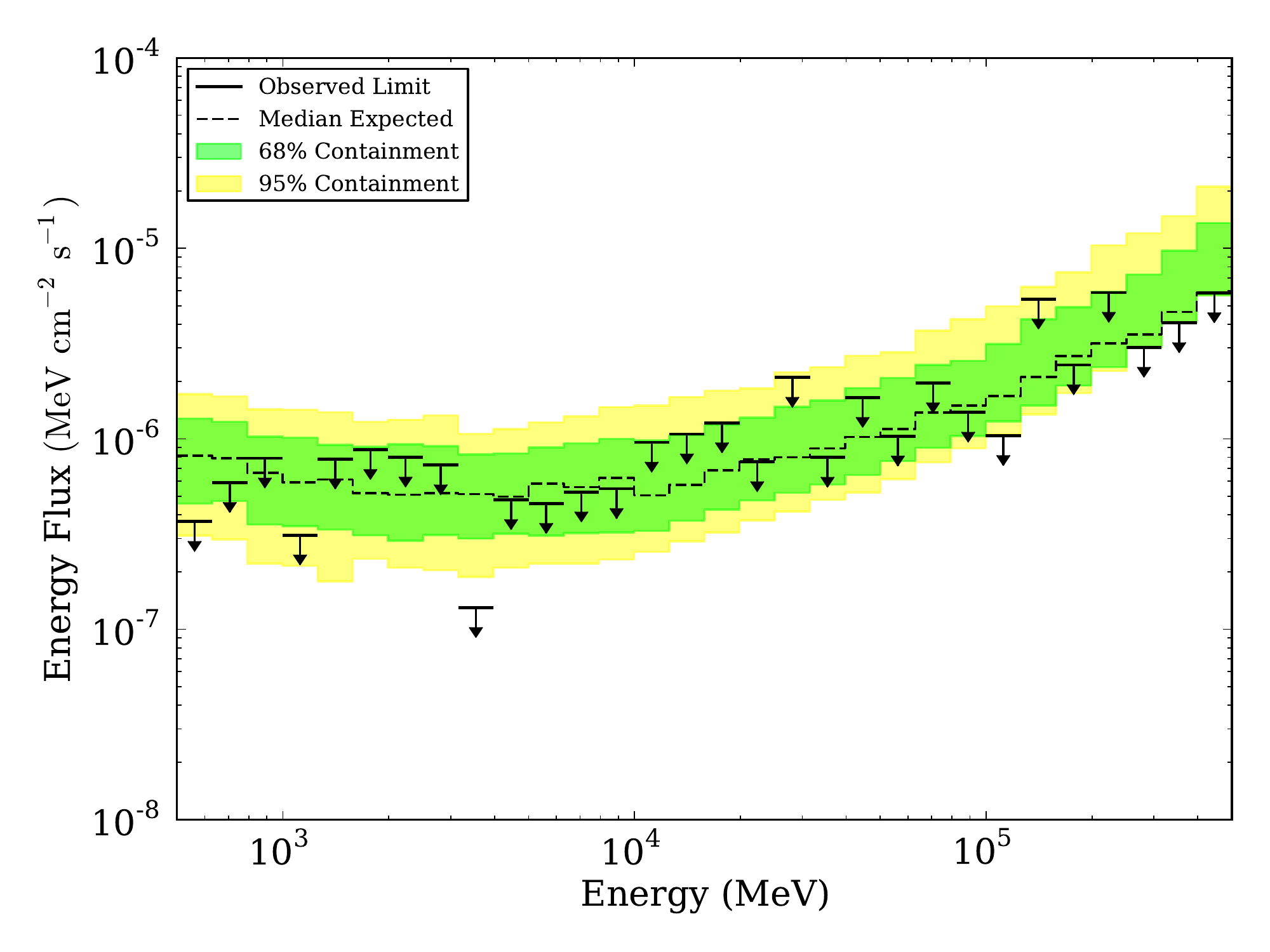}
\caption{Simulated 95\% (CL) upper limit on the gamma-ray flux associated with a {\tt sim-mean} profile at the {\tt HI} center, drawn from a Monte Carlo simulation of the LMC background-only model in the ROI with no DM contribution.  The expected 68\% (green) and 95\% (yellow) containment bands for the upper limits are also shown.  The upper limits are given as as function of the energy flux of the putative source in each bin. These limits depend on the assumed
spatial profile, but are independent of an assumed dark matter spectrum.  The nuisance parameters are constrained with a Gaussian prior with mean ${\boldsymbol {\hat{\theta} }}$ and width $10\delta{\boldsymbol \theta}$, where ${\boldsymbol{\hat{\theta}}}$ is the set of background parameters that maximize the likelihood in the broadband fit, and $\delta{\boldsymbol{\hat{\theta}}}$ are the uncertainties on those parameters.   See text for details.
\label{fig:covariance_demo}}
\end{figure}

In Figure~\ref{fig:covariance_demo}, we show simulated energy bin-by-bin 95\% CL exclusion limits for an energy flux from a dark matter signal with 
the morphology of the {\tt sim-mean} profile at the {\tt HI} center.   To construct these upper limits, we generated a Monte Carlo simulation of the LMC ROI drawn from the baryonic background-only model of Section~\ref{sec:baryons}, and applied the maximum likelihood fitting procedure to this pseudo-data.    We used 200 such iterations to construct the expected containment bands for the upper limits.

\subsubsection{Dark Matter Spectral Fitting}\label{subsec:dark_matter_fitting}

The final stage of our fitting procedure is to convert the energy bin-by-bin likelihood curve in flux into a likelihood 
curve in $\langle \sigma v\rangle$ for each dark matter spatial profile and spectrum.   For each dark matter spectrum
we scan over $\langle \sigma v\rangle$, extract the resulting expected flux $F_j$ in each energy bin, look up the log-likelihood of observing 
that flux value and sum these log-likelihoods over all the energy bins to get the log-likelihood curve:

\begin{equation}
\log\mathcal{L}({\boldsymbol \langle \sigma v\rangle},{\boldsymbol \mu},{\boldsymbol \theta} | \mathcal{D}) = \sum_{j} 
\log\mathcal{L}_{j}({\boldsymbol \langle \sigma v\rangle},{\boldsymbol \mu},{\boldsymbol \theta_{j}} | \mathcal{D}_j).\label{eq:loglike_combined}
\end{equation}
 
From this procedure, for each dark matter mass and channel we can calculate both the maximum likelihood cross section and the 95\% CL upper limit on the cross section.

\subsection{Systematic Uncertainties}\label{sec:systematic_errors}

In other {\it Fermi} LAT searches for dark matter annihilation, systematic uncertainties can be controlled by comparing observations of the signal region with observations from areas of the sky where the dark matter annihilation signal is expected to be greatly reduced, but which have similar baryonic backgrounds. However, as the LMC is a region with significant backgrounds from baryonic processes, and those backgrounds vary greatly with location within the ROI, we do not have easy access to such a control region. 

As a result, we must use the LMC itself to control for systematic uncertainties. As described in Section~\ref{sec:dm}, the direction of the center of the dark matter halo of the LMC is somewhat uncertain (the extremes in the possible locations differ by $\sim 1\fdg 5$), as is the dark matter density profile. However, even accounting for these uncertainties the contribution due to dark matter annihilation more than a few degrees away from the LMC bar is small.  This can be seen explicitly in the differential $J$-factors plotted in the left panel of Figure~\ref{fig:J}.  While a dark matter profile centered inside this region would result in a contribution to the gamma-ray flux in the region beyond $2-3^\circ$ from the LMC center, the fall-off in the differential $J$-factor away from the dark matter profile's center means that this flux would be small. As a result, we can use the LMC outside of $3^\circ$ from the center as a control to estimate the amount of TS variations we might expect in regions where any large deviation from our baseline baryonic model cannot be attributed to significant dark matter annihilation. As the background sources do vary across the LMC, this technique cannot estimate systematic uncertainties that only occur in the inner $3^\circ$ of the ROI, but we must accept this as a consequence of the complicated baryonic backgrounds in the LMC. For our purposes here, we define the control region as all of ROI more the $3^\circ$ from the average of the possible LMC centers (Table~\ref{tab:centers}),
\begin{equation}
(\ell,b) = (280\fdg 02,-33\fdg 13). \label{eq:center_loc}
\end{equation}
We take a two-step approach to control for systematic uncertainties inside the signal region: first we establish that the imperfect modeling of the baryonic backgrounds can induce a fake dark matter signal by studying the distribution of TS values as a function of dark matter mass and annihilation channel from the control region, and second we use the fitted numbers of signal and background events in the control region to estimate the level of systematic error.

\begin{figure}[th]
\includegraphics[width=0.5\columnwidth]{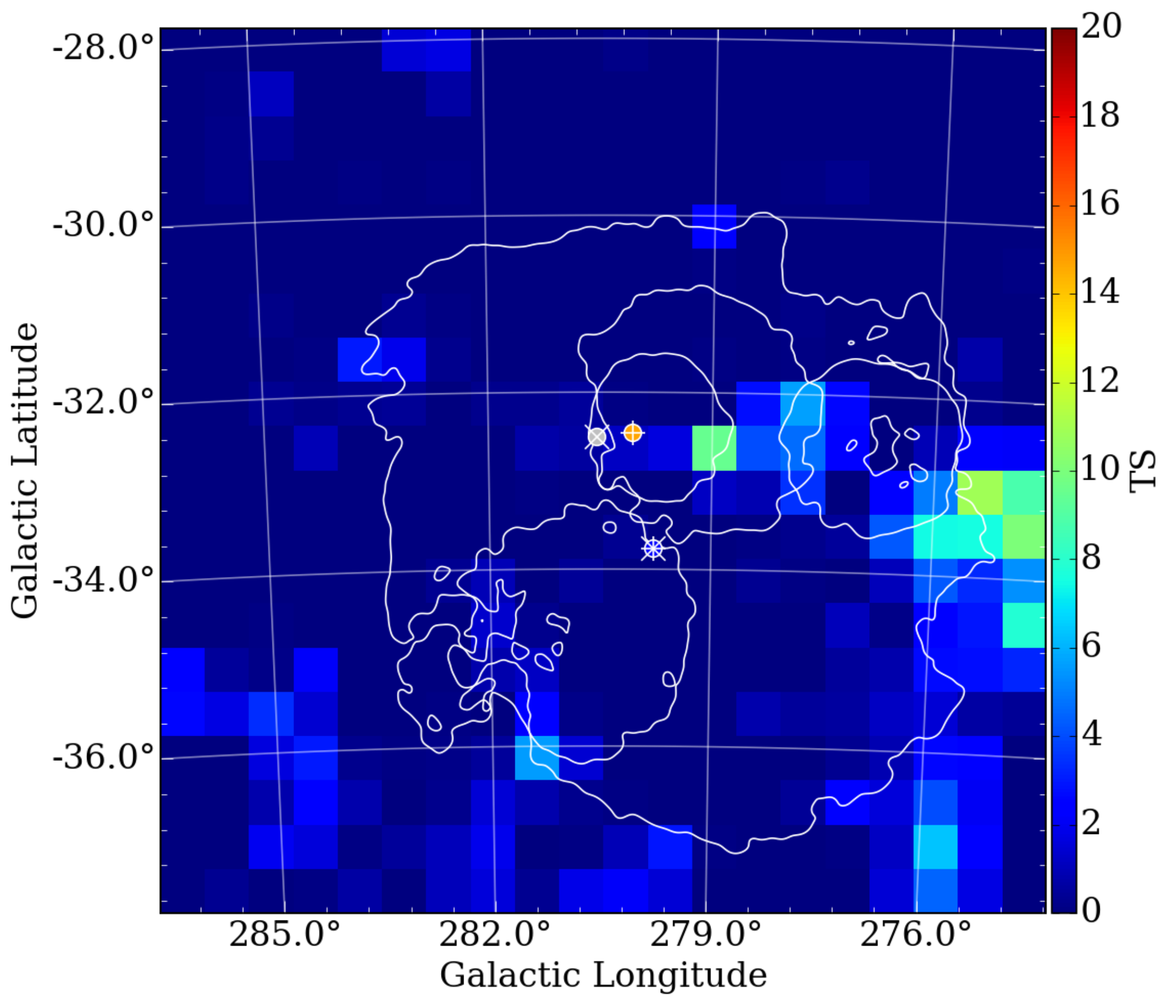}\includegraphics[width=0.5\columnwidth]{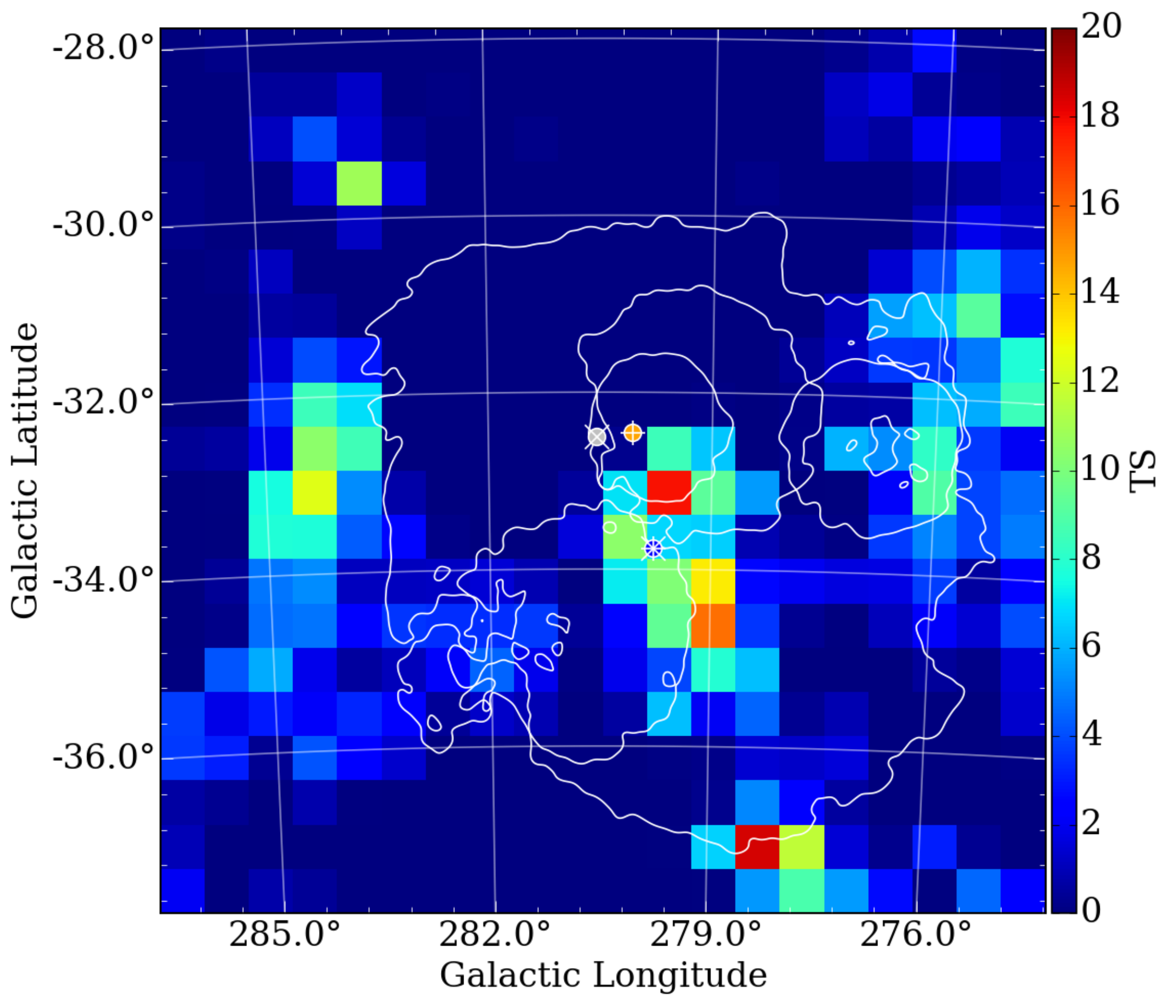}
\caption{Maximum TS value as a function of location, assuming the {\tt sim-mean} profile centered at each point in a grid of Galactic latitude and longitude with $0\fdg 5\times 0\fdg 5$ spacing across the ROI, and 50~GeV dark matter annihilating in the $b\bar{b}$ channel. Left: TS values for Monte Carlo generated pseudo-data drawn from the background-only baryonic model. Right: TS values from the fit to real LAT data. Also shown for reference are the smoothed background models and our three likely dark matter centers: {\tt stellar} (white circle with $\times$ cross), {\tt outer} (orange circle with $+$ cross), and {\tt HI} (blue circle with $\rlap{+}{\times}$ cross). \label{fig:TS_map}}
\end{figure}

\begin{figure}[th]
\includegraphics[width=0.5\columnwidth]{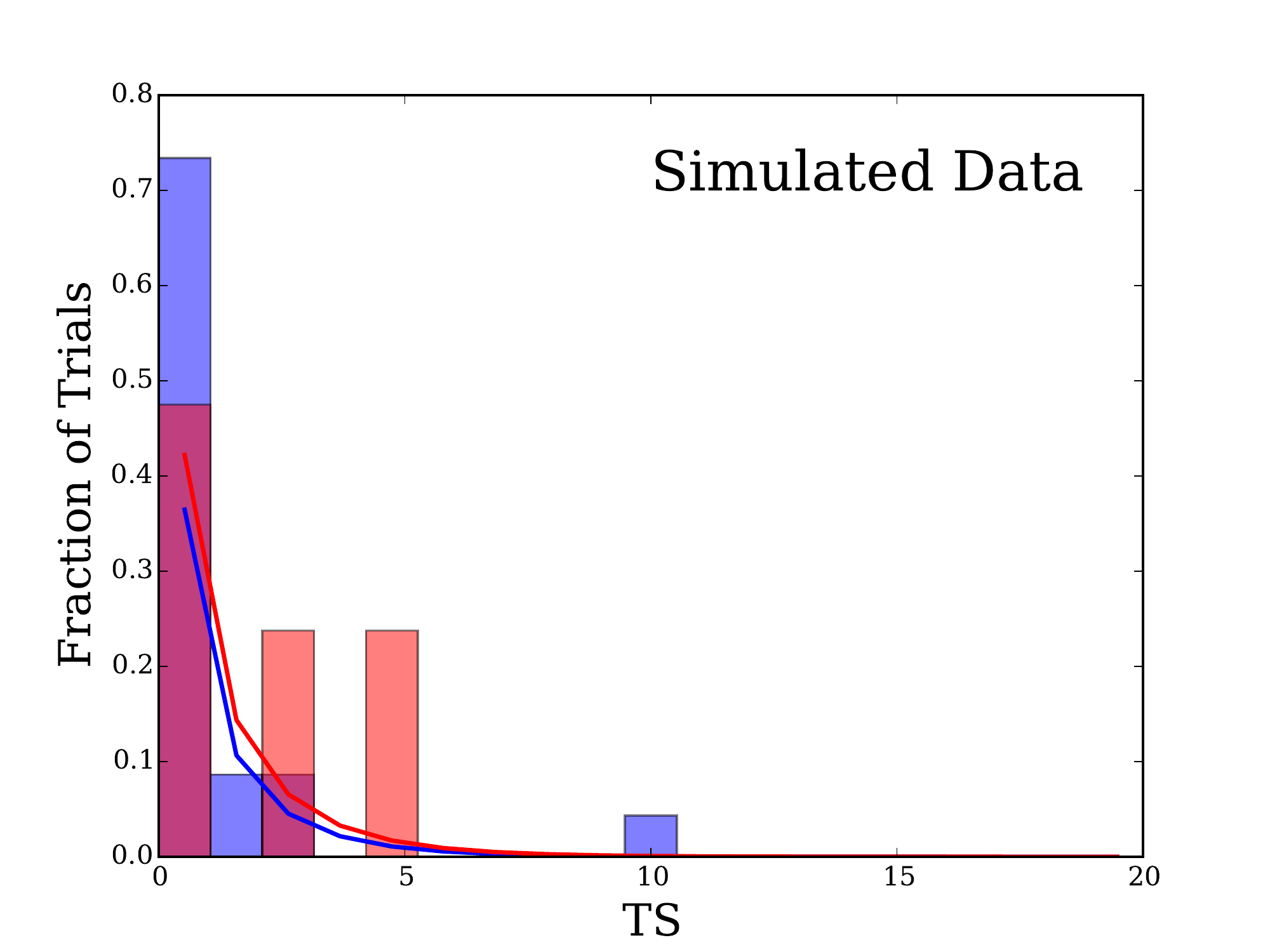}\includegraphics[width=0.5\columnwidth]{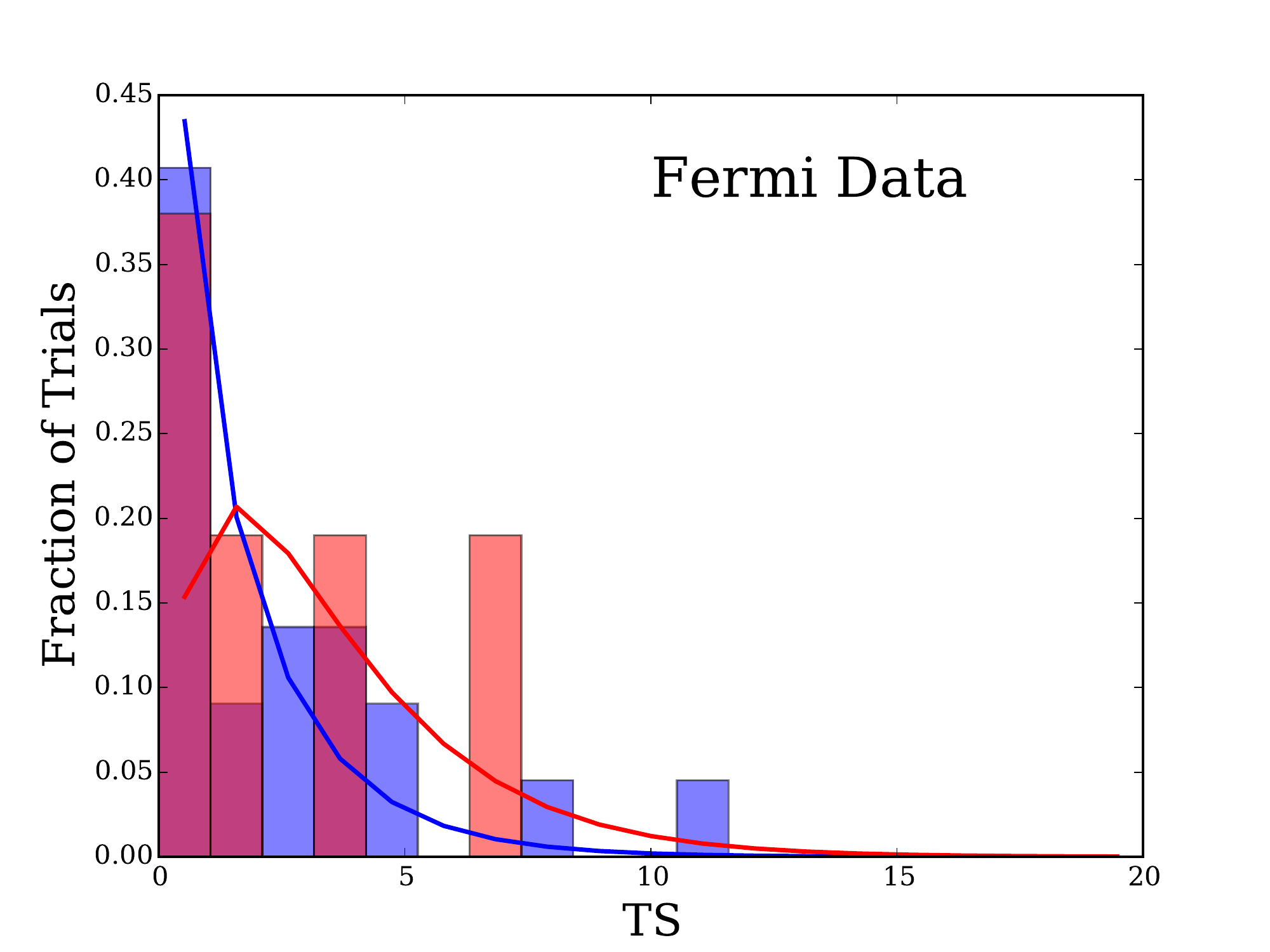}
\caption{Normalized histograms of TS values assuming the {\tt sim-mean} profile centered at points across the ROI, and 50~GeV dark matter annihilating in the $b\bar{b}$ channel. TS values of profiles centered within $3^\circ$ of the average LMC center (the signal region) are shown in red, TS values from profiles in the control region are in blue. Lines show the $\chi^2$ distribution with the best-fit number of degrees of freedom. Left: TS values from fits to Monte Carlo-generated pseudo-data drawn from the background-only baryonic model. Right: TS values from fits to real LAT data. \label{fig:TS_histogram}}
 \end{figure}

In Figure~\ref{fig:TS_map}, we show the TS value as a function of dark matter profile center location over the entire ROI for the {\tt sim-mean} profile assuming 50~GeV dark matter annihilating into $b\bar{b}$, both fit to data and also to pseudo-data generated from the background-only model. In Figure~\ref{fig:TS_histogram}, we show histograms of TS values for the 50~GeV $b\bar{b}$ channel, separated into fits of {\tt sim-mean} profiles centered within $3^\circ$ of our center point, and those outside of this region for both the pseudo-data drawn from our background model as well as our fits to LAT data. As expected, the TS values for the presence of dark matter in pseudo-data drawn from the background-only model and then fit to that same model are consistent with a $\chi^2$ distribution with one degree of freedom, both in the control and signal regions. However, the real TS values in the control regions have significantly more large ($\gtrsim 10$) TS values than one would expect from a sample drawn from a $\chi^2$ distribution with one degree of freedom.

As can be seen in the right-hand panel of Figure~\ref{fig:TS_histogram}, the distribution of TS values inside the signal region is markedly different from that in the control region.  However, it is also true that the statistics are larger in the signal region than the control region.   Therefore, mismodeling of the baryonic backgrounds at a given fractional level would result in a higher significance ({\it i.e.}, larger TS values) in the signal region for any systematically induced signal.  This exercise indicates that the strict statistics-only interpretation of the TS distribution (and the limits on the dark matter cross section one would infer from that) must be modified to include the systematic errors that are present when comparing our model to real data. 

We therefore add a second step to our approach to ensure we are fully taking into account the systematic uncertainties, by directly calculating the ratio of signal events to background events (which estimates the possible level of systematic error) and comparing to the statistical error, which depends on the square root of the number of background events. This must be done carefully, for as we move from the signal region to the control, we are moving from the center of the LMC to the outskirts. While we do not expect the dark matter signal to be present in the control region, we also do not expect the total rate of baryonic background gamma rays to be as high in this region. Therefore, while we want to estimate the systematic error in the control region in order to apply that to the signal region, we cannot proceed by simply evaluating the total number of ``signal-like'' or ``background-like'' gamma-ray events in the control region. The equivalent numbers in the signal region would be much larger.   We have therefore adopted a technique to estimate the ``effective background',' {\it i.e.}, the background that overlaps with the signal~\cite{2014JCAP...10..023A}. This technique relies on the {\em ansatz} that the systematic uncertainty scales with the background, and accounts for the fact that not all of the background in the ROI overlaps with the signal distribution.

When testing a particular dark matter model ({\it i.e.}, mass, spectrum, spatial profile, and center of profile location), 
given the normalized signal and background models, $P_{{\rm sig,}i}({\bf \mu}) = \lambda_{{\rm sig,}i}({\bf \mu}) / \sum_{k} \lambda_{{\rm sig,}k}({\bf \mu})$ and $P_{{\rm bkg,}i}({\bf \theta}) = \lambda_{{\rm bkg,}i}({\bf \theta}) / \sum_{k} \lambda_{{\rm bkg,}k}({\bf \theta})$, we can estimate the effective background by calculating the likelihood fit covariance matrix element for the signal size ({\it e.g.}, starting from Eq.~28 in Ref.~\cite{Cowan2011}) in the approximation that the background is much larger than the signal, giving:
\begin{equation}
b_{\rm eff} = \frac{N}{ \left( \sum_{k}\frac{P_{{\rm sig,}k}^{2}({\bf \mu})}{P_{{\rm bkg,}k}({\bf \theta})} \right) - 1},\label{eq:beff}
\end{equation}
where the summation runs over all pixels in the ROI and all the energy bins and $N$ is the total number of events in the ROI. We note that the definition of $b_{\rm eff}$ given here differs slightly from the definition used in Ref.~\cite{2014JCAP...10..023A}. The two definitions give very similar results in the case of large backgrounds and small signals.  The definition used here is more general and has a few useful properties.  First, if the only free parameters in the fit are the overall normalization of the signal and background components, then in the limit that the signal is much smaller than the background the statistical uncertainty on the number of signal counts will be $\delta N_{sig} \simeq \sqrt{b_{\rm eff}}$.  Second, if the signal and background models are totally degenerate ($P_{{\rm sig,}k}({\bf \mu}) = P_{{\rm bkg,}k}({\bf \theta})$ for all $k$), then the term in the summation will be equal to 1 and $b_{\rm eff}$ will diverge, indicating that we have little power to distinguish signal from background. If this were the case, the statistical errors for the likelihood fit would be extremely large, corresponding to an upper limit on the cross section sufficient to generate all of the measured signal through dark matter annihilation.  Finally, if the signal and background models differ significantly the term in the summation will be much greater than 1 and $b_{\rm eff}$ will be proportionally less than $N$. That is, the statistical uncertainty on the signal will correspond to an effective background that is much less than the total number of background events in the ROI. As we will discuss below, by quantifying the systematic uncertainties of the background modeling as a percentage of $b_{\rm eff}$, we are able to account for those uncertainties in the likelihood fitting procedure and include them in our DM constraints.

\begin{figure}[ht]
\includegraphics[width=0.3\columnwidth]{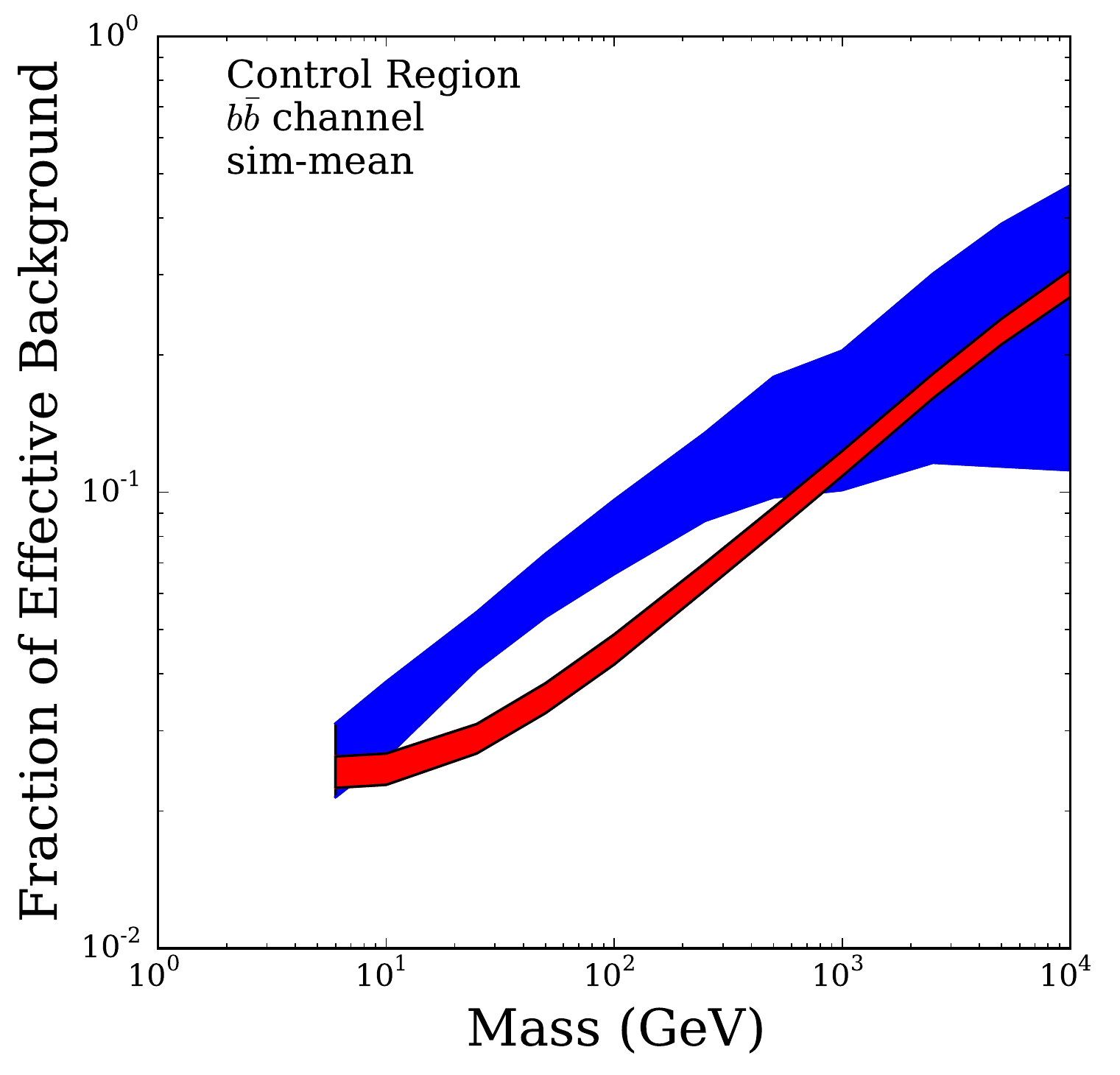}\includegraphics[width=0.3\columnwidth]{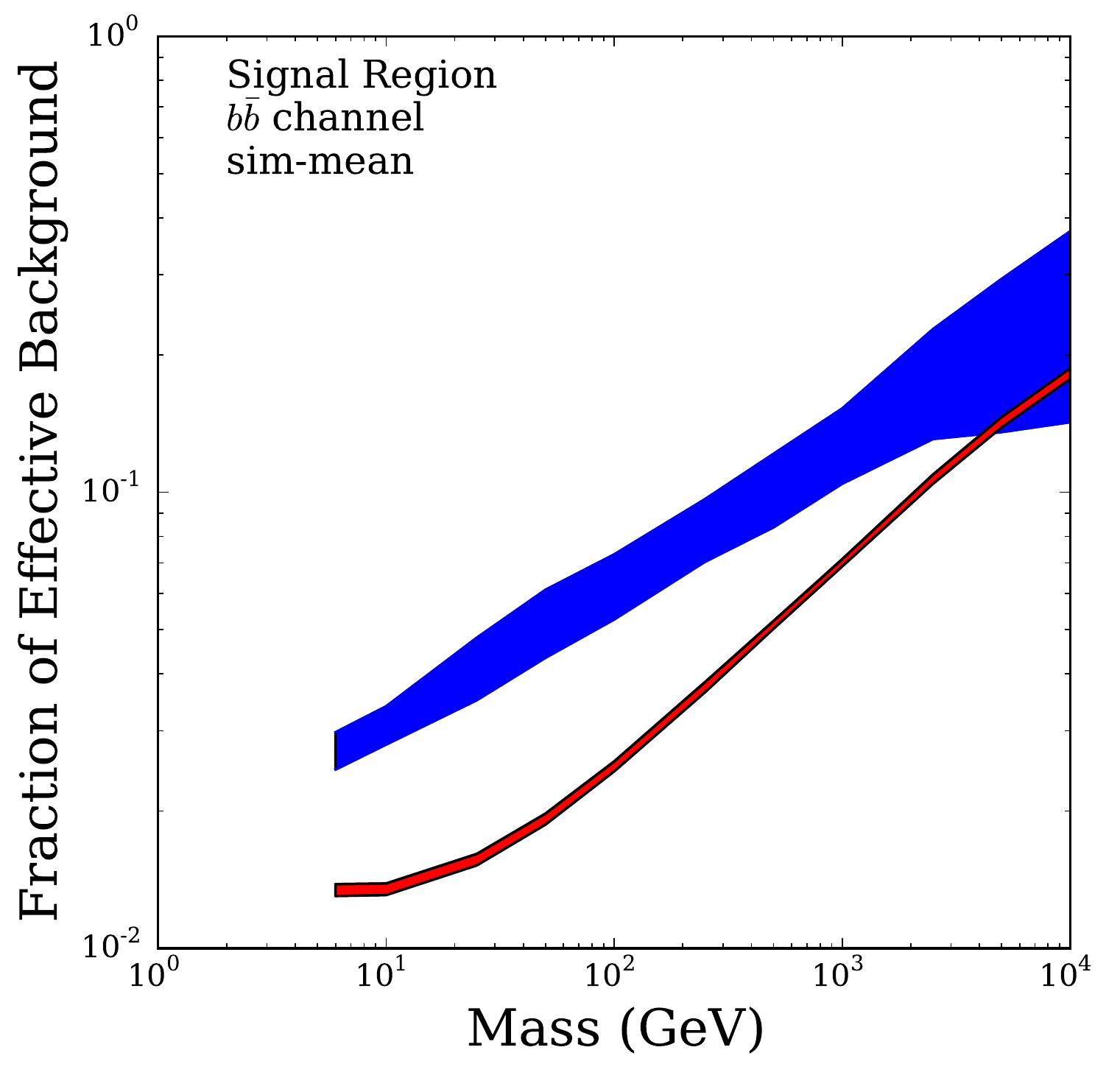}\includegraphics[width=0.3\columnwidth]{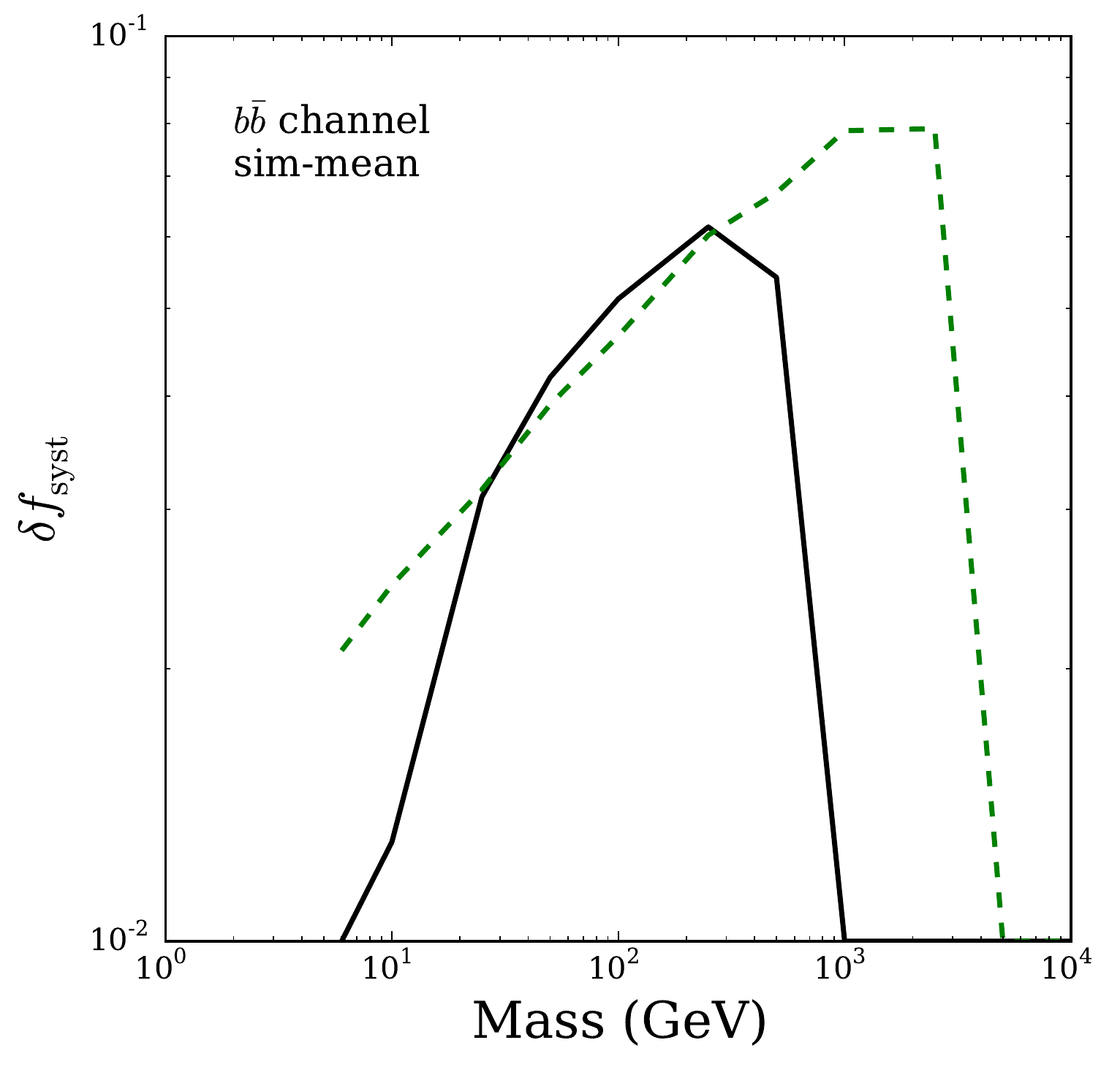}

\caption{Left: Plot of $f_{\rm sig}$ (blue) and $f_{\rm stat}$ (red) for the {\tt sim-mean} profile annihilating into the $b\bar{b}$ channel in the control region. Shaded regions span the 84\%-95\% CL of the distributions. Center: As the left plot, but for the signal region. Right: $\delta f_{\rm syst}$ for the control region (black) and signal region (green, dashed). \label{fig:fsyst}}
\end{figure}

Again, for each particular dark matter model, we can calculate the number of signal events $N_{\rm sig}$, given by assuming the annihilation cross section is the maximum likelihood estimate from our fitting procedure.   We then can define the ratio of the signal to the effective background as $f_{\rm sig}$, and the estimate of the statistical uncertainty $\delta f_{\rm stat}$ in terms of the effective background:
\begin{eqnarray}
f_{\rm sig} = & \frac{N_{\rm sig}}{b_{\rm eff}}, & \label{eq:fsig} \\
\delta f_{\rm stat} = & \frac{\delta N_{sig}}{b_{\rm eff}} \simeq &  {b_{\rm eff}}^{-1/2} \label{eq:fstat}.
\end{eqnarray}
From the width of the distribution of $f_{\rm sig}$ for the trials in the control region, where we do not expect to detect any signal, we can estimate the total (statistical + systematic) uncertainty.  In Figure~\ref{fig:fsyst}, we show the distributions of $f_{\rm sig}$ and $\delta f_{\rm stat}$ for the control region, assuming the {\tt sim-mean} profile and annihilation into $b\bar{b}$. We plot the 84\% to 95\% enclosure of $f_{\rm sig}$ and $\delta f_{\rm stat}$, obtained by fitting the gamma-ray data to a {\tt sim-mean} profile scanned across possible center locations in the ROI outside the control region.  For comparison, we also show these quantities for the signal region, though of course we cannot use those results to estimate systematic errors. When $f_{\rm stat} < f_{\rm sig}$, the total error is dominated by systematic uncertainties.  If our fitting procedure allowed for negative signals we could take a simple measure of the width such as the root-mean-square of the distribution.   However, our fitting procedure only allows for positive signals and approximately half of the trials have $ f_{\rm sig} = 0$.   We therefore define for each profile our estimate of the systematic error as the difference (taken in quadrature) of the $1\sigma$ (84\% CL) enclosure of the total error estimate and the statistical error estimate from the control region:
\begin{equation}
\delta f_{\rm syst}^2 = f_{\rm sig}^2(84\%) - \delta f_{\rm stat}^2(84\%). \label{eq:fsyst}
\end{equation}
The distribution of $\delta f_{\rm syst}$ for the {\tt sim-mean} profile is shown in the right panel of Figure~\ref{fig:fsyst}. Again we show the equivalent quantity derived from the error estimates in the signal region, but this is for reference only. The distribution of $\delta f_{\rm syst}$ is evaluated separately for each choice of profile and annihilation channel. In our later calculations, we will place a lower limit on the systematic error, $\delta f_{\rm syst} > 0.01$, to include some level of systematic error even in the annihilation spectra that are statistics dominated.

These data-driven calculations of the systematic error are then included in our derivations of the upper limits on the cross section.  In calculating the likelihood for each choice of dark matter profile, mass, and annihilation spectrum we must account for the possibility that events originate with a source in the background but their distribution is not accurately described by the background model.   To do this we assume that the fitted cross section is the sum of the cross section of the true annihilation signal $\langle \sigma v\rangle_{\rm ann.}$ and an additional cross section induced by potential systematic biases $x_{\rm syst}$.  That latter cross section is drawn from a Gaussian distribution centered at zero with standard deviation $\langle \sigma v\rangle_{\rm syst}$ set by $n_{\rm syst} = \delta f_{\rm syst} b_{\rm eff}$ (converted to a cross section through appropriate factors of the exposure, $J$-factor, and $dN_\gamma/dE_\gamma$). That is, we replace the likelihood function Eq.~\eqref{eq:bin_by_bin_likelihood} with
\begin{equation}
{\cal L}(\langle\sigma v\rangle_{\rm ann.} ) \to {\cal L}(\langle\sigma v\rangle_{\rm ann.} + x_{\rm syst}) \times \frac{1}{\sqrt{2\pi}\langle \sigma v\rangle_{\rm syst}} e^{-x_{\rm syst}^2/2\langle \sigma v\rangle_{\rm syst}^2}. \label{eq:likely_syst}
\end{equation}
Where $x_{\rm syst}$ is the nuisance parameter representing systematic uncertainties that can induce a false signal or mask a true signal, distributed according to our estimate of the systematic error on the cross section.
To include the range of systematic uncertainties in our Monte Carlo estimations for the predicted limits (that is, in the fits to pseudo-data used to derive the expected limit bands), we allow the Gaussian prior in Eq.~\eqref{eq:likely_syst} to be centered not at zero, as in the fits to real data, but to be centered at a non-zero $\mu_{\rm syst}$ obtained from sampling the distribution of $\langle \sigma v\rangle_{\rm syst}$ in the control region for that density profile.

\begin{figure}[th]
\includegraphics[width=0.4\columnwidth]{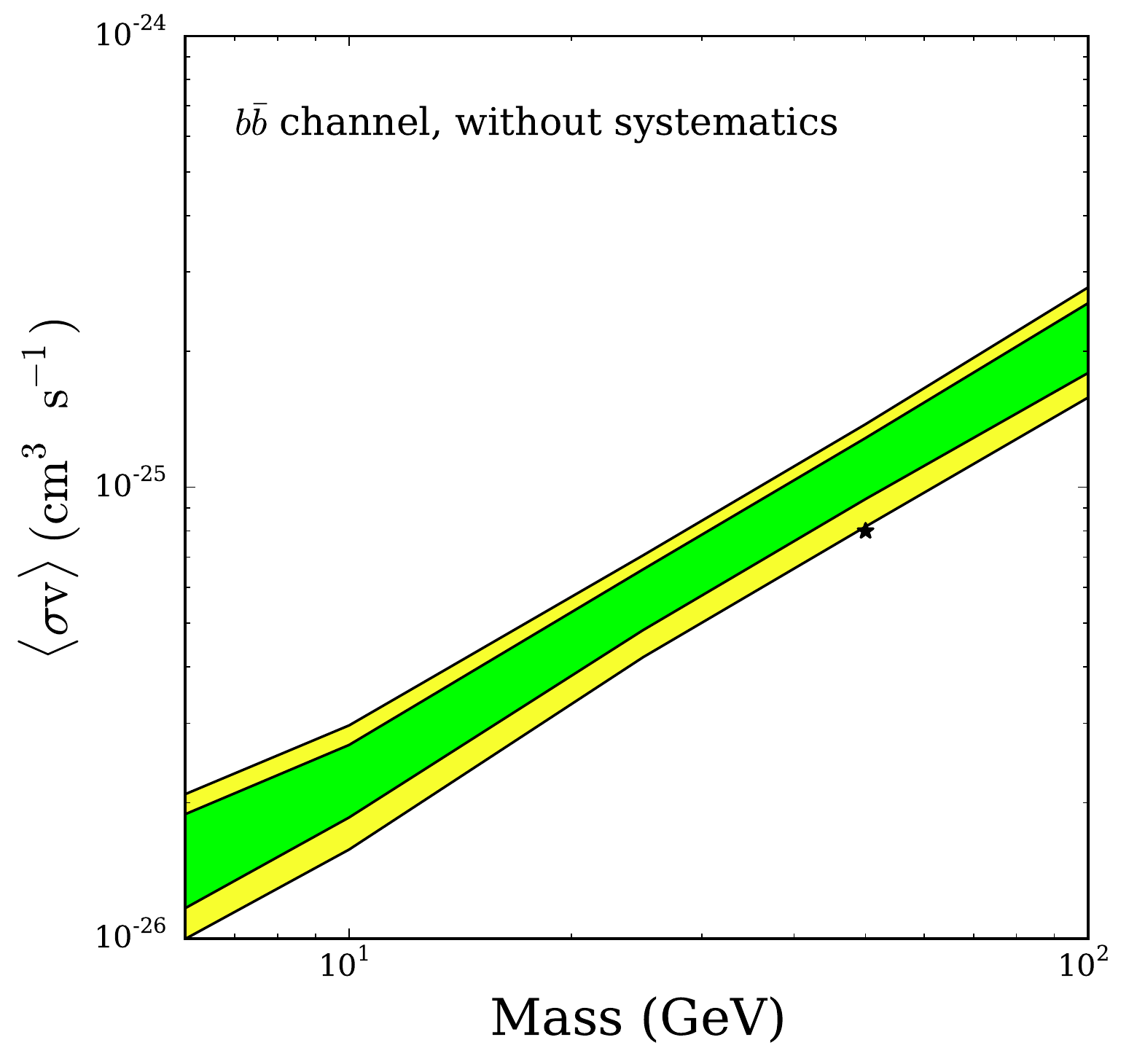}\includegraphics[width=0.4\columnwidth]{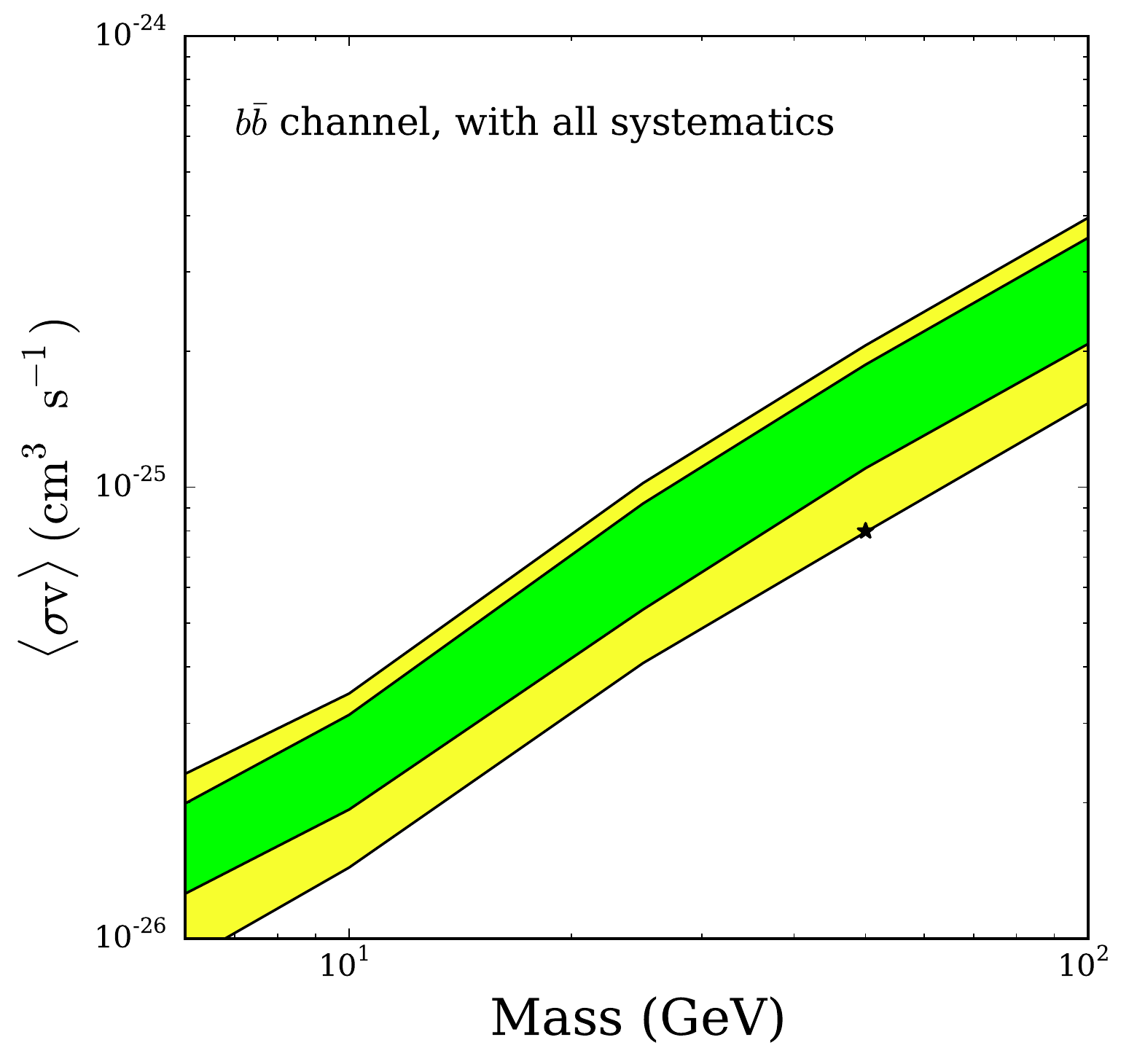}

\caption{Coverage study for dark matter signal injected into LMC background model.  84\% (green) and 95\% (yellow) containment bands for the 95\% CL upper limit on the annihilation cross section $\langle \sigma v \rangle$ in the $b\bar{b}$ channel for the {\tt sim-mean} profile at the {\tt HI} center, as a function of dark matter mass, drawn from 200 iterations of Monte Carlo-generated pseudo-data of the LMC baryonic backgrounds with an injected signal of 50~GeV dark matter annihilating into $b\bar{b}$ with a cross section of $\langle \sigma v\rangle = 8 \times 10^{-26}$~cm$^3$/s.    Left: Predicted exclusion curves in the $b\bar{b}$ channel without including systematic uncertainties ({\it i.e.}, statistical errors only).   Right: Predicted exclusion curves in $b\bar{b}$ after applying the systematic error corrections Eqs.~\eqref{eq:beff} to \eqref{eq:likely_syst}. Star indicates injected dark matter parameter point. 
\label{fig:inject_test_syst}}
\end{figure}

In Figure~\ref{fig:inject_test_syst}, we demonstrate the coverage of our upper limit calculations, as we show the expected exclusion curves for dark matter injected at the {\tt HI} center with a {\tt sim-mean} profile, both with and without the systematic effects.  The injected signal (assuming $50$~GeV dark matter annihilating into $b\bar{b}$ with $\langle \sigma v\rangle = 8 \times 10^{-26}$~cm$^3$/s) should lie below the 95\% CL upper limit on the cross section in 95\% of pseudo-experiments -- for that specific choice of dark matter mass. As can be seen in Figure~\ref{fig:inject_test_syst}, the actual signal point falls below the 95\% CL upper limit very nearly 95\% of the time in both cases.  More notably, the resulting upper limits when including systematic uncertainties are weaker than we would have obtained without including the systematic error, and the widths of the $1\sigma$ and $2\sigma$ error bands are larger as well.  While these facts are perhaps unfortunate from the standpoint of placing the best possible bounds or detecting an unambiguous signal of dark matter, our method is conservative as indicated by this coverage study, and so we are confident that any bounds we place will be robust.

\section{Constraints on Dark Matter}\label{sec:constraints}

We are now able to set constraints on the annihilation of dark matter into Standard Model particles that result in gamma rays after decays and hadronization. We report the 95\% CL upper limit on the annihilation cross section $\langle \sigma v\rangle$ for each channel of interest Er.~\eqref{eq:channels}, including all systematic effects discussed in the previous section. 
We show results for three profiles that span the expected range of dark matter in the LMC: the average profile fit to simulation {\tt sim-mean}, the averaged NFW profile {\tt nfw-mean}, and the averaged isothermal profile {\tt iso-mean}. This subset encompasses a reasonable range of plausible dark matter profiles for the LMC.

\begin{figure}[ht]
\includegraphics[width=0.5\columnwidth]{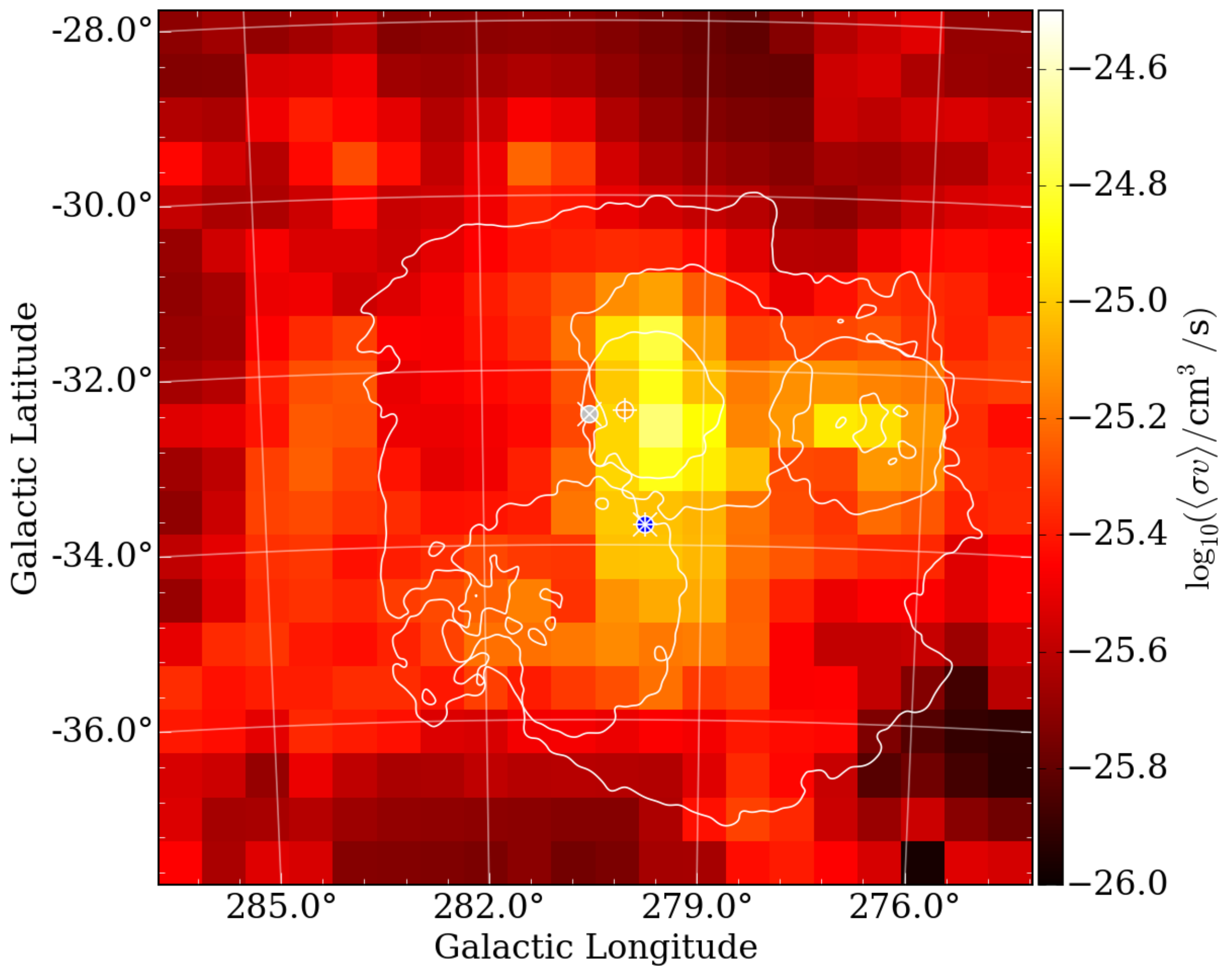}\includegraphics[width=0.48\columnwidth]{./envmean_gridscan_syst_TS_plot.pdf}
\includegraphics[width=0.5\columnwidth]{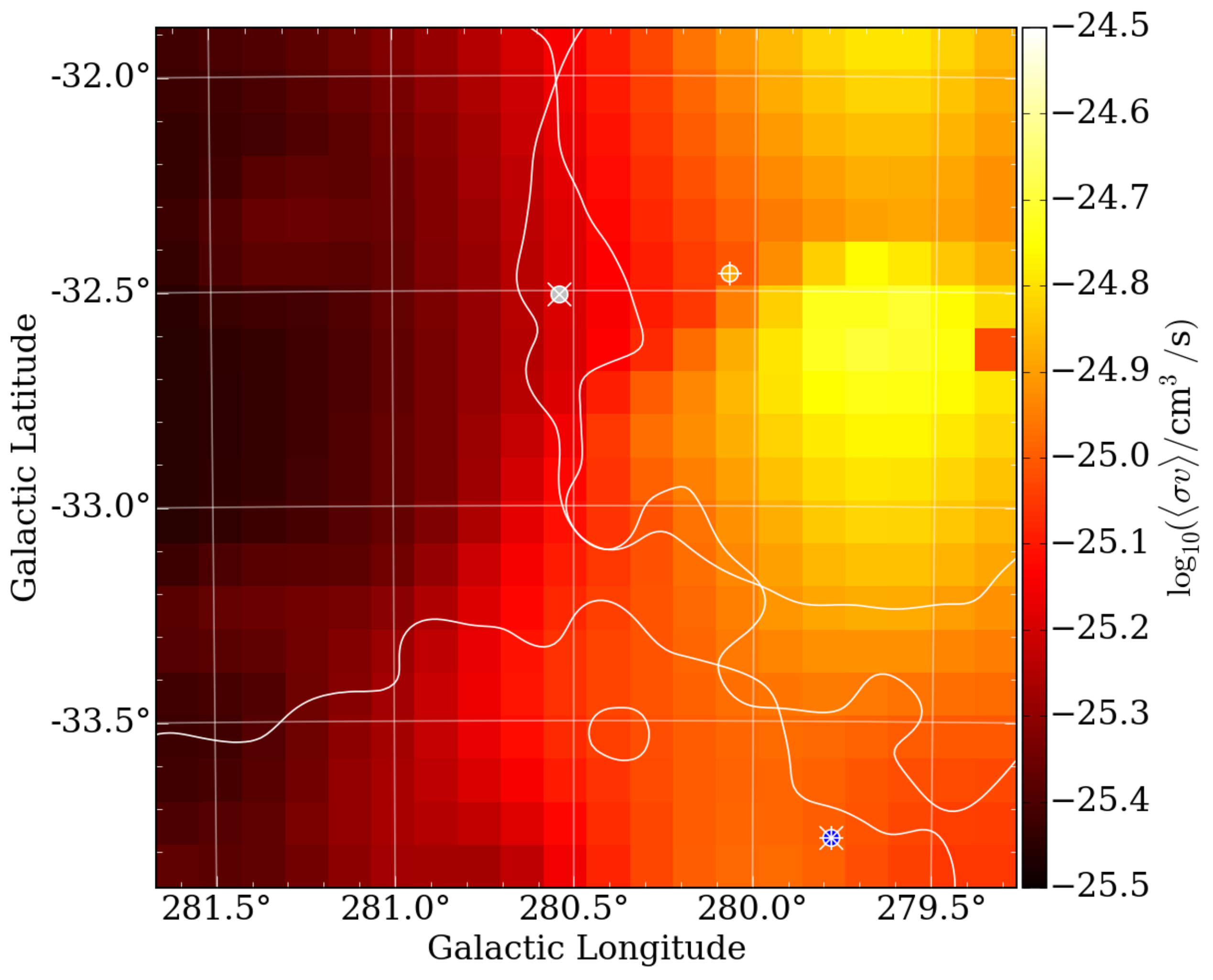}\includegraphics[width=0.48\columnwidth]{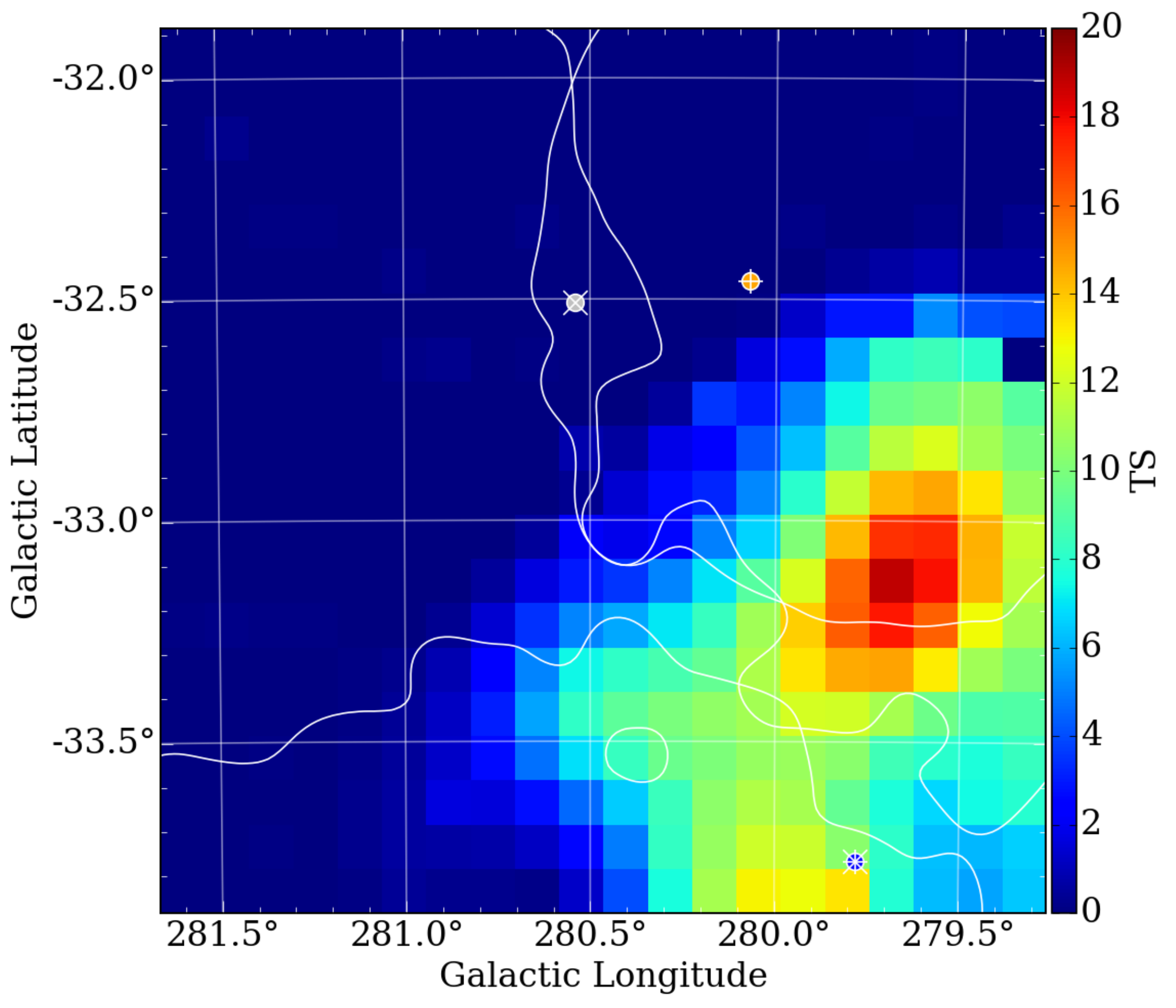}
\caption{Upper Left: 95\% CL upper bound on the annihilation of 50~GeV dark matter into $b\bar{b}$, assuming the {\tt sim-mean} profile, as a function of profile center across the entire ROI. Upper Right: TS for an additional component of 50~GeV dark matter annihilating into $b\bar{b}$, assuming the {\tt sim-mean} profile.
Lower Row: Cross section limits and TS values for the inner $4^\circ \times 4^\circ$ region of the LMC. 
Smoothed LMC background components are shown in white, along with three likely dark matter centers: {\tt stellar} (white circle with $\times$ cross), {\tt outer} (orange circle with $+$ cross), and {\tt HI} (blue circle with $\rlap{+}{\times}$ cross); the grid spacing is $0\fdg 5\times 0\fdg 5$ for the upper plots and $0\fdg 2\times 0\fdg 2$ for the lower plots. \label{fig:envmean_map}}
\end{figure}

\begin{figure}[ht]
\includegraphics[width=0.5\columnwidth]{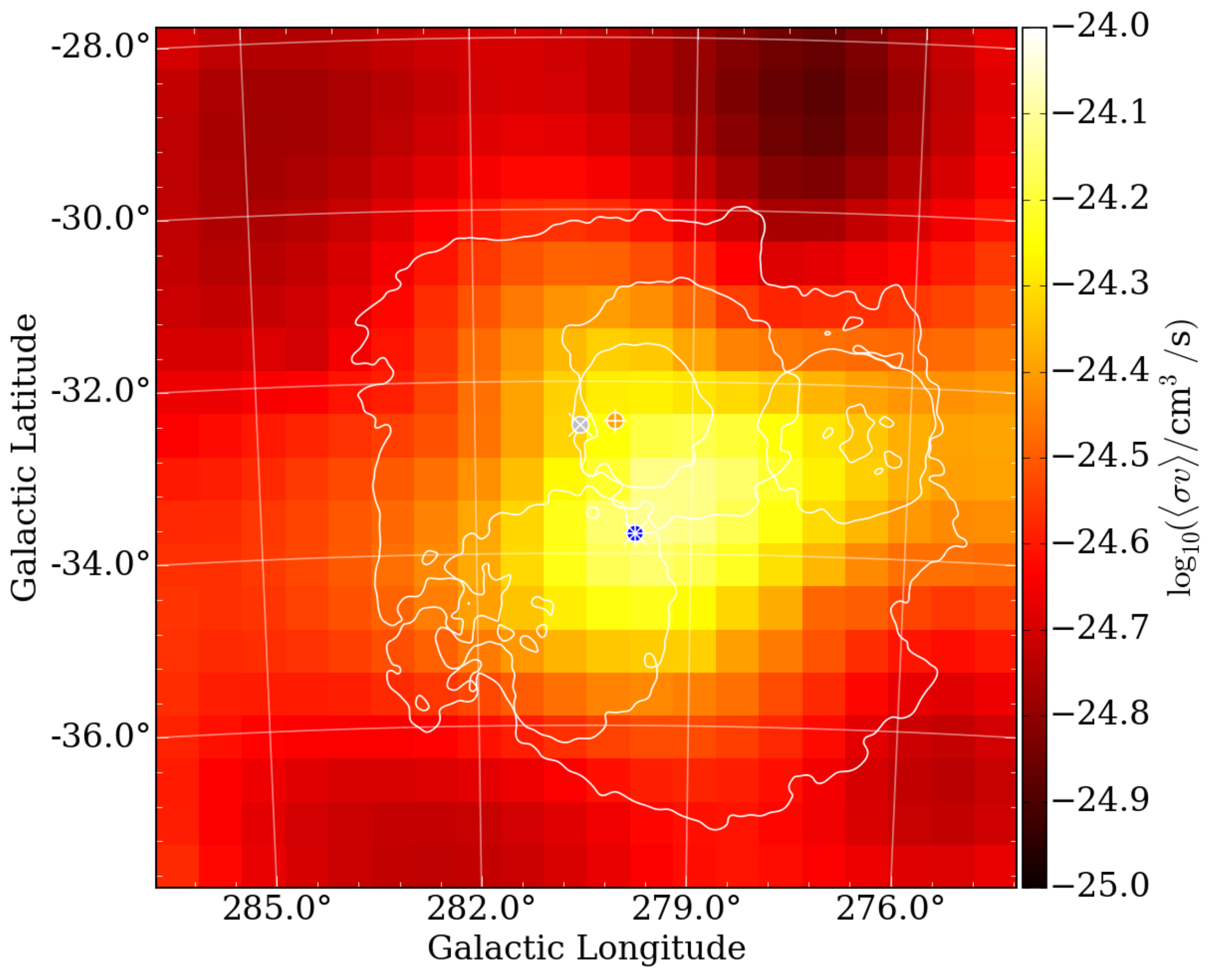}\includegraphics[width=0.48\columnwidth]{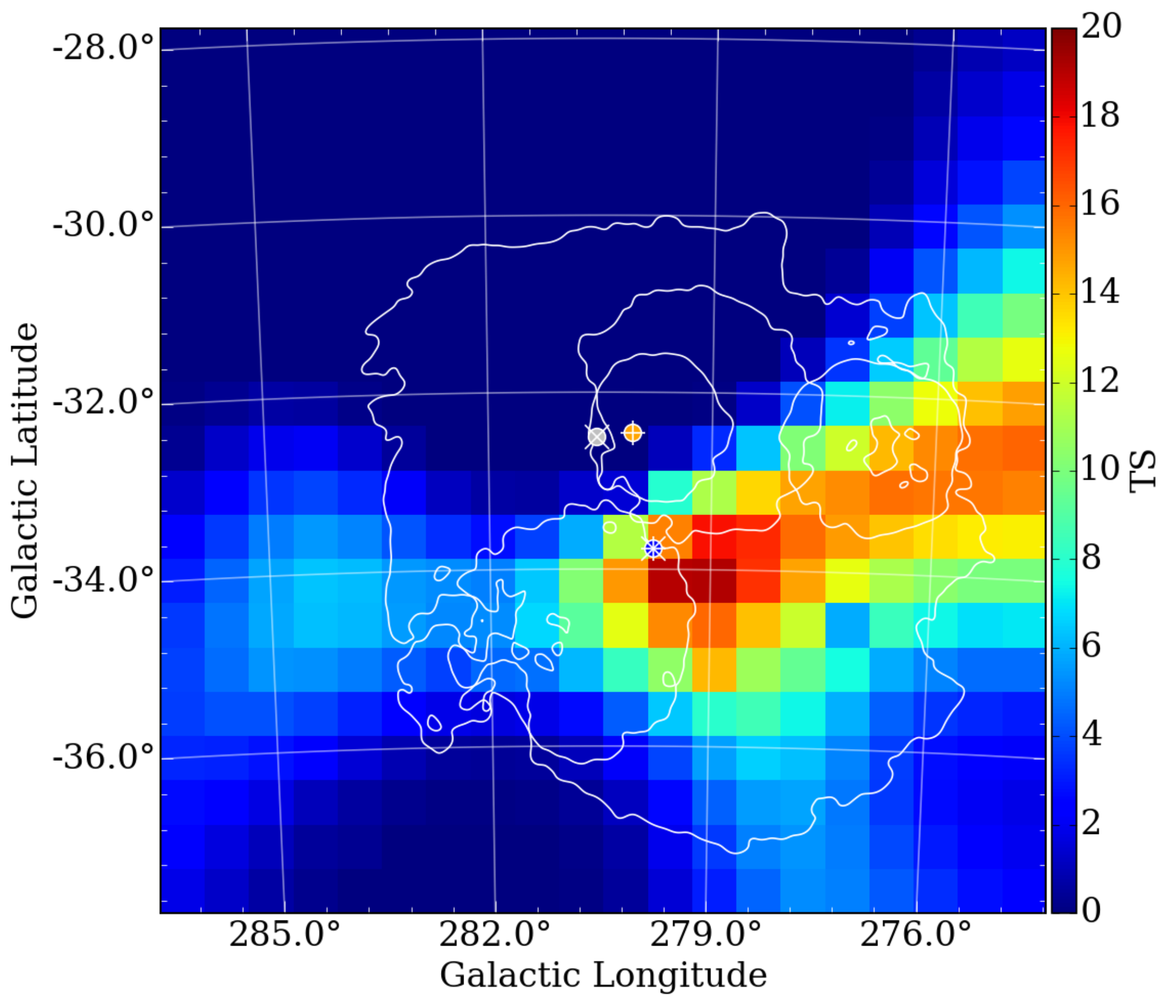}
\includegraphics[width=0.5\columnwidth]{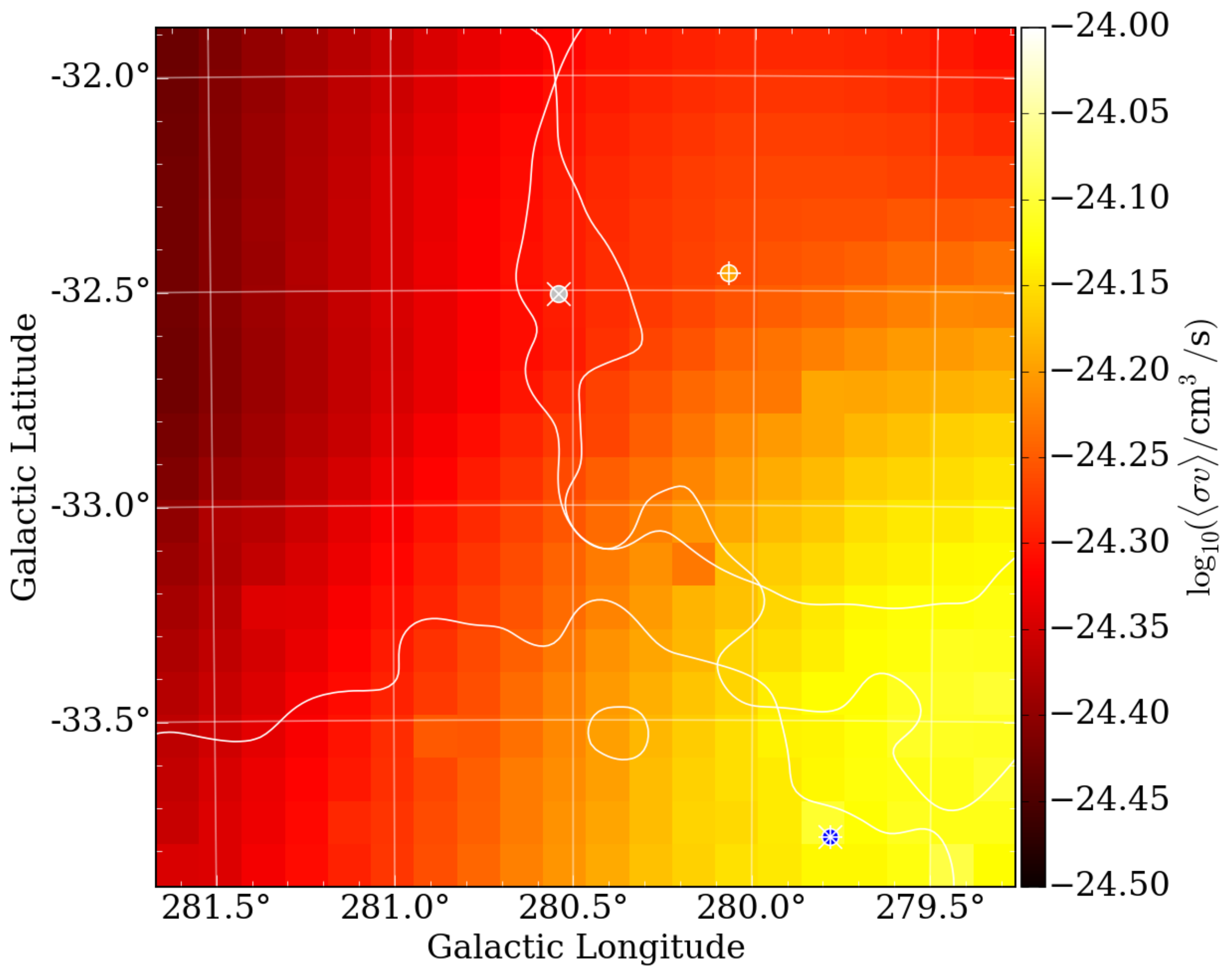}\includegraphics[width=0.48\columnwidth]{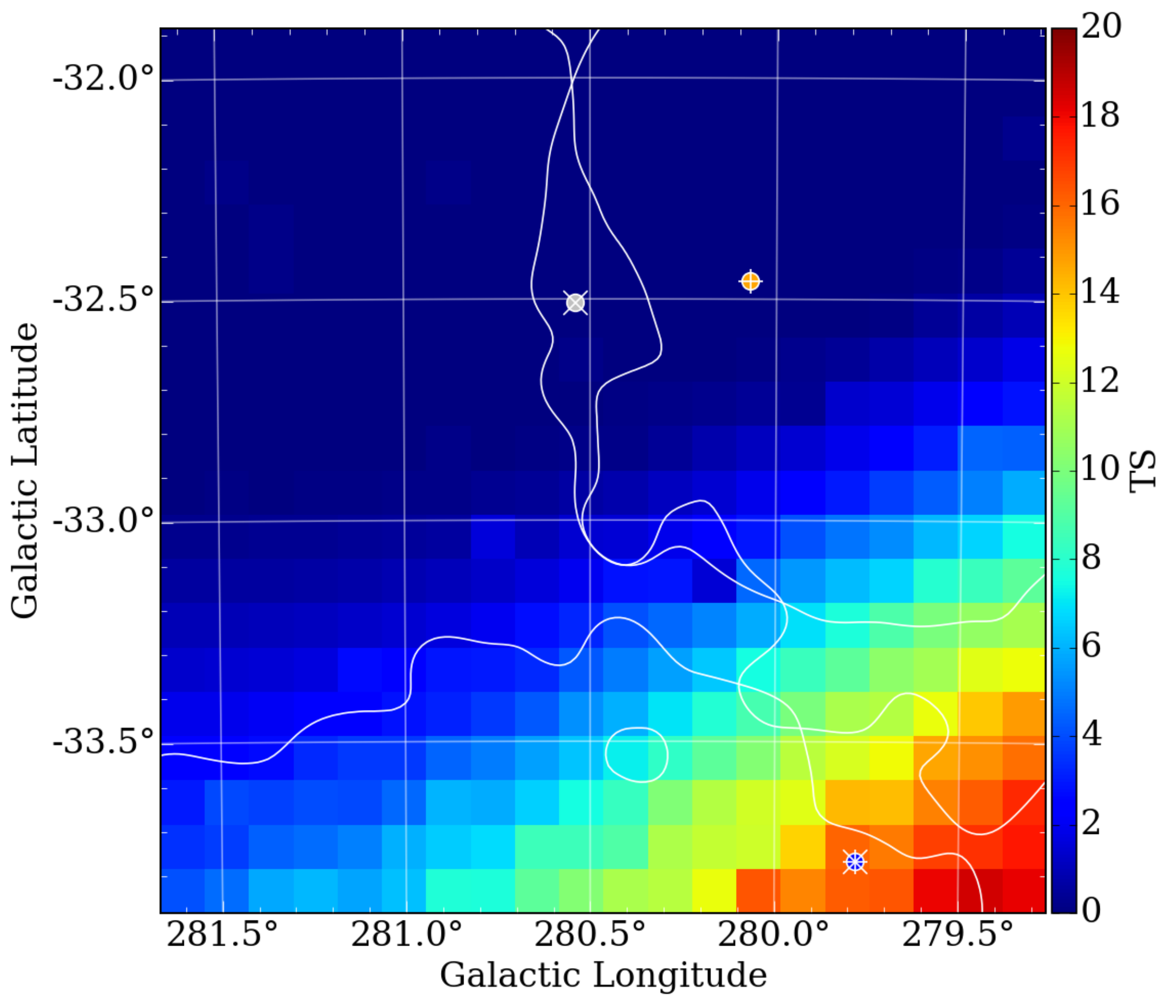}
\caption{Upper Left: 95\% CL upper bound on the annihilation of 50~GeV dark matter into $b\bar{b}$, assuming the {\tt iso-mean} profile, as a function of profile center across the entire ROI. Upper Right: TS for an additional component of 50~GeV dark matter annihilating into $b\bar{b}$, assuming the {\tt iso-mean} profile.
Lower Row: Cross section limits and TS values for the inner $4^\circ \times 4^\circ$ region of the LMC. 
Smoothed LMC background components are shown in white, along with three likely dark matter centers: {\tt stellar} (white circle with $\times$ cross), {\tt outer} (orange circle with $+$ cross), and {\tt HI} (blue circle with $\rlap{+}{\times}$ cross);  the grid spacing is $0\fdg 5\times 0\fdg 5$ for the upper plots and $0\fdg 2\times 0\fdg 2$ for the lower plots.\label{fig:isomin_map}}
\end{figure}

\begin{figure}[ht]
\includegraphics[width=0.5\columnwidth]{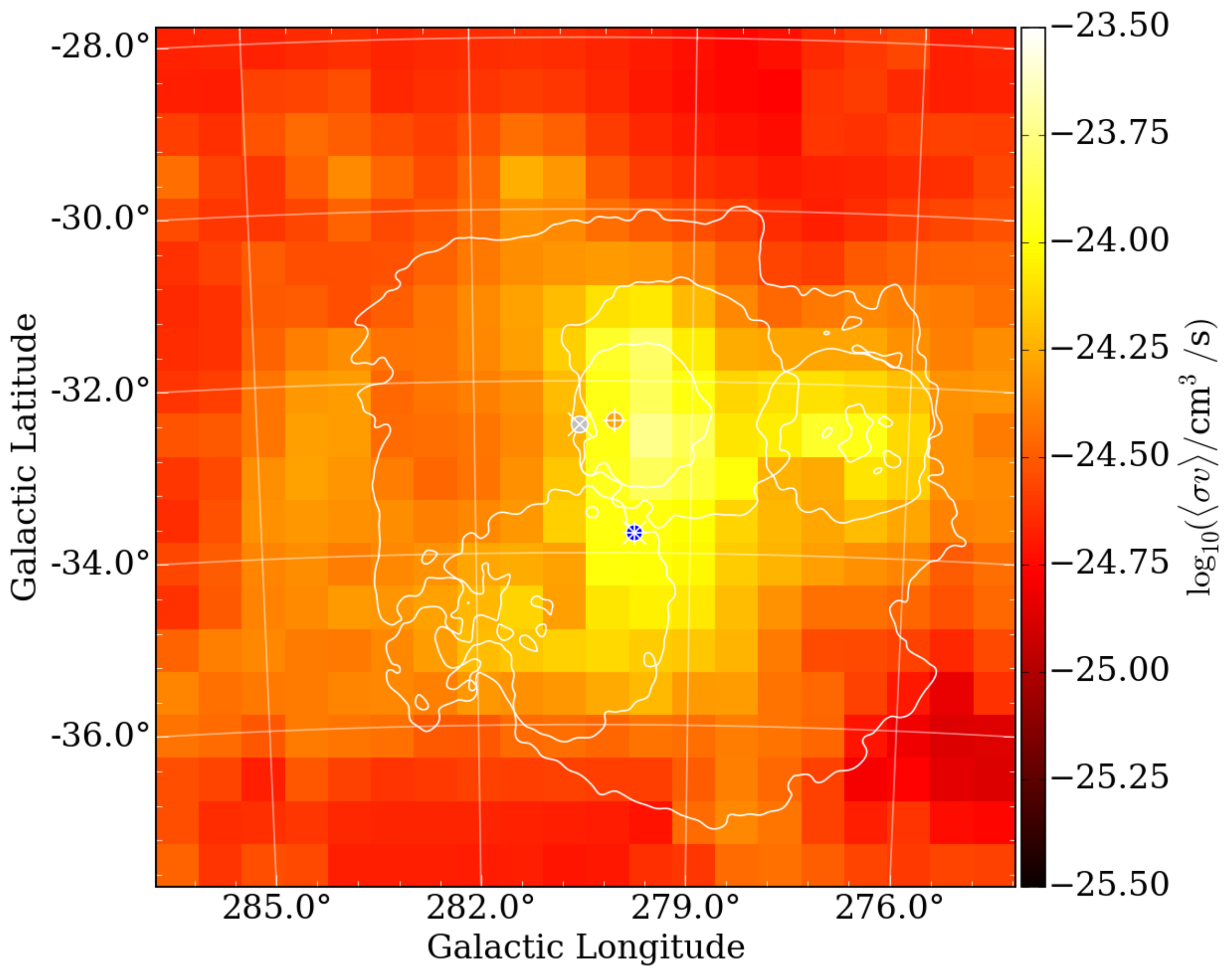}\includegraphics[width=0.48\columnwidth]{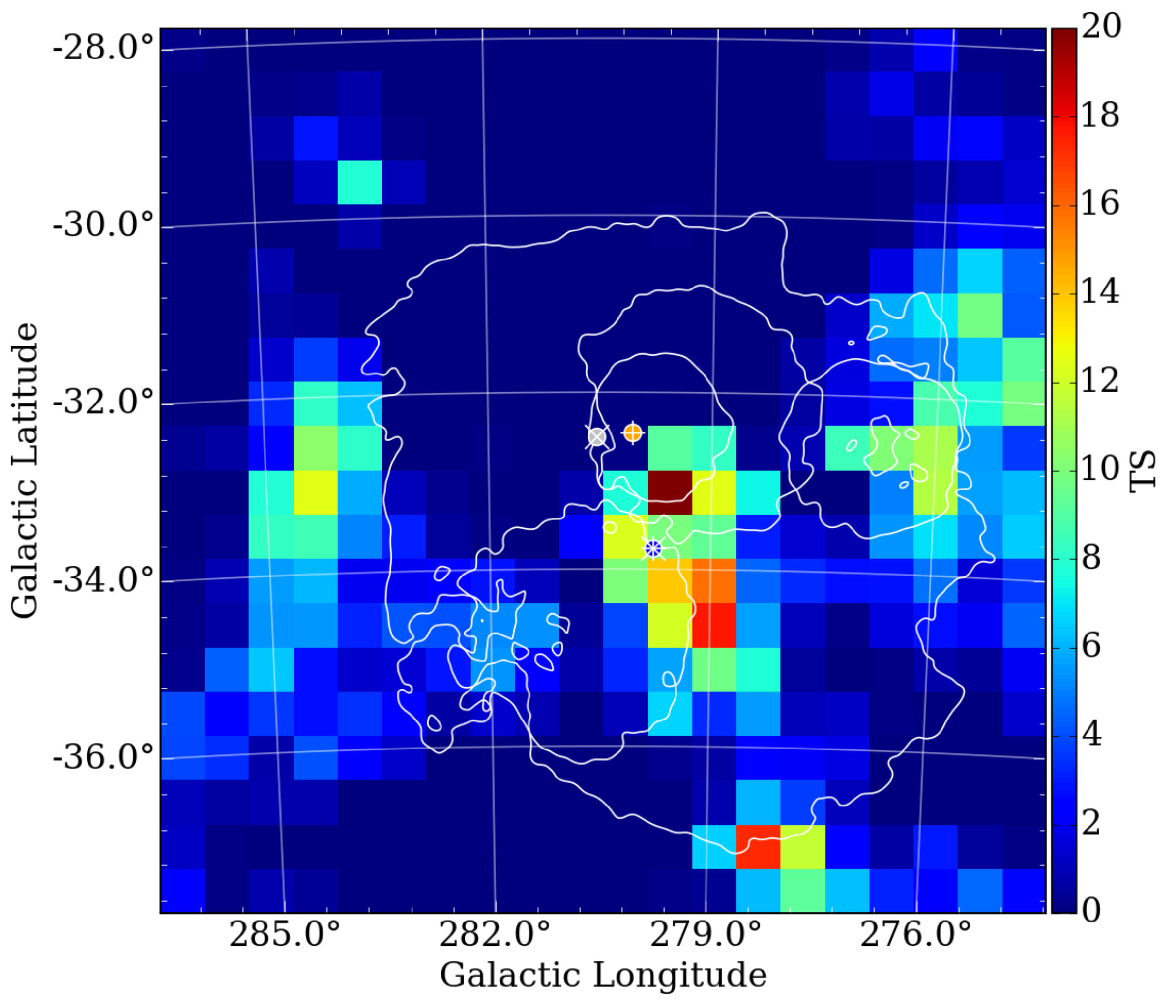}
\includegraphics[width=0.5\columnwidth]{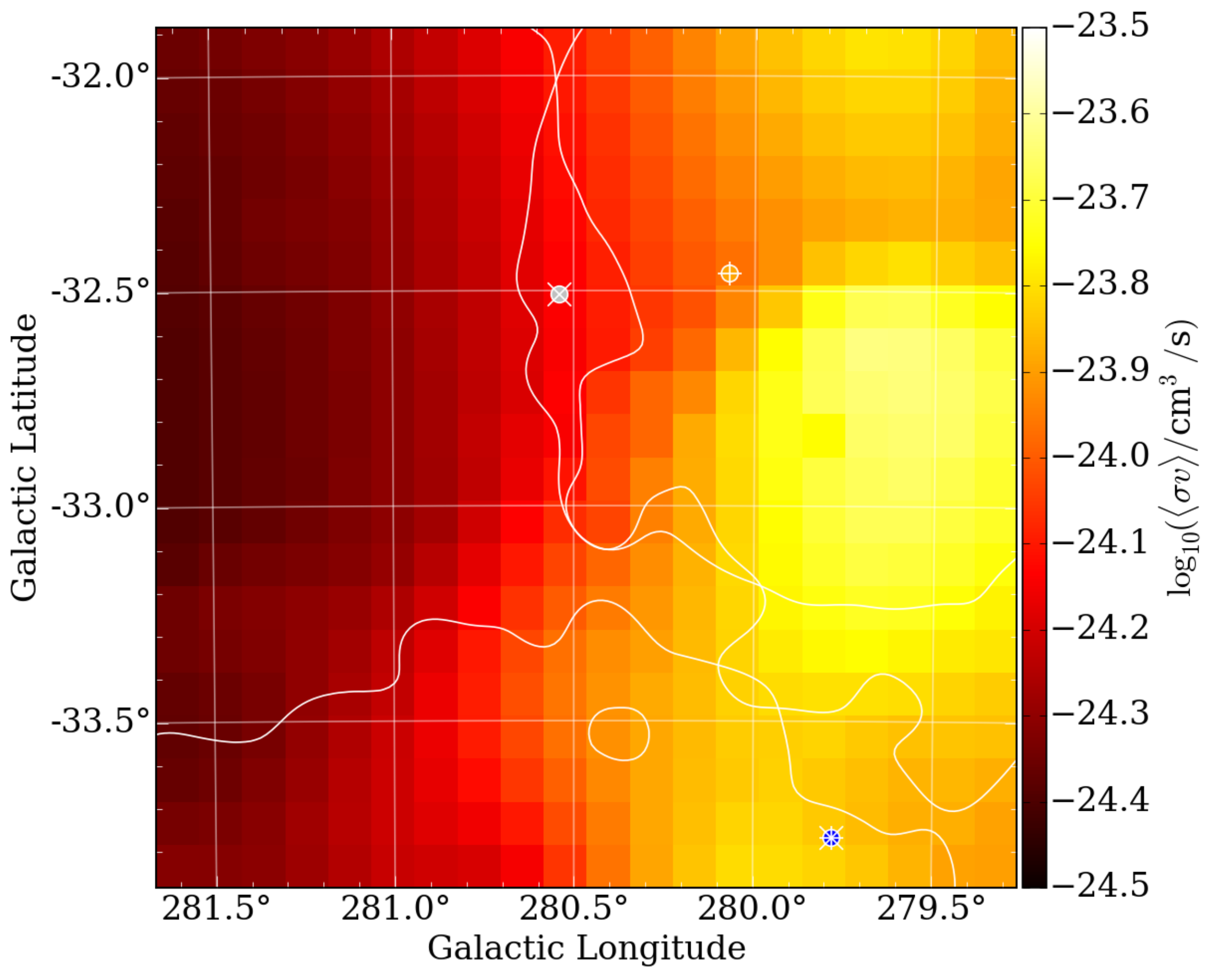}\includegraphics[width=0.48\columnwidth]{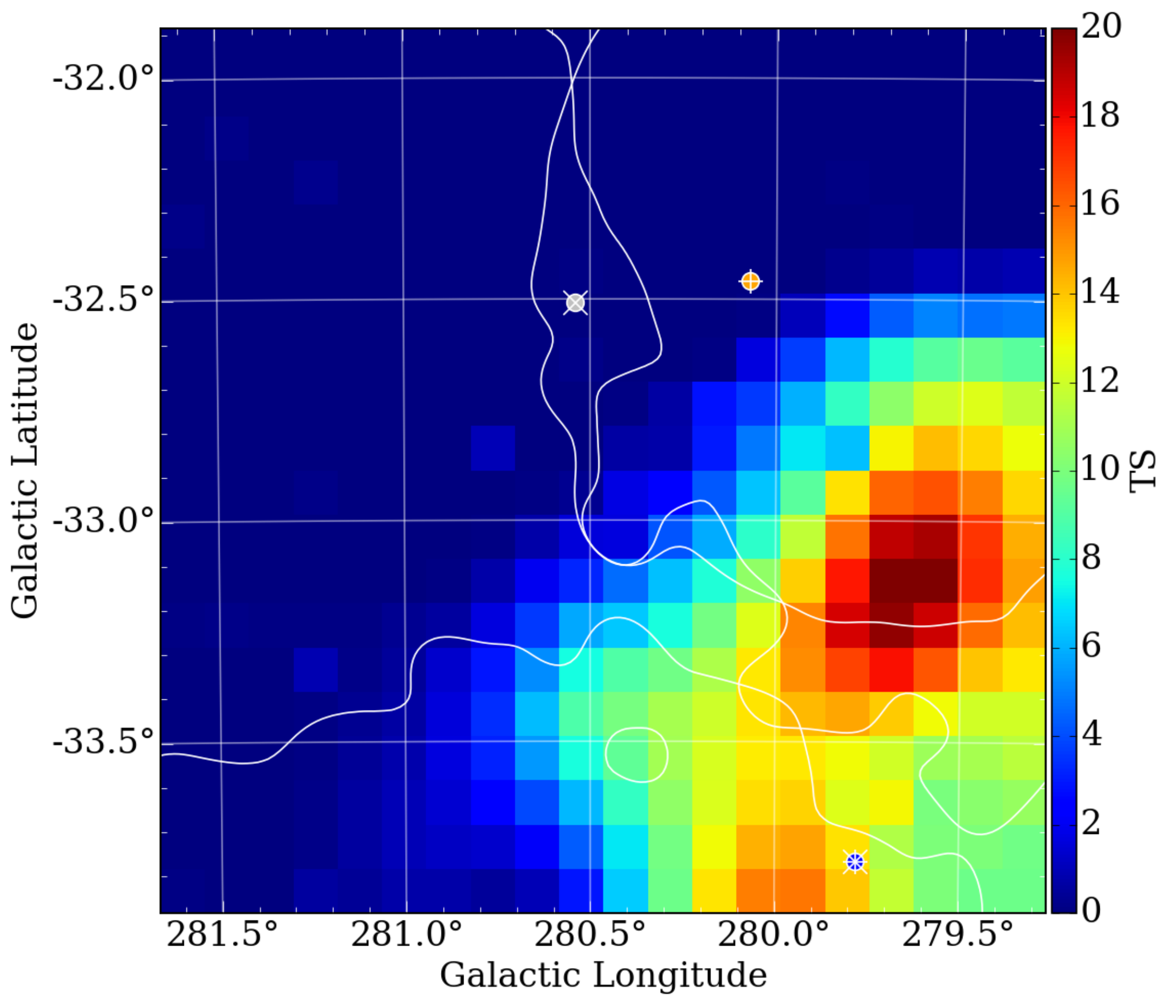}
\caption{Upper Left: 95\% CL upper bound on the annihilation of 50~GeV dark matter into $b\bar{b}$, assuming the {\tt nfw-mean} profile, as a function of profile center across the entire ROI. Upper Right: TS for an additional component of 50~GeV dark matter annihilating into $b\bar{b}$, assuming the {\tt nfw-mean} profile.
Lower Row: Cross section limits and TS values for the inner $4^\circ \times 4^\circ$ region of the LMC. 
Smoothed LMC background components are shown in white, along with three likely dark matter centers: {\tt stellar} (white circle with $\times$ cross), {\tt outer} (orange circle with $+$ cross), and {\tt HI} (blue circle with $\rlap{+}{\times}$ cross);  the grid spacing is $0\fdg 5\times 0\fdg 5$ for the upper plots and $0\fdg 2\times 0\fdg 2$ for the lower plots.\label{fig:nfwmin_map}}
\end{figure}

As Figures~\ref{fig:envmean_map}, \ref{fig:isomin_map}, and \ref{fig:nfwmin_map} indicate (for the representative choice of 50~GeV dark matter annihilating into $b\bar{b}$), the bounds on annihilation cross sections for all masses and channels are stronger for the {\tt star} and {\tt outer} centers. We therefore concentrate on the weaker limits derived from assuming dark matter located at the {\tt HI} center. This does match our theoretical prejudice, which would have located the dark matter near the dynamical center of the H\,{\sc i} gas.  (Since the H\,{\sc i} can cool, it would be more likely to trace the current dynamical center.)  We note that all of these figures contain a broad region relatively close to the {\tt HI} center where the fitting procedure indicates some amount of dark matter annihilation with a TS of 15--20. We will discuss this excess shortly, but for now we make two comments. First, note that the shape of the excess in the TS map should {\it not} be misinterpreted as the shape of some spatial region with excess gamma-ray emission over the background model. The profiles of the dark matter $J$-factors extend out $1^\circ-2^\circ$ from the center (especially for the isothermal profiles, which do not have a cusp). The TS map shown indicates only the preference of the fit for the center location. Second, notice that the regions of high TS do not always align with the weakest upper limits on the annihilation cross section. If the background counts are large over a significant region, then the upper limits on $\langle \sigma v\rangle$ are weakened, even if there is no strong preference for dark matter annihilation over background in that region.

\begin{figure}

\includegraphics[width=0.33\columnwidth]{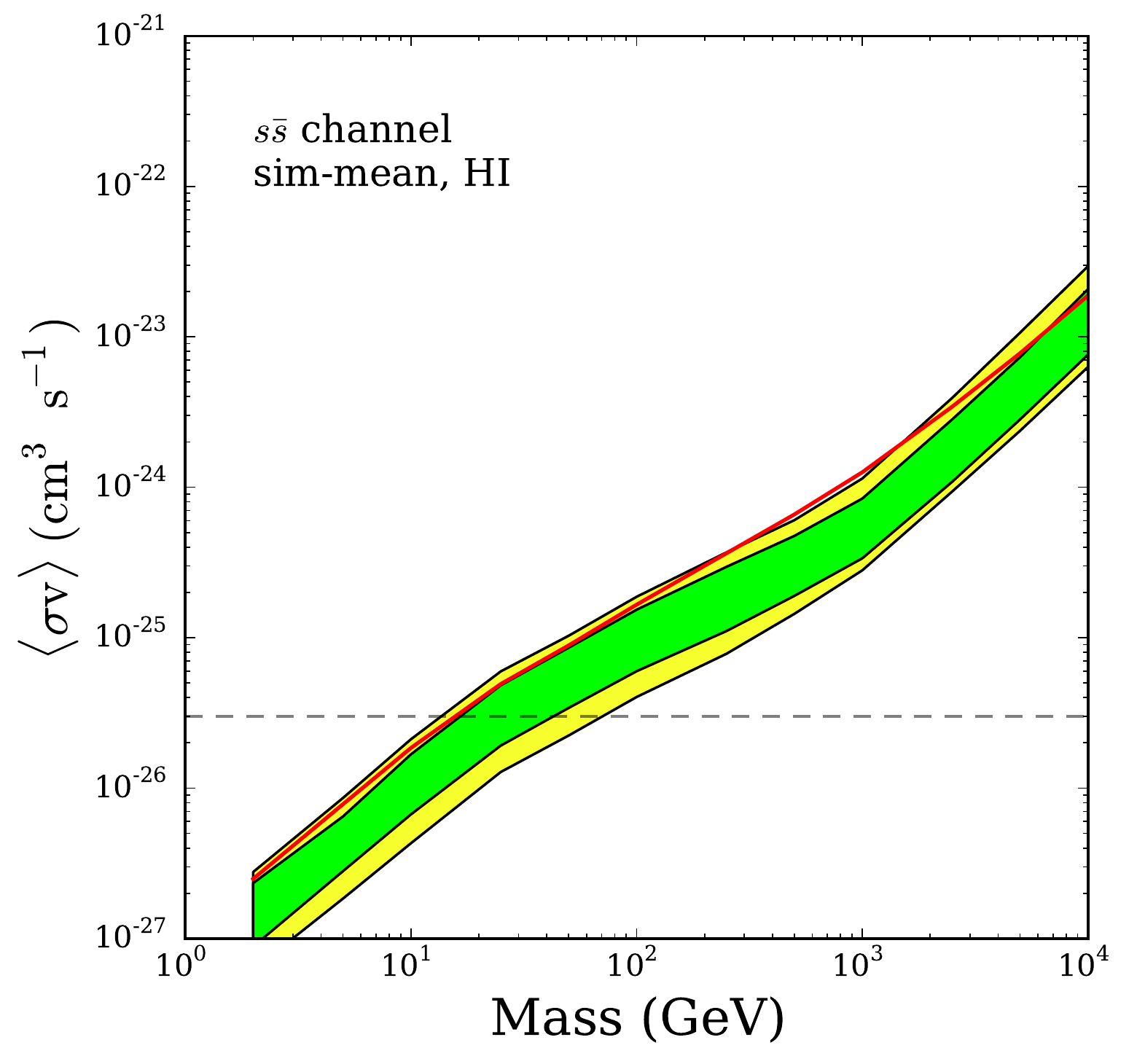}\includegraphics[width=0.33\columnwidth]{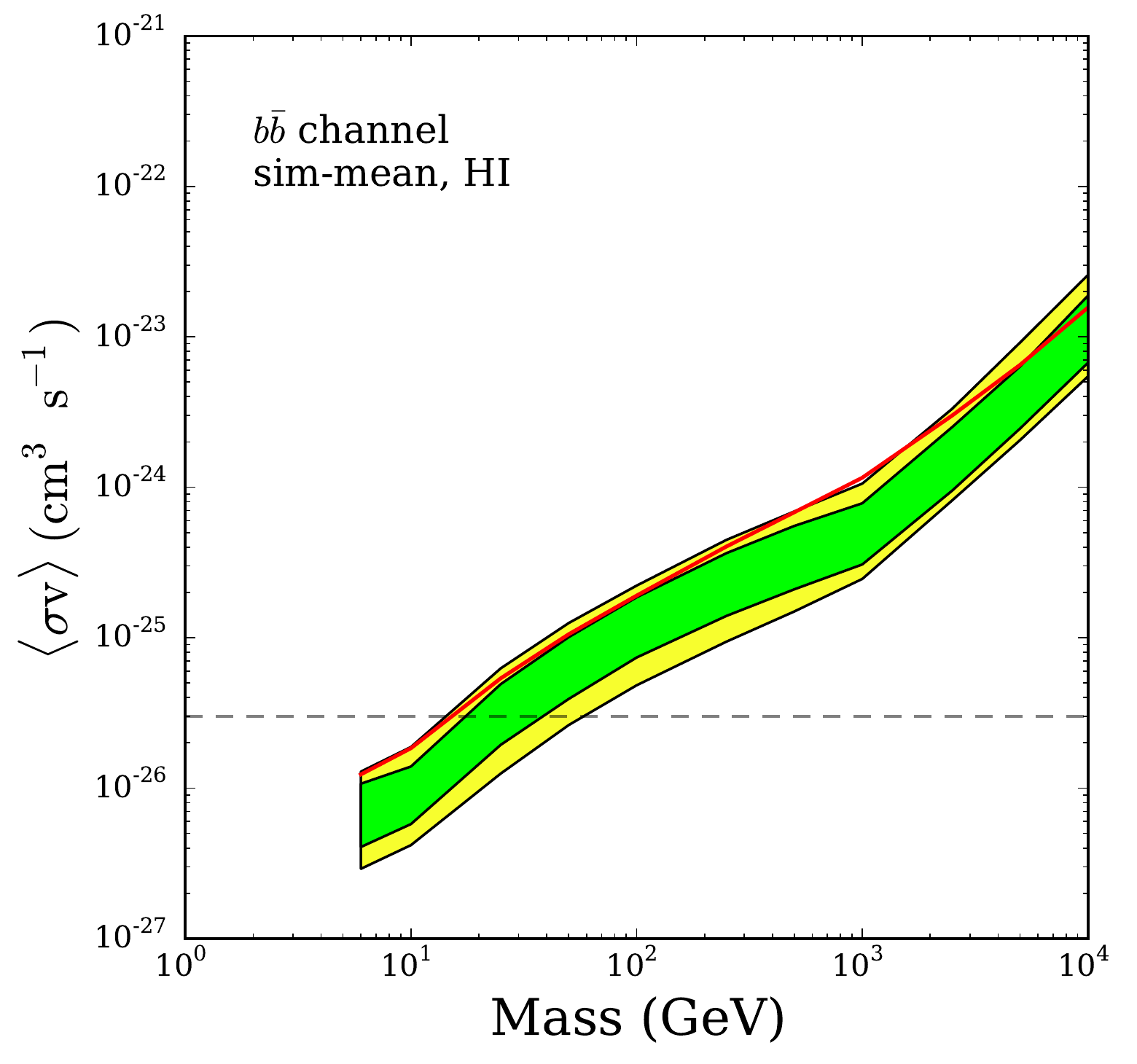}\includegraphics[width=0.33\columnwidth]{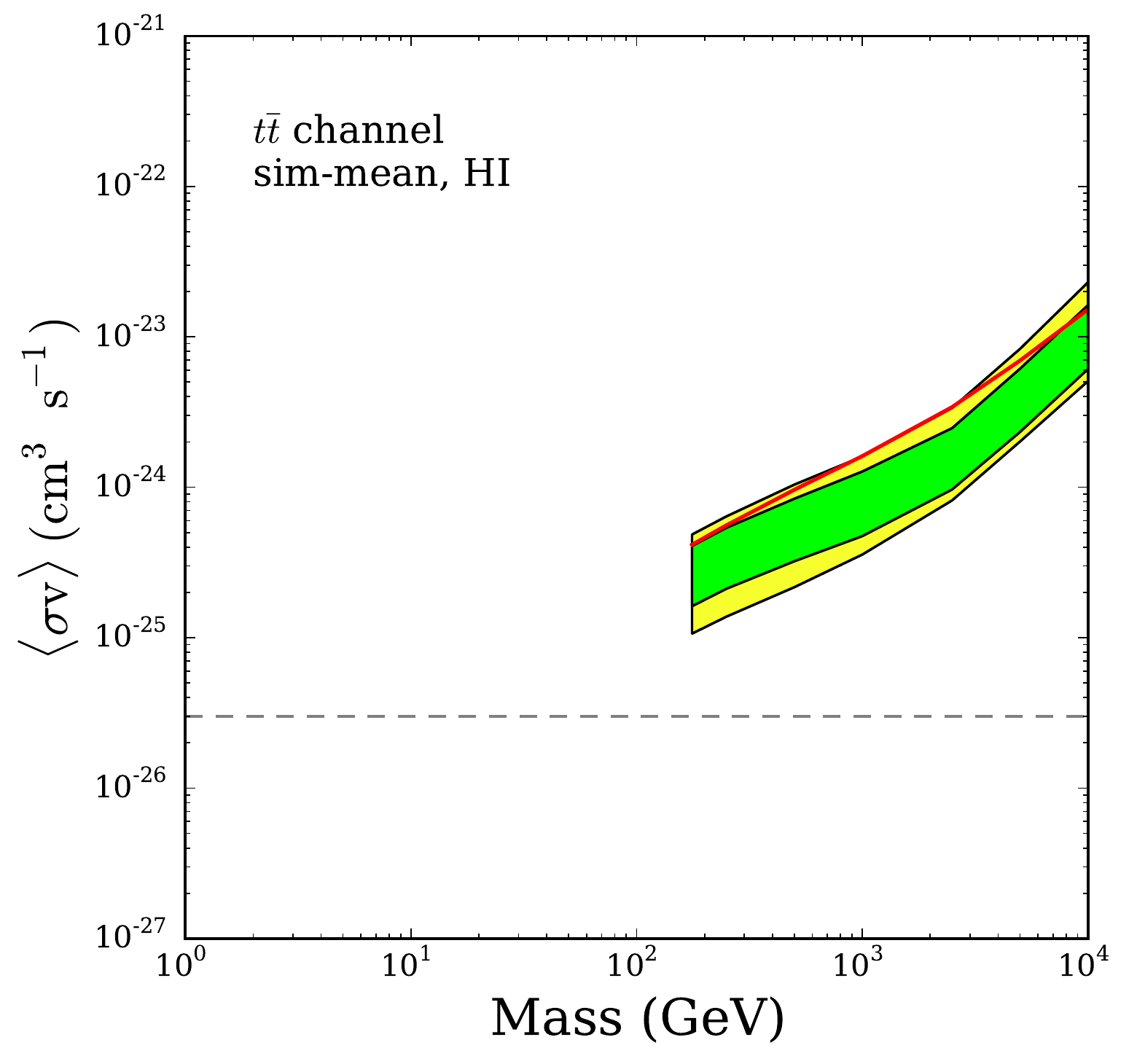}
\includegraphics[width=0.33\columnwidth]{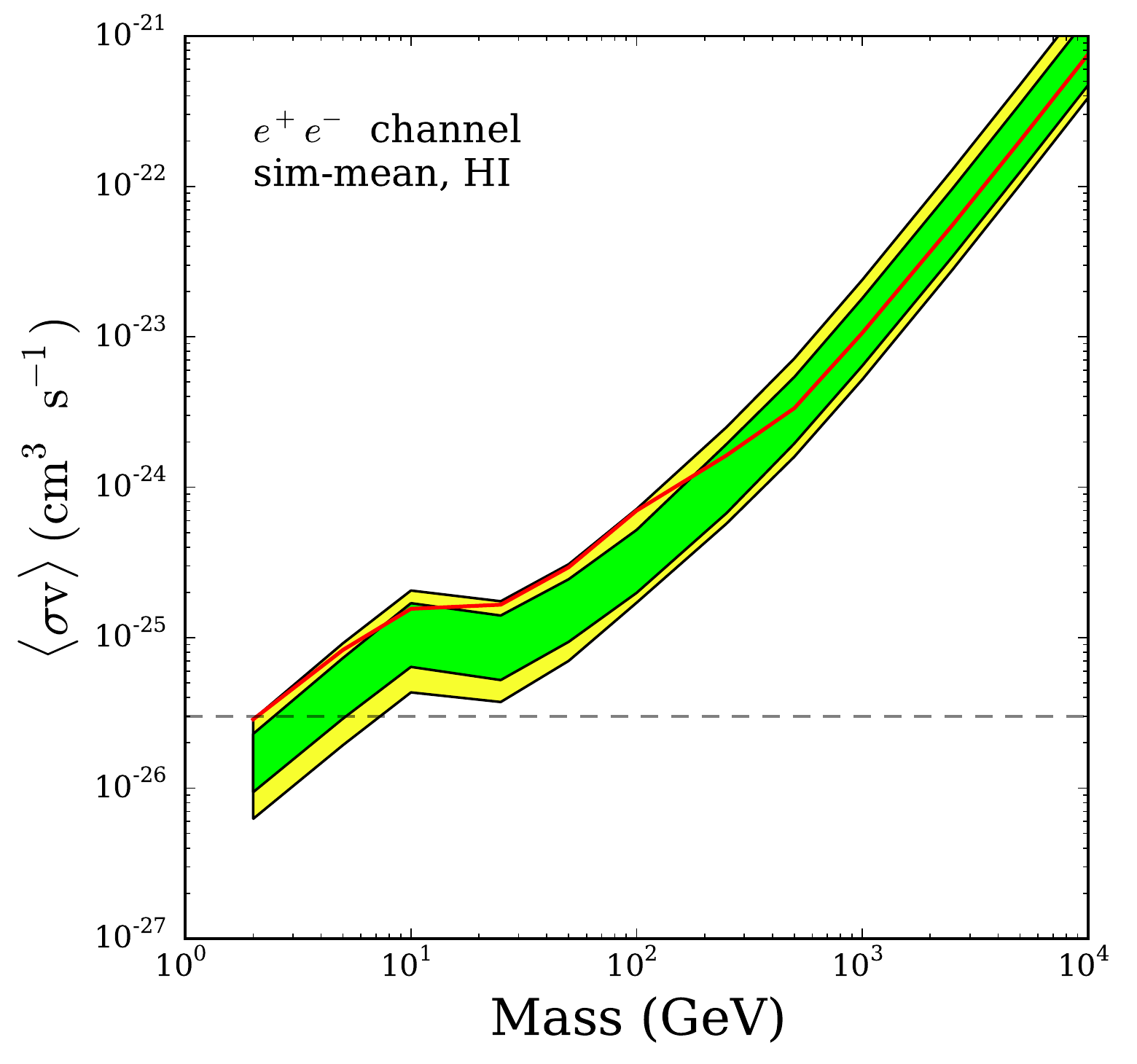}\includegraphics[width=0.33\columnwidth]{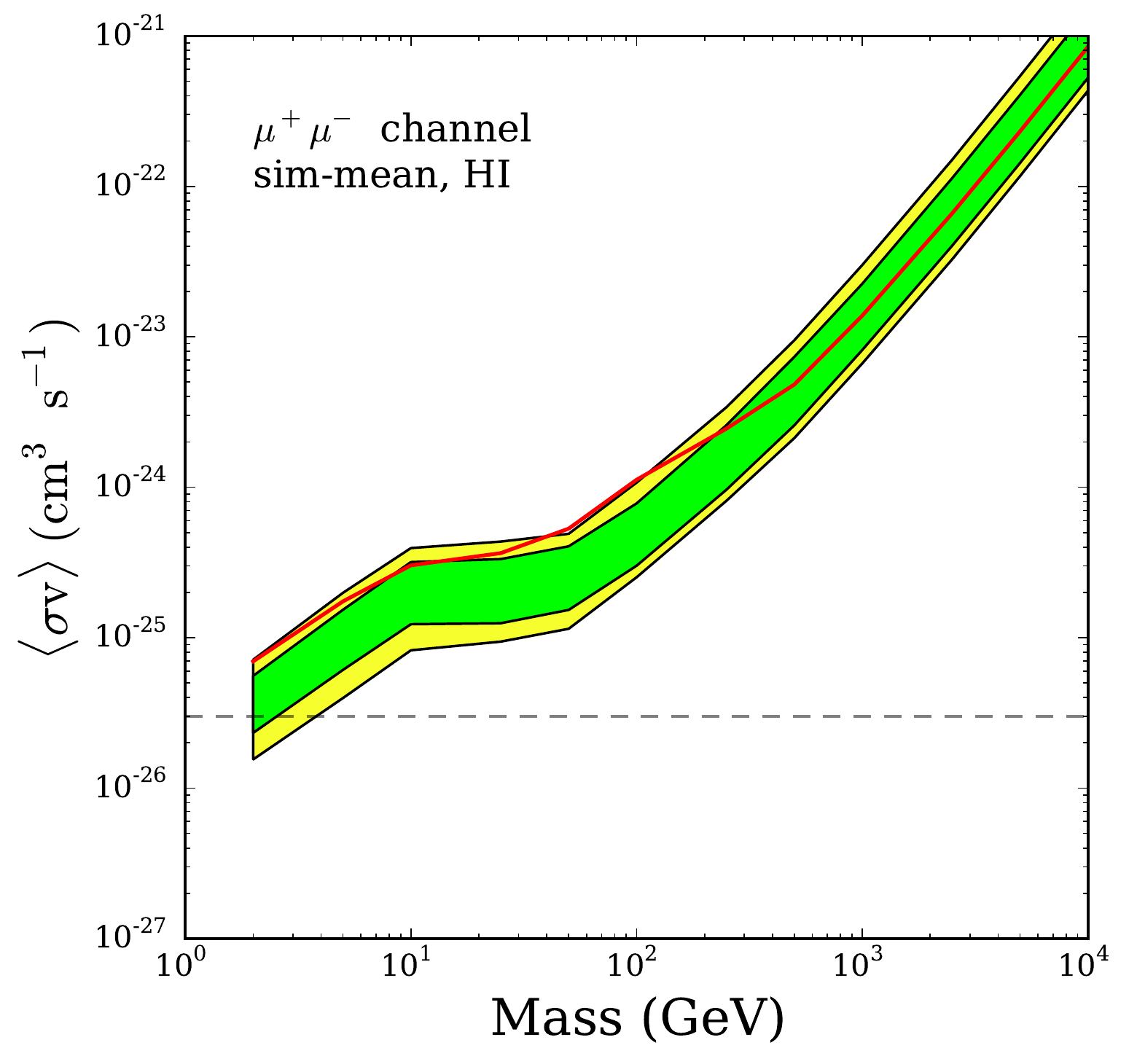}\includegraphics[width=0.33\columnwidth]{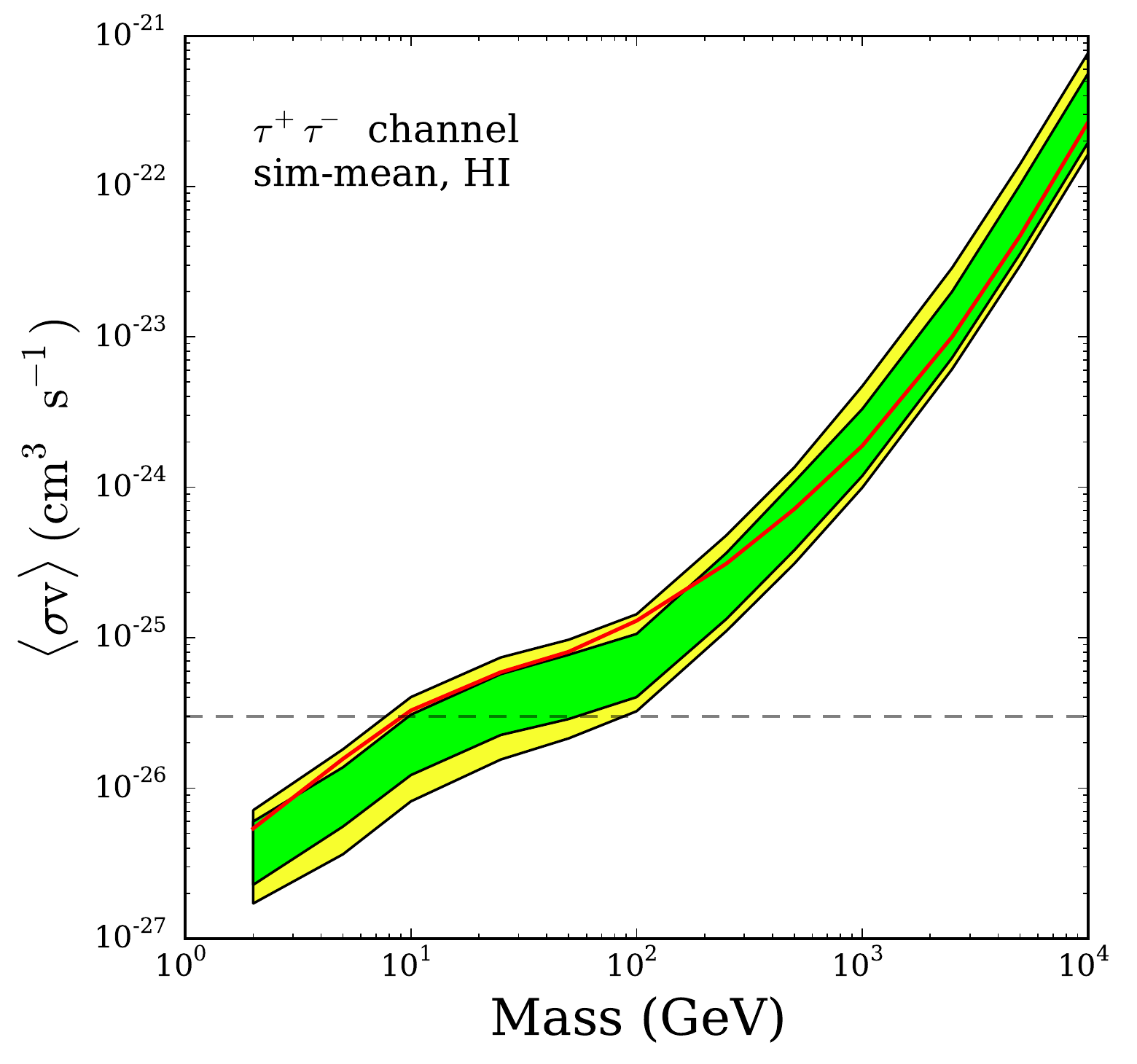}
\includegraphics[width=0.33\columnwidth]{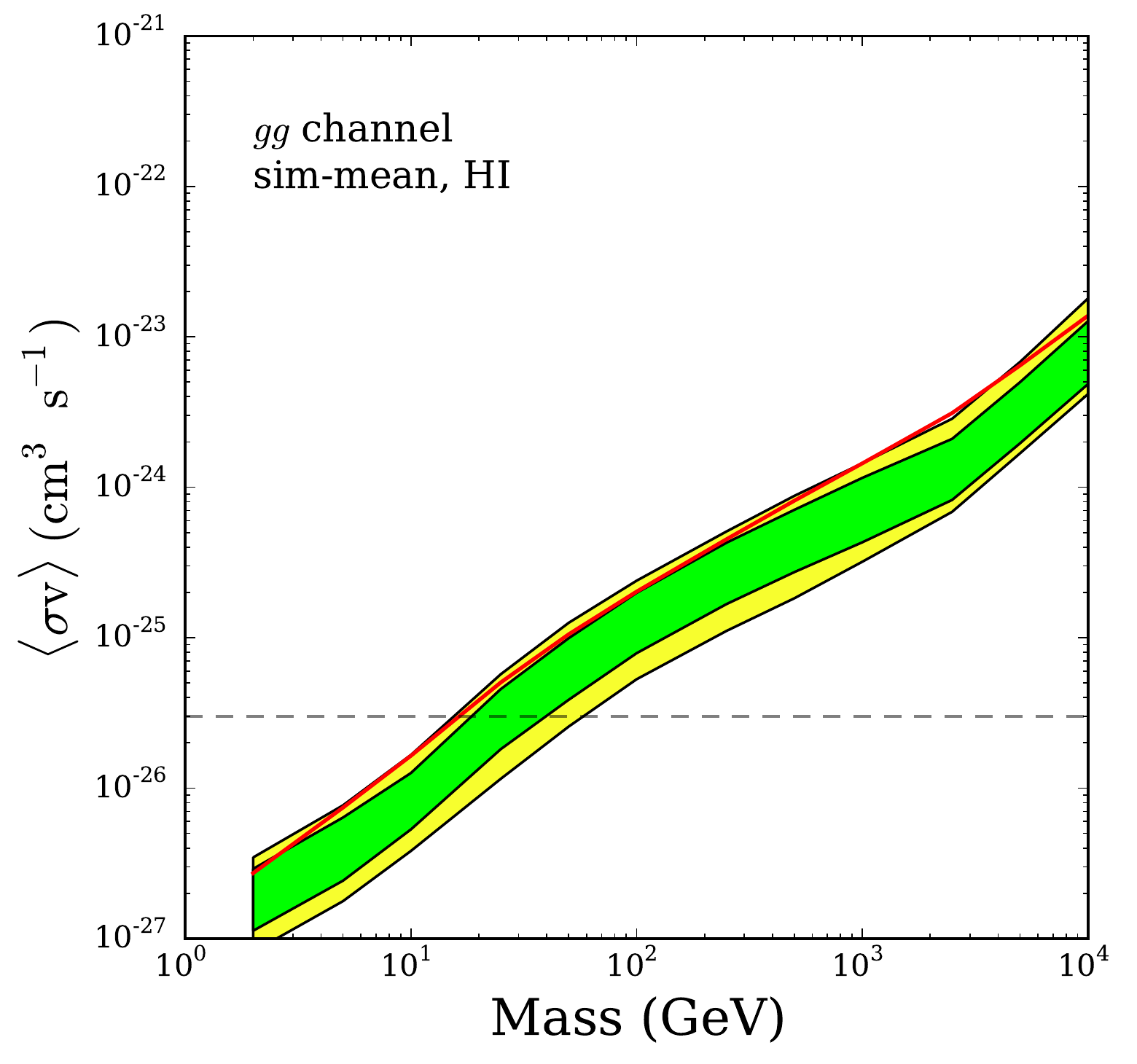}\includegraphics[width=0.33\columnwidth]{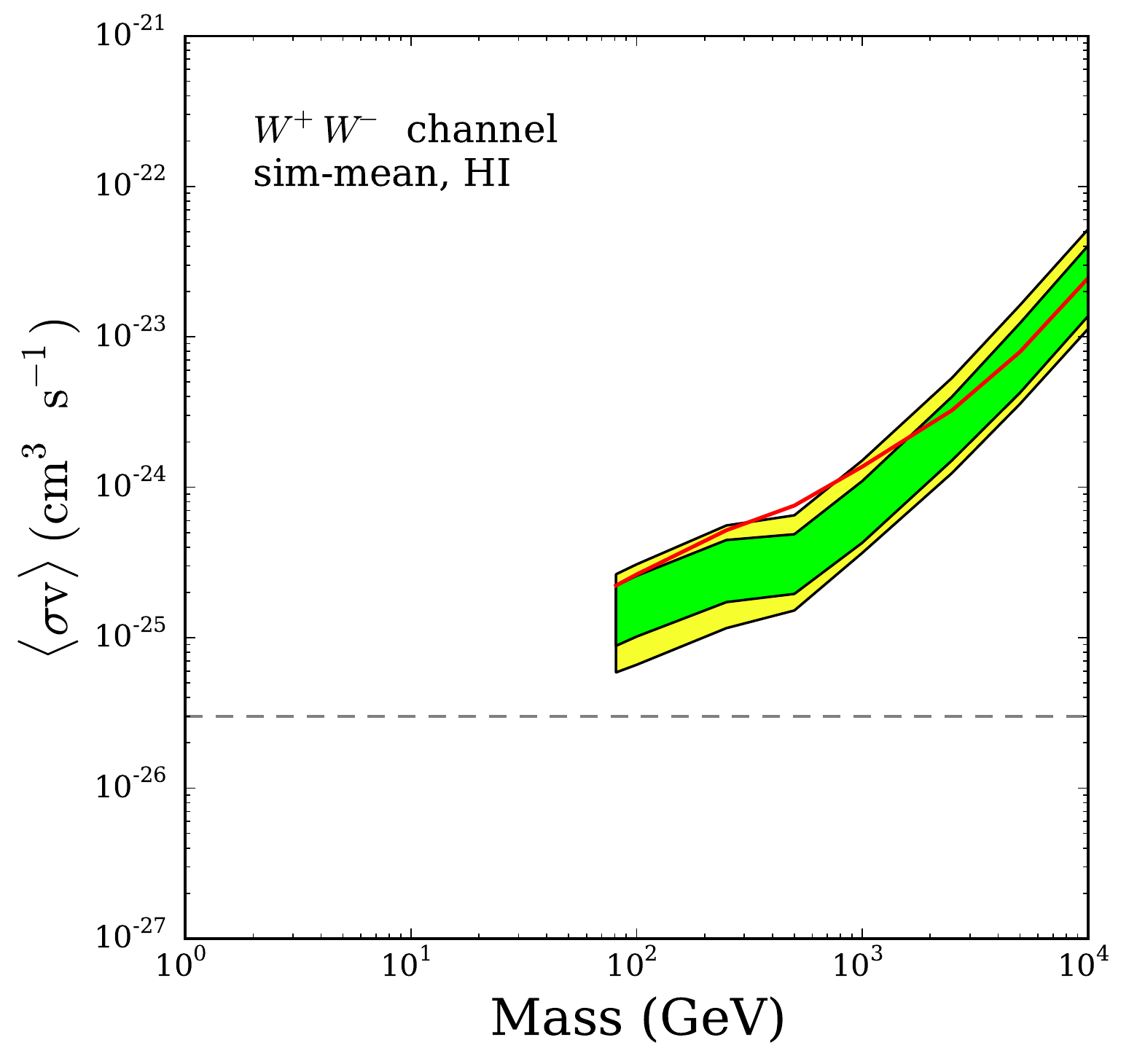}
\caption{Upper limits on $\langle \sigma v\rangle$ for the indicated annihilation channels (red), as a function of dark matter mass, assuming the {\tt sim-mean} profile located at the {\tt HI} center. Also shown are the 84\% (green) and 95\% (yellow) containment bands of the upper limit drawn from background only simulations.   The horizontal dashed line shows the canonical thermal relic cross section. \label{fig:envmean_results}}
\end{figure}

\begin{figure}

\includegraphics[width=0.33\columnwidth]{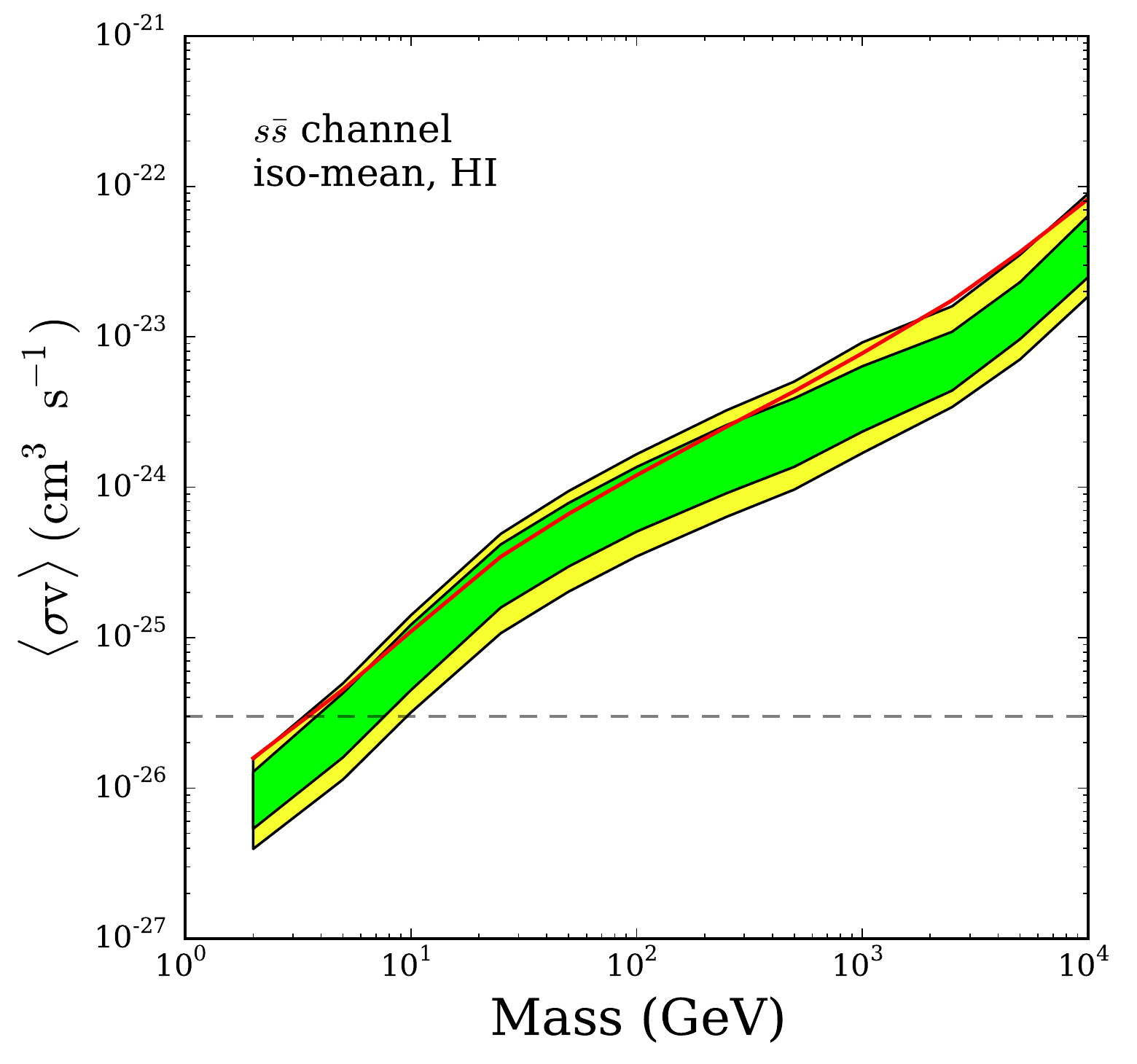}\includegraphics[width=0.33\columnwidth]{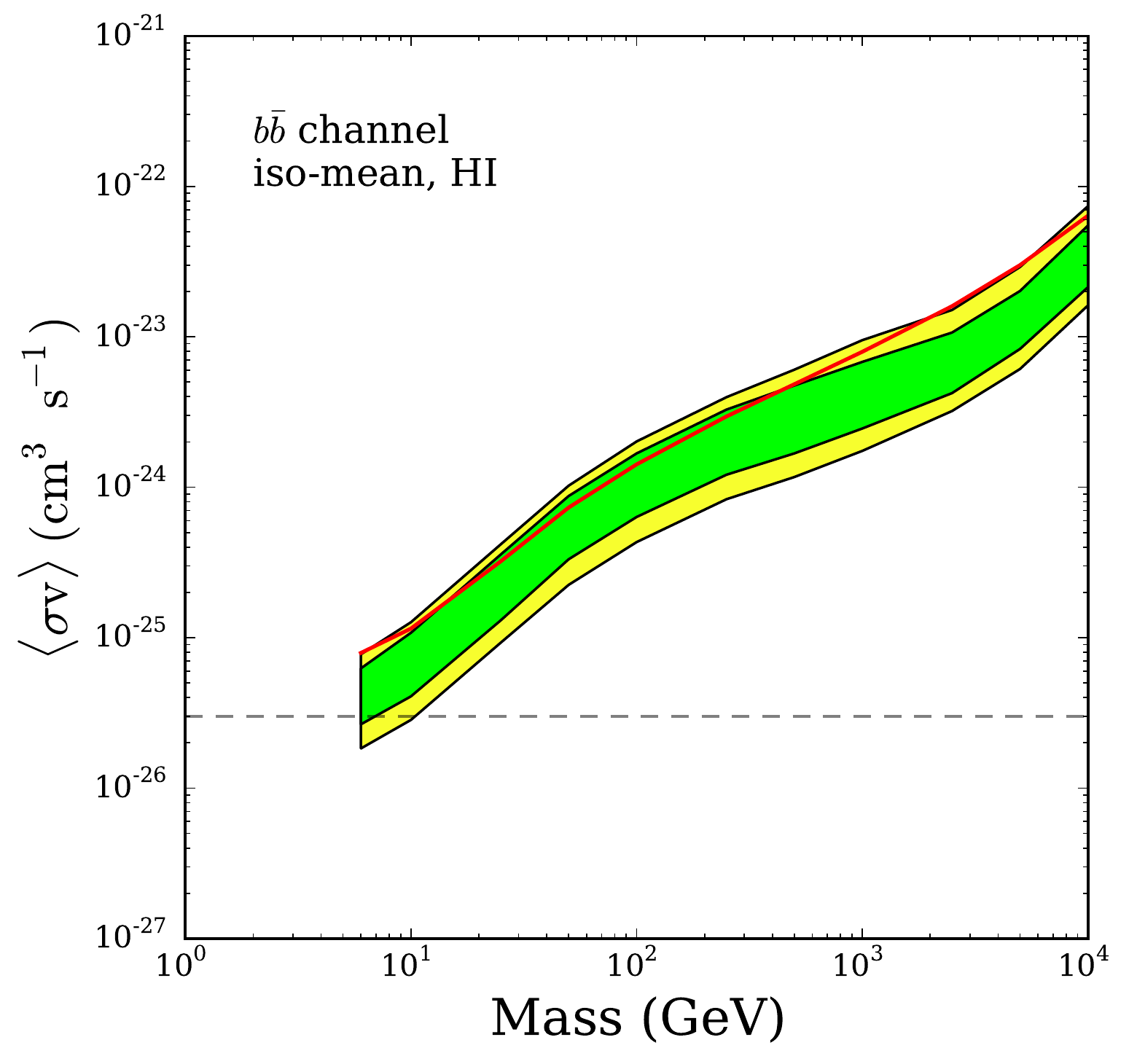}\includegraphics[width=0.33\columnwidth]{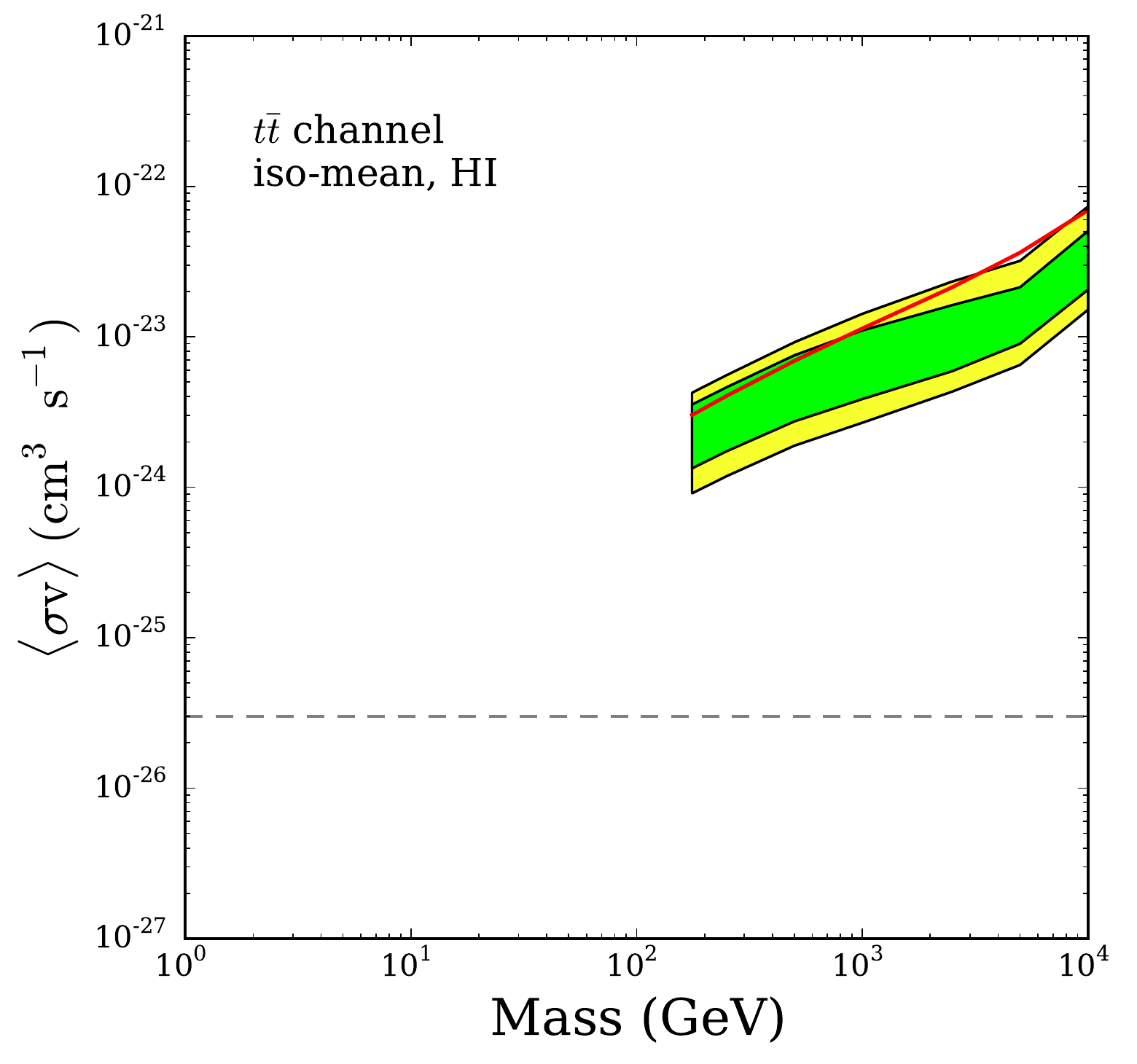}
\includegraphics[width=0.33\columnwidth]{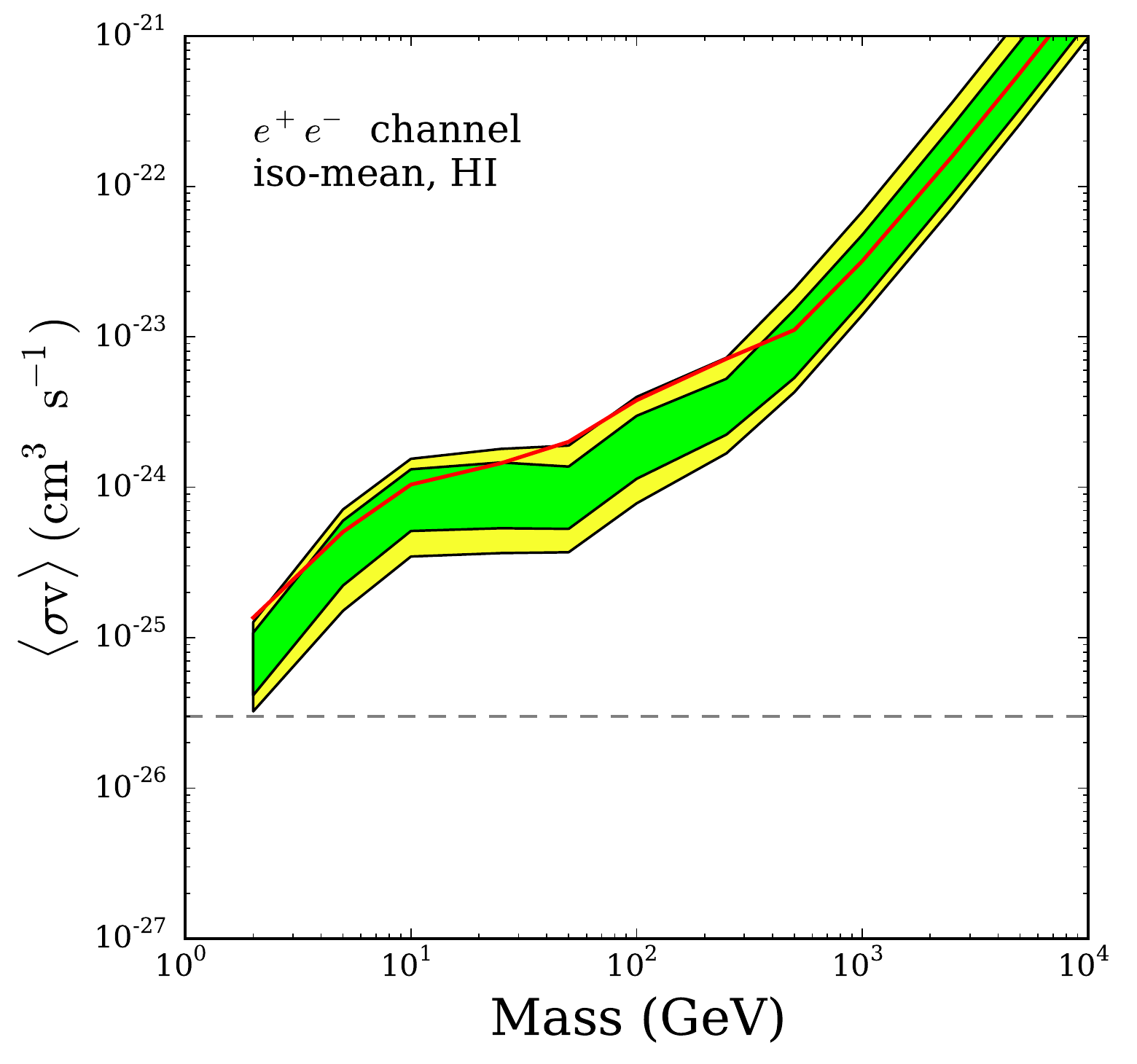}\includegraphics[width=0.33\columnwidth]{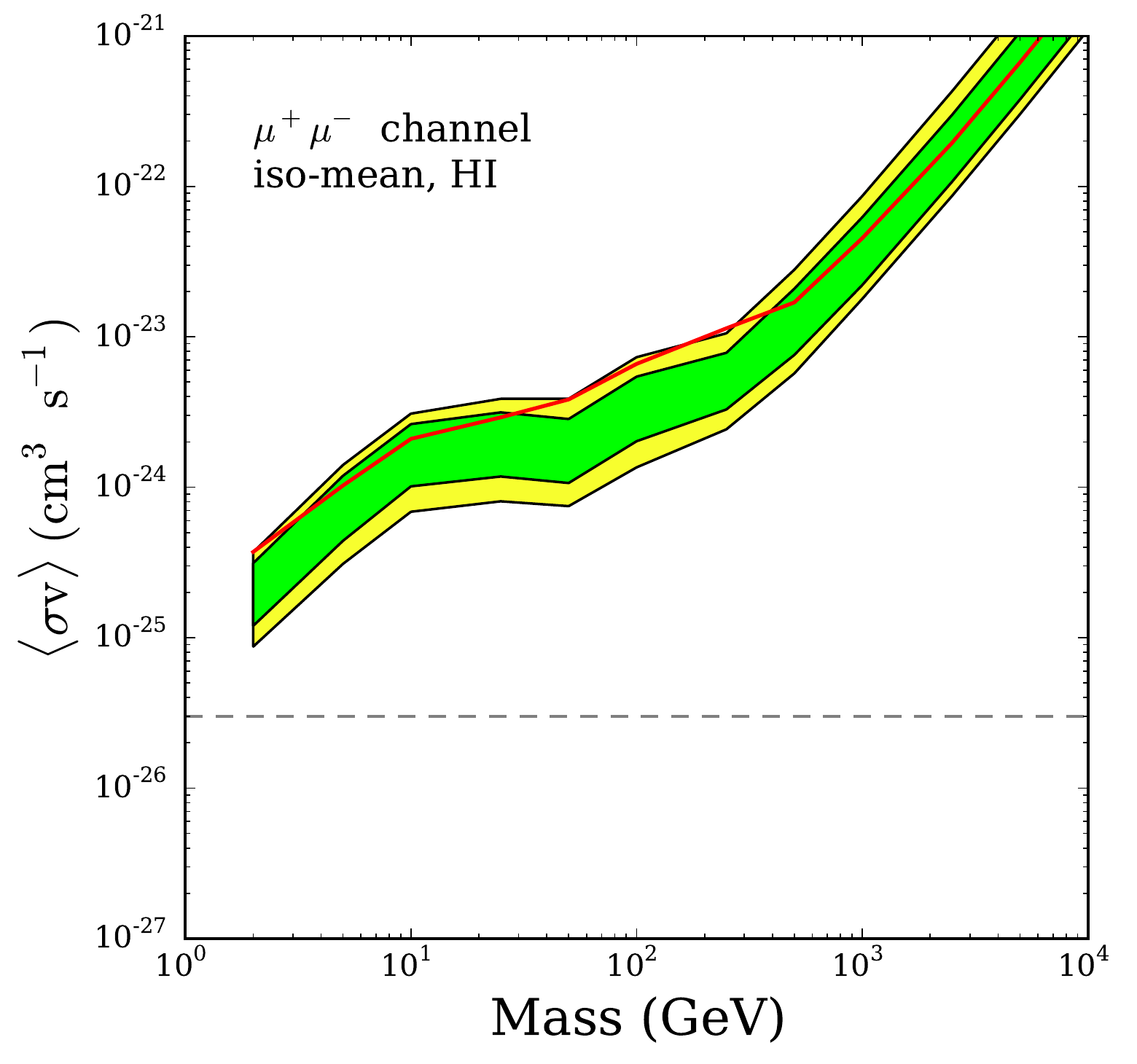}\includegraphics[width=0.33\columnwidth]{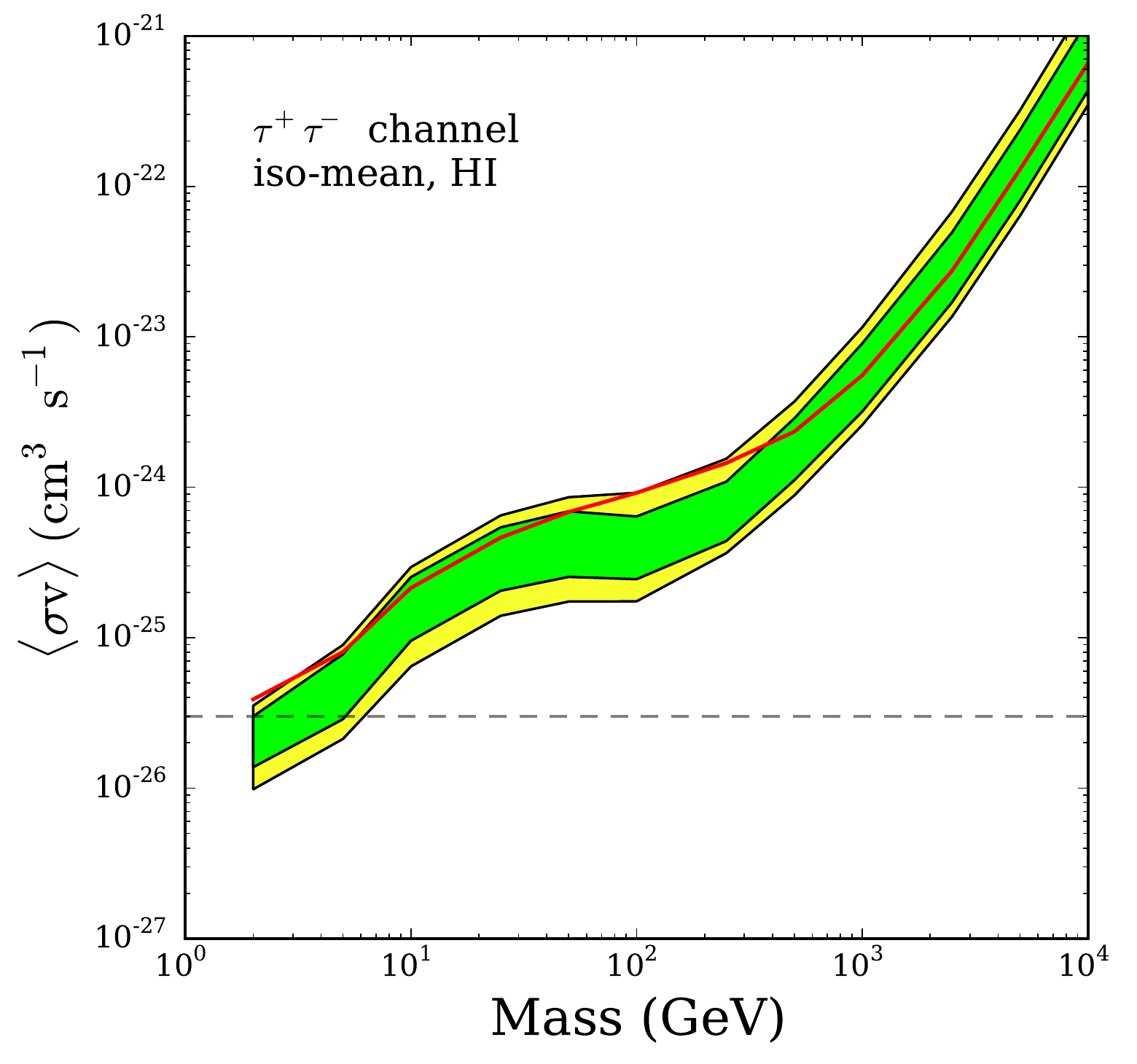}
\includegraphics[width=0.33\columnwidth]{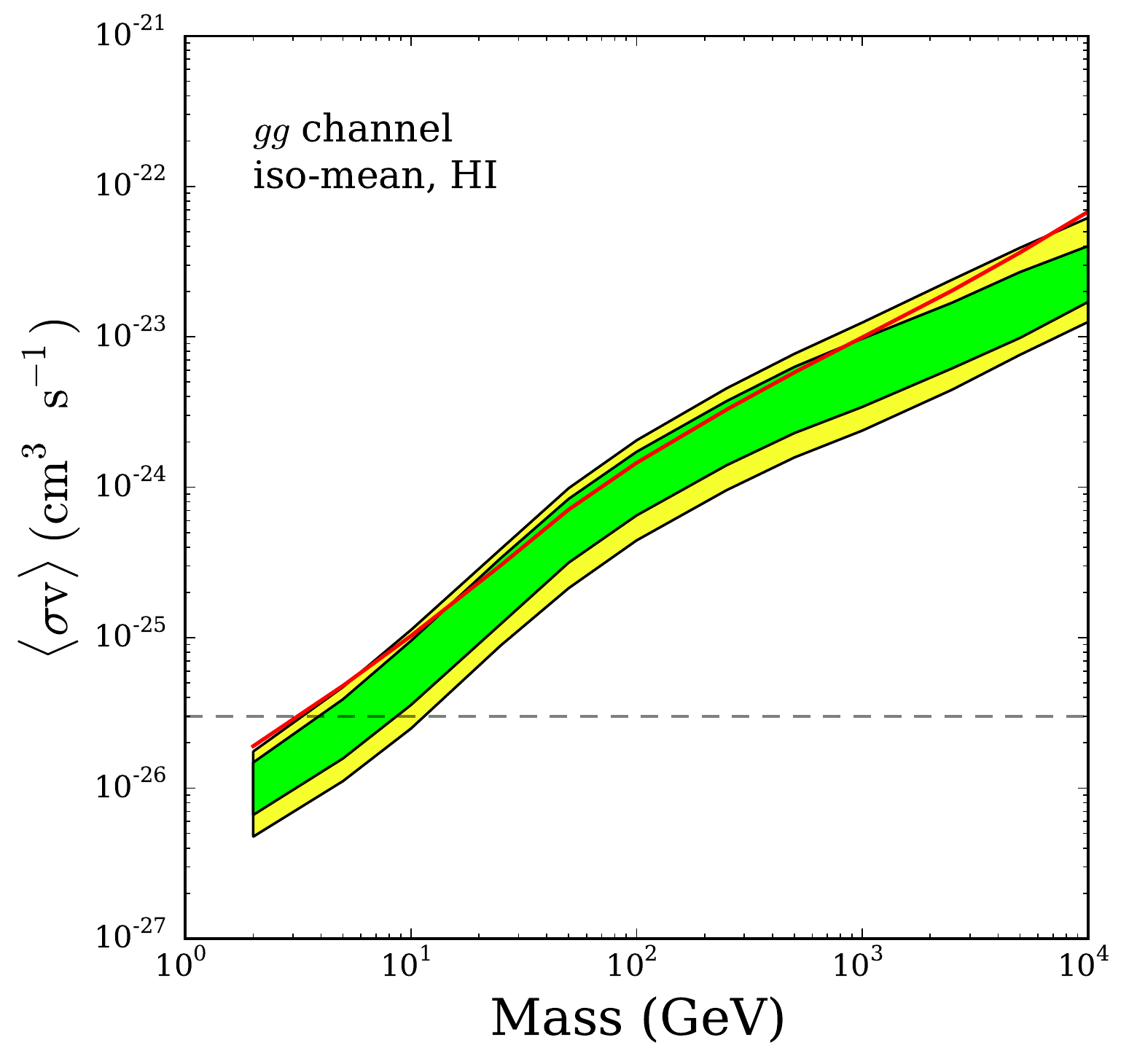}\includegraphics[width=0.33\columnwidth]{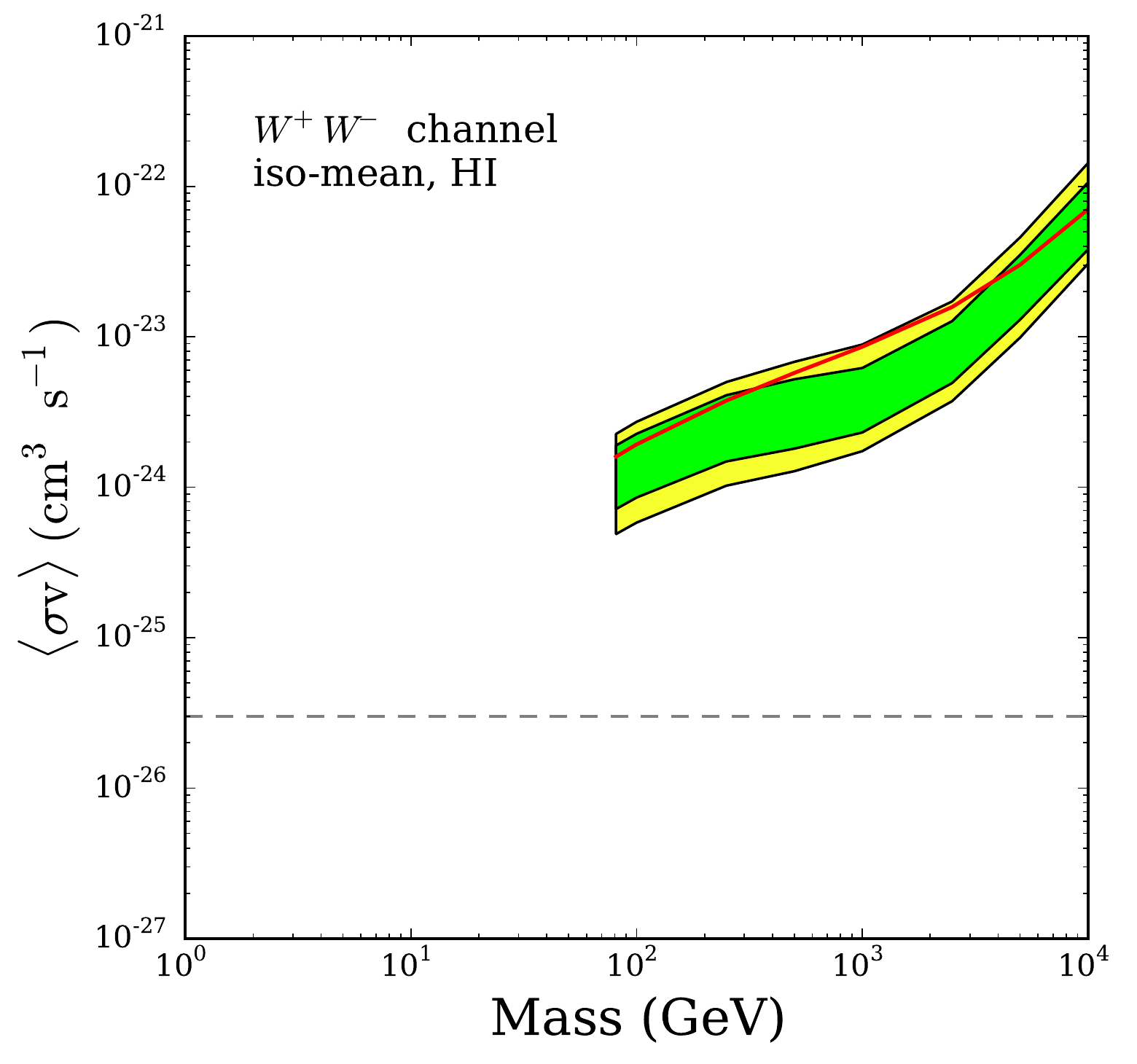}
\caption{Upper limits on $\langle \sigma v\rangle$ for the indicated annihilation channels (red), as a function of dark matter mass, assuming the {\tt iso-mean} profile located at the {\tt HI} center.  Also shown are the 84\% (green) and 95\% (yellow) containment bands of the upper limit drawn from background only simulations.  The horizontal dashed line shows the canonical thermal relic cross section. \label{fig:isomin_results}}
\end{figure}

\begin{figure}

\includegraphics[width=0.33\columnwidth]{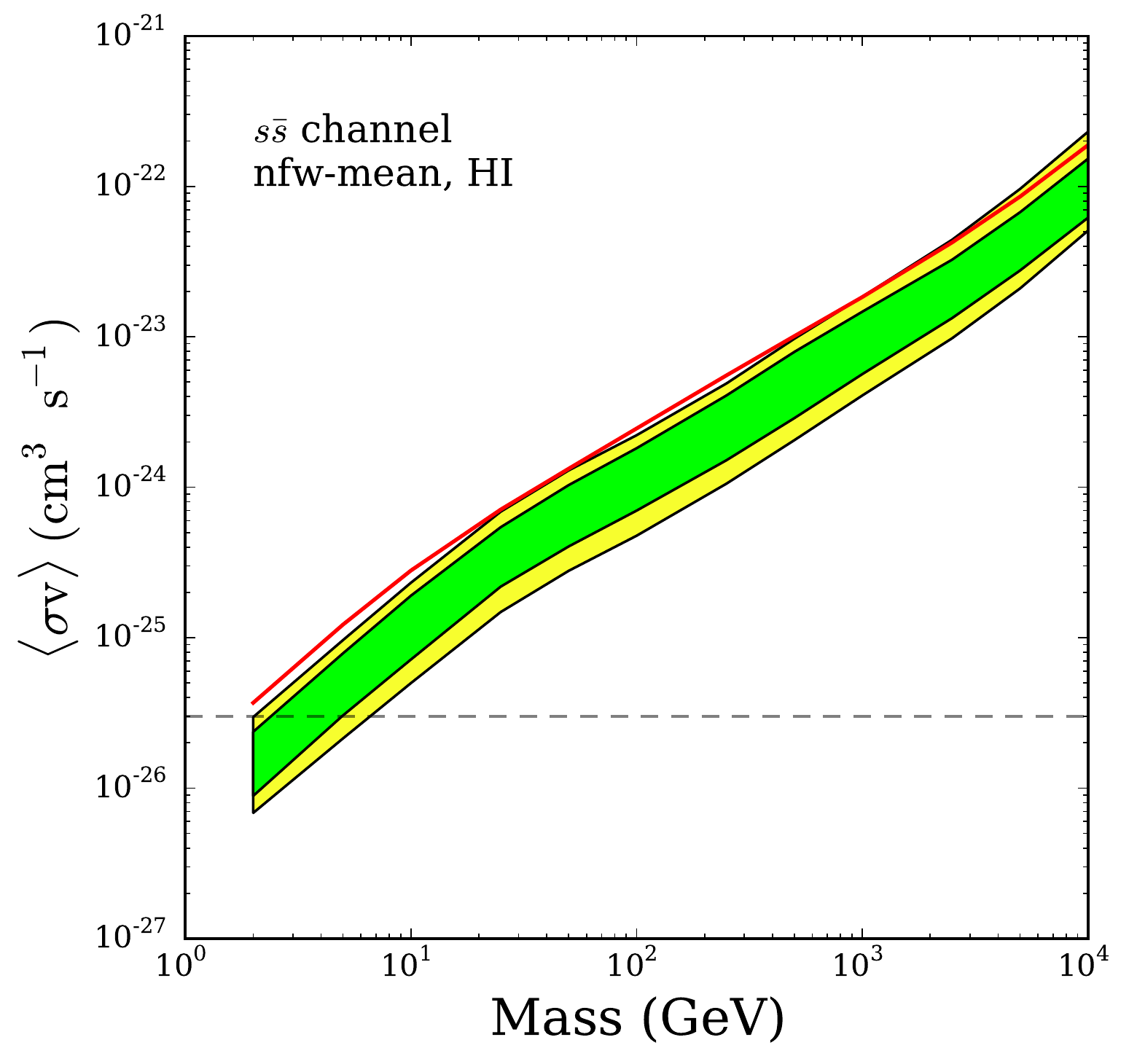}\includegraphics[width=0.33\columnwidth]{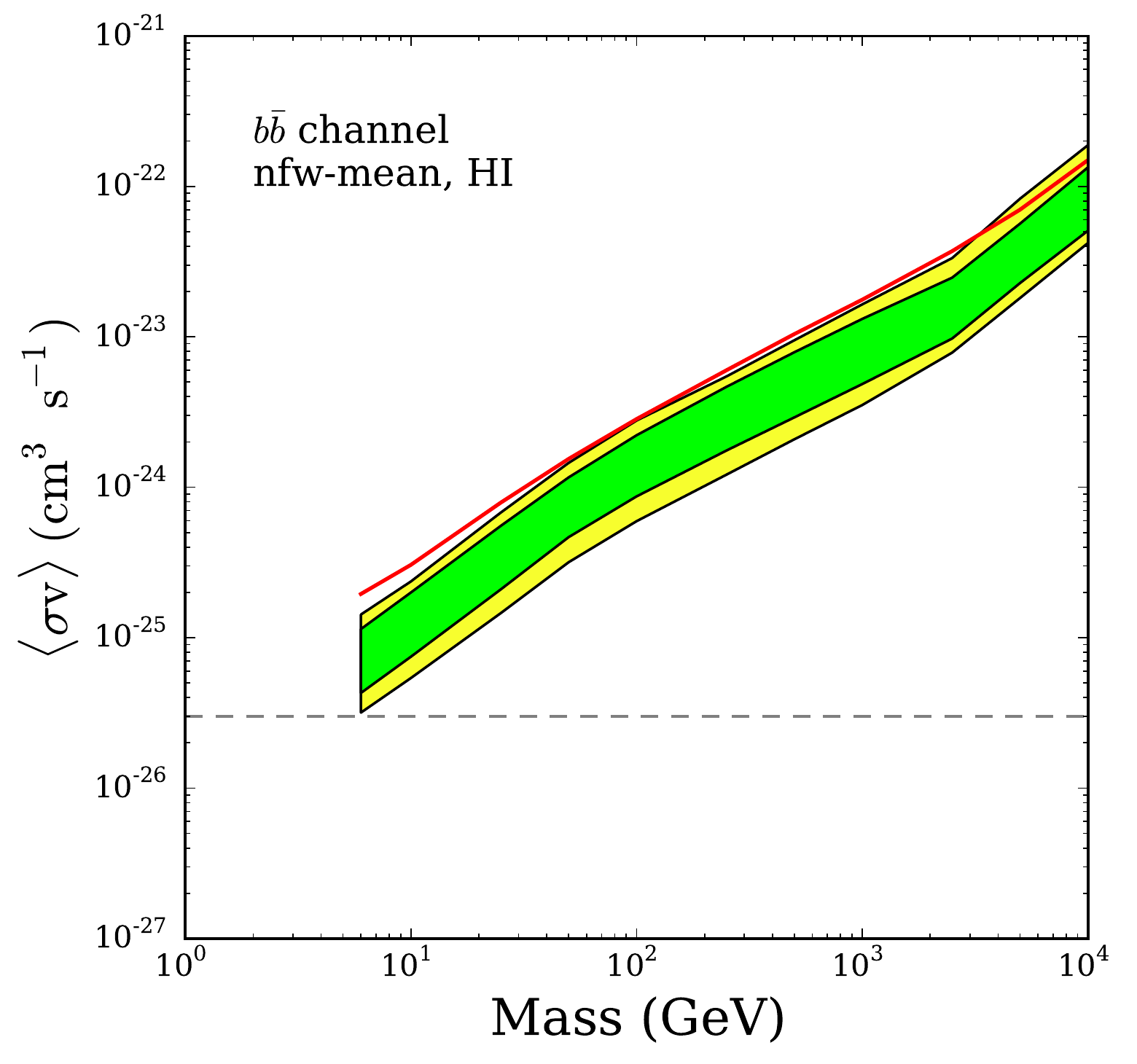}\includegraphics[width=0.33\columnwidth]{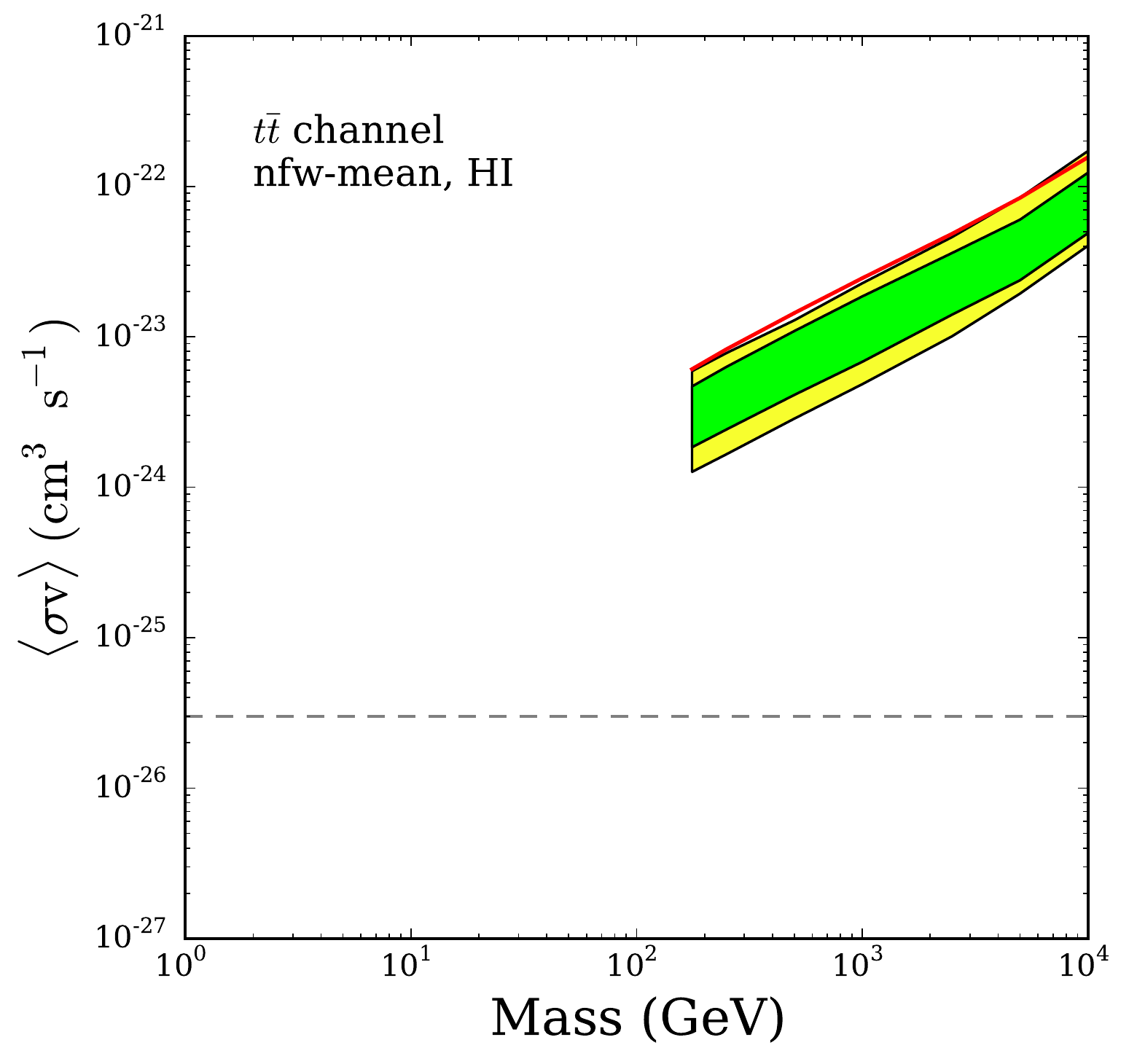}
\includegraphics[width=0.33\columnwidth]{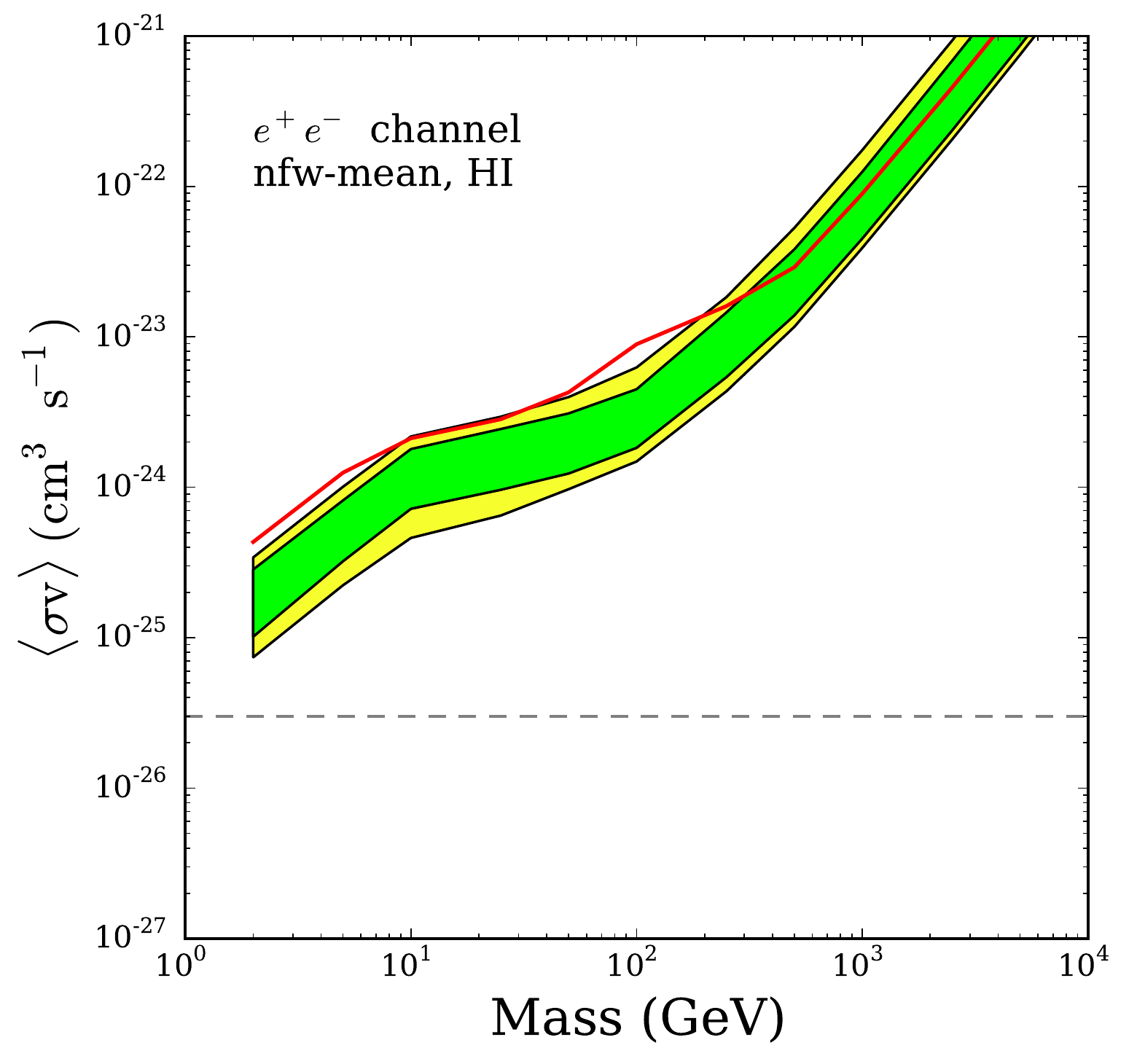}\includegraphics[width=0.33\columnwidth]{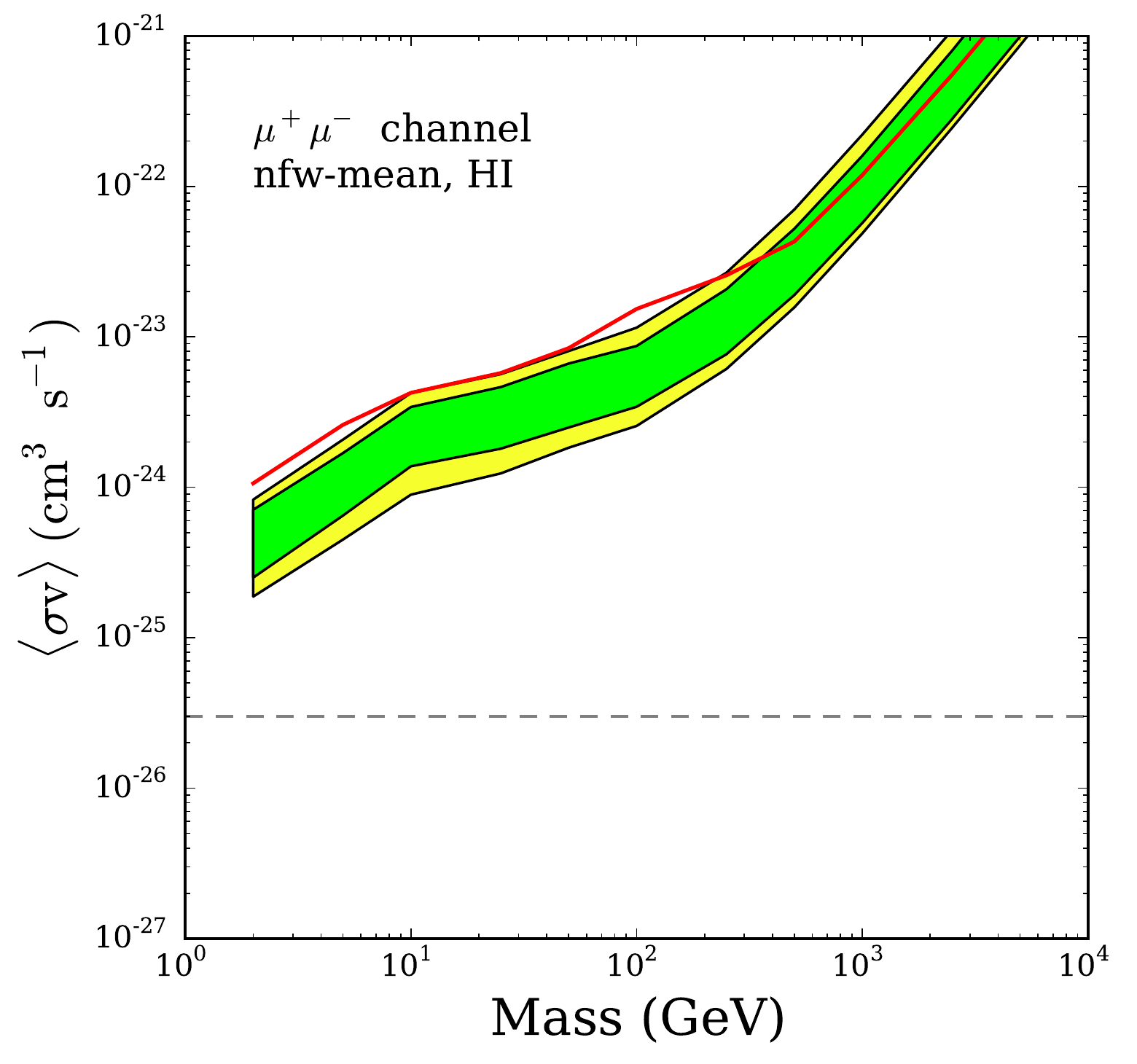}\includegraphics[width=0.33\columnwidth]{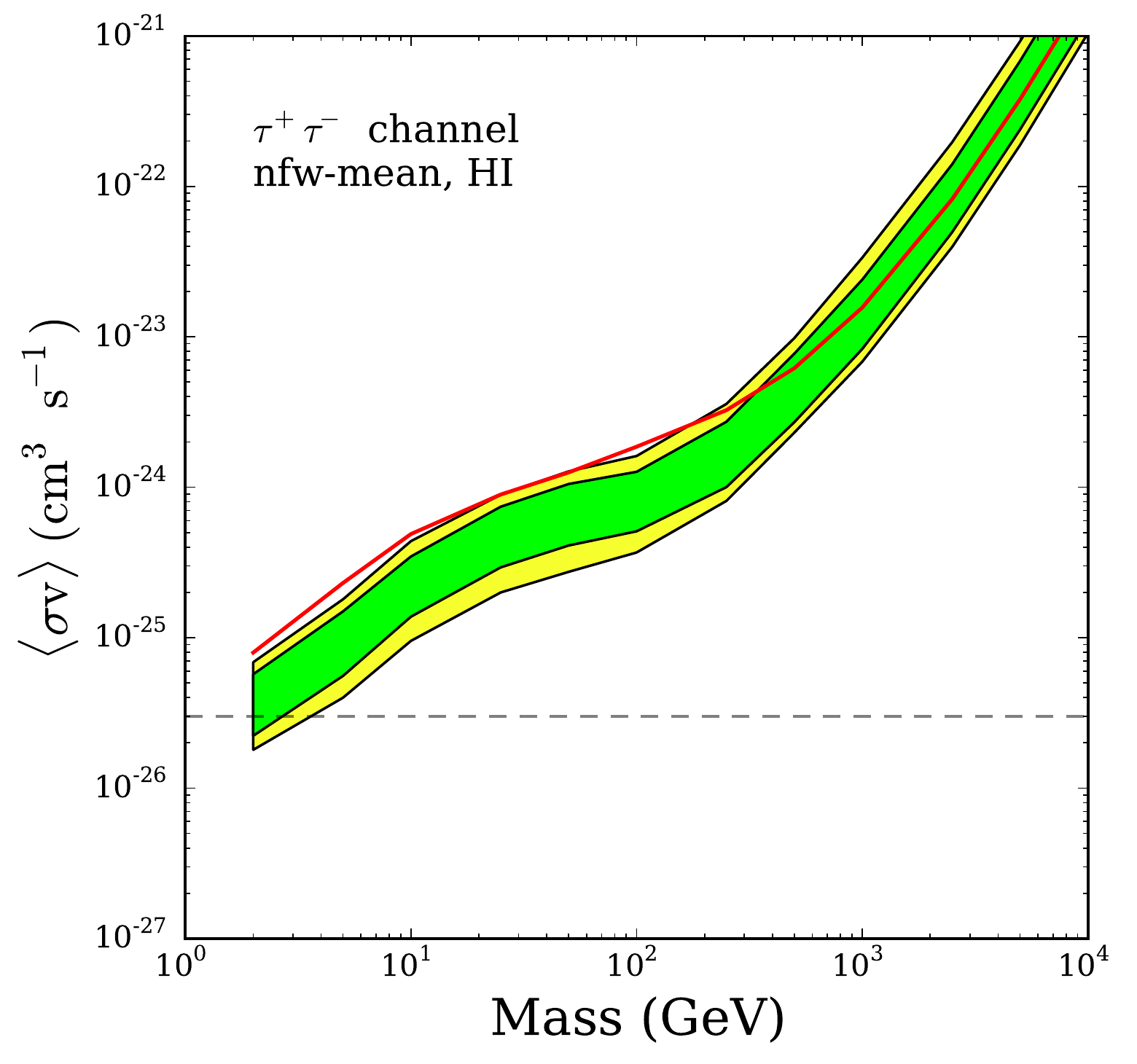}
\includegraphics[width=0.33\columnwidth]{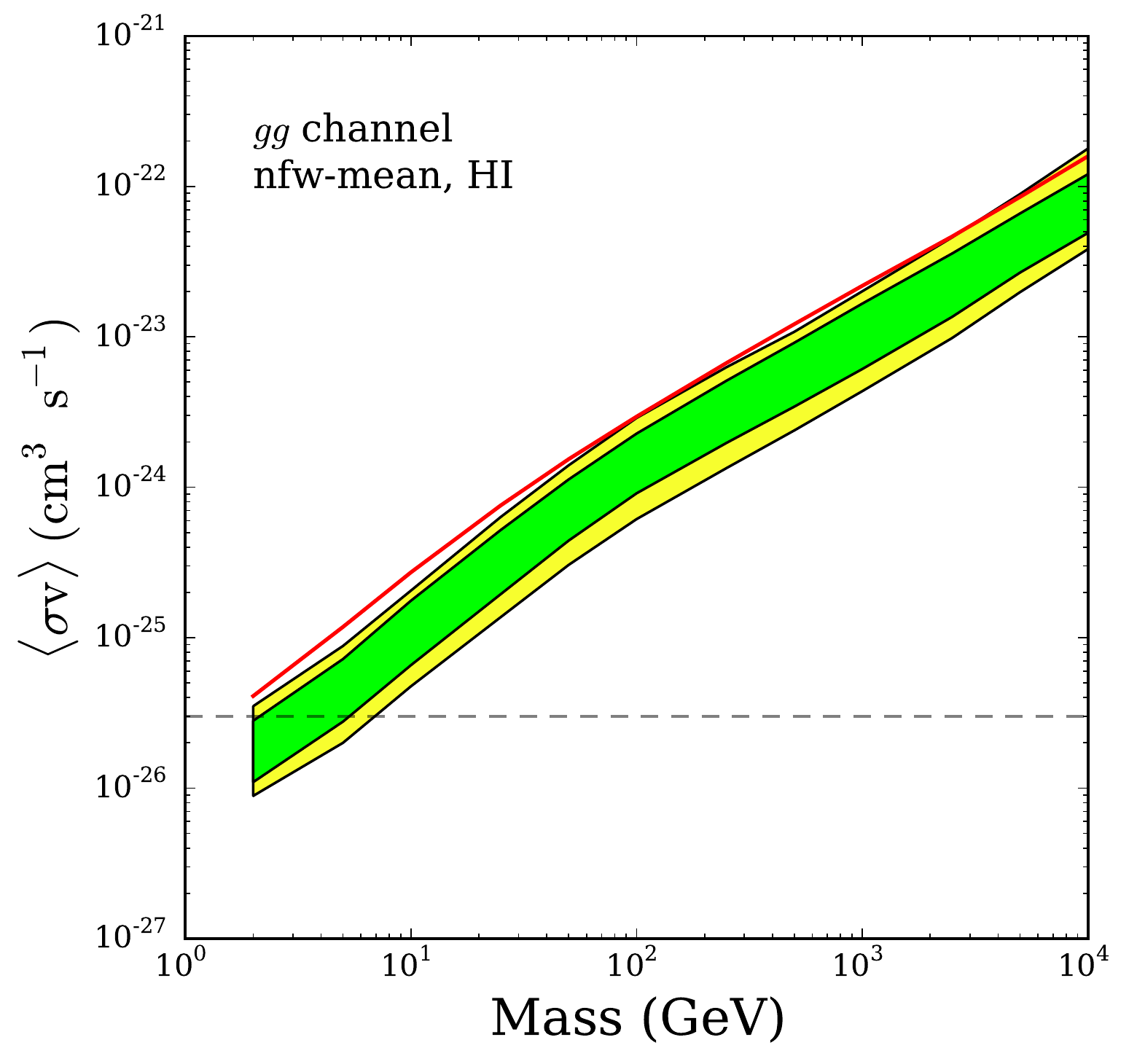}\includegraphics[width=0.33\columnwidth]{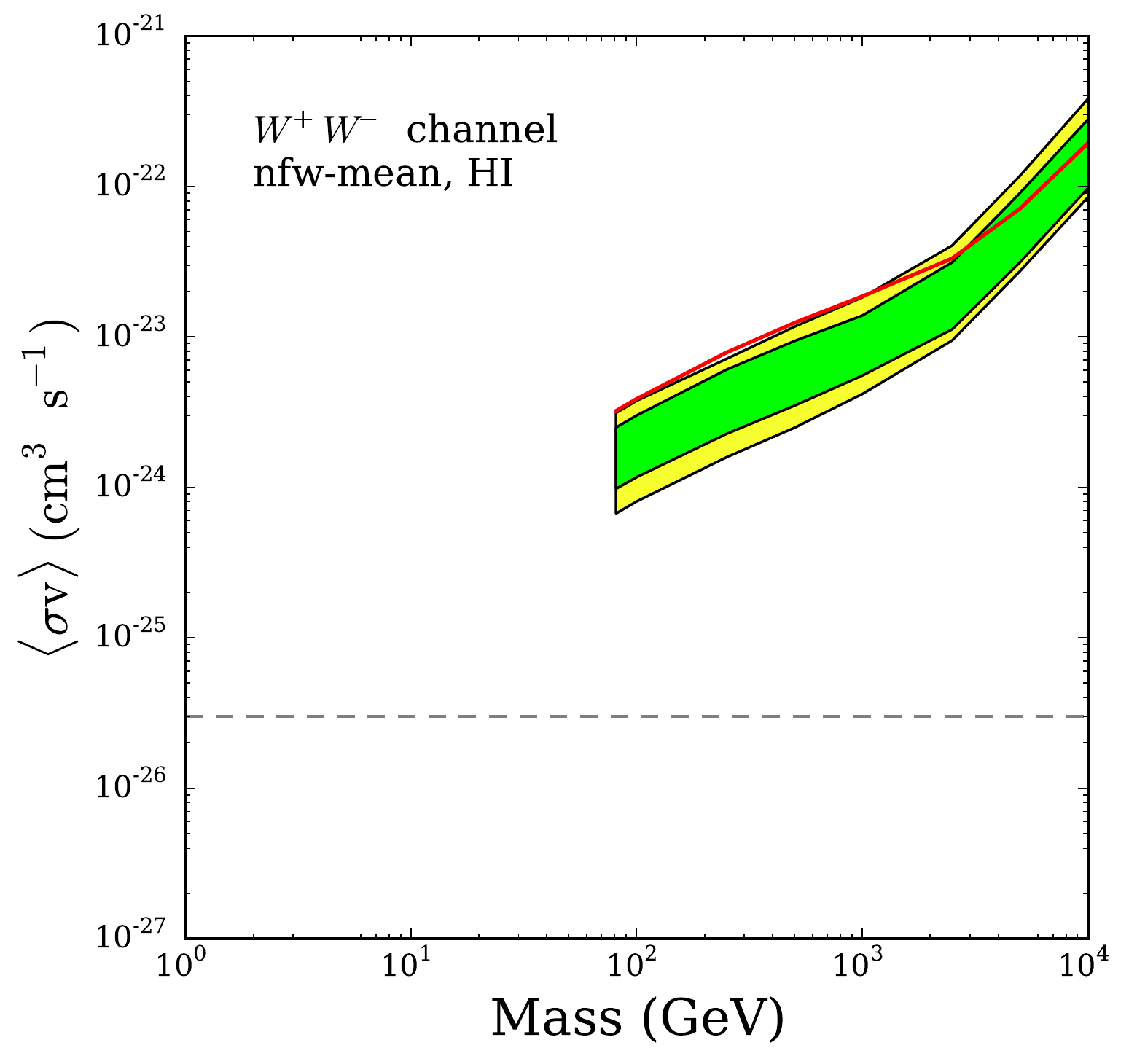}
\caption{Upper limits on $\langle \sigma v\rangle$ for the indicated annihilation channels (red), as a function of dark matter mass, assuming the {\tt nfw-mean} profile located at the {\tt HI} center.  Also shown are the 84\% (green) and 95\% (yellow) containment bands of the upper limit drawn from background only simulations.  The horizontal dashed line shows the canonical thermal relic cross section. \label{fig:nfwmin_results}}
\end{figure}

Focusing now on the upper limits on the annihilation cross sections for the most conservative of our three initial choices for the center of the LMC dark matter profile -- the {\tt HI} center -- we show the 95\% CL upper bounds for each annihilation channel for the {\tt sim-mean} profile in Figure~\ref{fig:envmean_results}, the {\tt iso-mean} profile in Figure~\ref{fig:isomin_results}, and the {\tt nfw-mean} profile in Figure~\ref{fig:nfwmin_results}. On first glance, the most obvious feature of these constraints is the sharp kink at dark matter masses corresponding to a spectrum with gamma rays in the $0.1-2$~GeV range (see Figure~\ref{fig:annihilation_spectra}). For dark matter annihilation channels with this type of spectrum, the fit is dominated by systematic uncertainties. This is demonstrated in Figure~\ref{fig:fsyst} for the $b\bar{b}$ channel. Here dark matter with masses between $\sim10-300$~GeV have $\delta f_{\rm syst} \gtrsim 3\%$. This would correspond to different dark matter mass ranges for other annihilation channels, due to the differences in spectra. The result is a marked weakening of the constraints relative to the regions that are statistics dominated. While unusual compared to the familiar shape of a statistics-dominated exclusion plot, these results are not unexpected for our search. We are confronting the fact that gamma rays of a few hundred MeV to a few GeV are the generic expectation of both dark matter annihilation and a wide variety of baryonic backgrounds.

With that comment, we now turn to the results themselves. We see that of the three profiles considered, the strongest constraints for all channels come from the {\tt sim-mean} profile, which is unsurprising, as this profile has a larger $J$-factor in the inner few degrees than the NFW profile, which is more concentrated than the relatively diffuse isothermal profile (see Figure~\ref{fig:J}). For the {\tt sim-mean} profile, we can exclude the canonical thermal cross section for dark matter up to 10~GeV in the $b\bar{b}$ channel. This compares favorably with the bounds set by the {\it Fermi} LAT analysis of the dwarf spheroidal galaxies using Pass 7 data~\cite{Ackermann:2012nb}. The {\tt iso-mean} and {\tt nfw-mean} profiles place significantly weaker bounds on dark matter annihilation. For example, these profiles do not rule out the thermal cross section at any mass for the $b\bar{b}$ channels.

Given the range of dark matter profiles, there will no doubt be some question as to which set of bounds should be used as ``the'' constraint from the LMC. The weakest bound comes from the isothermal profile, and can be used as the maximally conservative choice. We stress however that we have no particular reason to expect that the dark matter in the LMC is so broadly distributed as in the isothermal profiles, and the interpretation of the stellar and gas rotation curves to which this profile was fit was done under assumptions to minimize the LMC dark matter content, and thus provides the weakest possible bounds. Simulations of galaxies of similar mass and luminosity as the LMC suggest much more cuspy profiles.
However, reducing this uncertainty requires resolving the central 1~kpc of more galaxies to better test cored versus cuspy density profiles; so more data are needed, possibly from the {\it GAIA} survey. Given the power of the limits we have obtained, it is clear that such an effort has the potential to achieve very high sensitivity to dark matter annihilation. We will discuss possible improvements that would merit a reanalysis of the LMC in our conclusions.

\subsection{Statistical Excess} 

Before concluding, we return to a discussion of the slight dark matter-like excess in the dark matter profiles we considered. This excess manifests as a TS for dark matter of 15-20, in the channels and masses that correspond to gamma-ray emission primarily in the hundreds of MeV to GeV energy range. These excesses are clearly seen in Figures~\ref{fig:envmean_results}, \ref{fig:isomin_results}, and \ref{fig:nfwmin_results}, in which the observed upper limit on $\langle \sigma v\rangle$ is $1-2\sigma$ outside the expected constraints in this range of masses and channels, after systematic uncertainties are taken into account.  Before considering systematic uncertainties, this excess is $4-5\sigma$, depending on the choice of dark matter profile. As can be seen in the scans over center locations for the different profiles, the excess is located near to the center of the H\,{\sc i} gas rotation curve, but the ``center'' moves significantly as the profile is varied (compare, {\it e.g.}, Figs.~\ref{fig:envmean_map},\ref{fig:isomin_map} and \ref{fig:nfwmin_map}), and the required cross section (in the $b\bar{b}$ or $\tau^+\tau^-$ channels) is ${\cal O}(10^{-26}~\mbox{cm$^3$/s})$ for the {\tt sim-mean} profile.

The {\it most probable} explanation for this observed excess is an additional non-dark matter component that was not included in our background model. Though the region of interest is not particularly unusual in terms of gas density, and does not contain an unusual number of identified supernova remnants, it is at the intersection of multiple background components. Furthermore, as the iterative procedure used to build the background model looked for Gaussian components with TS values of 25 or larger, it is not surprising that a non-Gaussian component of TS$ < 20$ would not be added in. Furthermore, gamma rays of sub-GeV energies are the expected spectrum from CRs colliding with gas. Therefore, the likely explanation is that there is a slightly higher flux of CRs near the {\tt HI} center, which we have not modeled correctly, and as a result our bounds are weaker than expected.   Unresolved point sources in the LMC may also contribute to the observed excess.

\begin{figure}[ht]
\includegraphics[width=0.5\columnwidth]{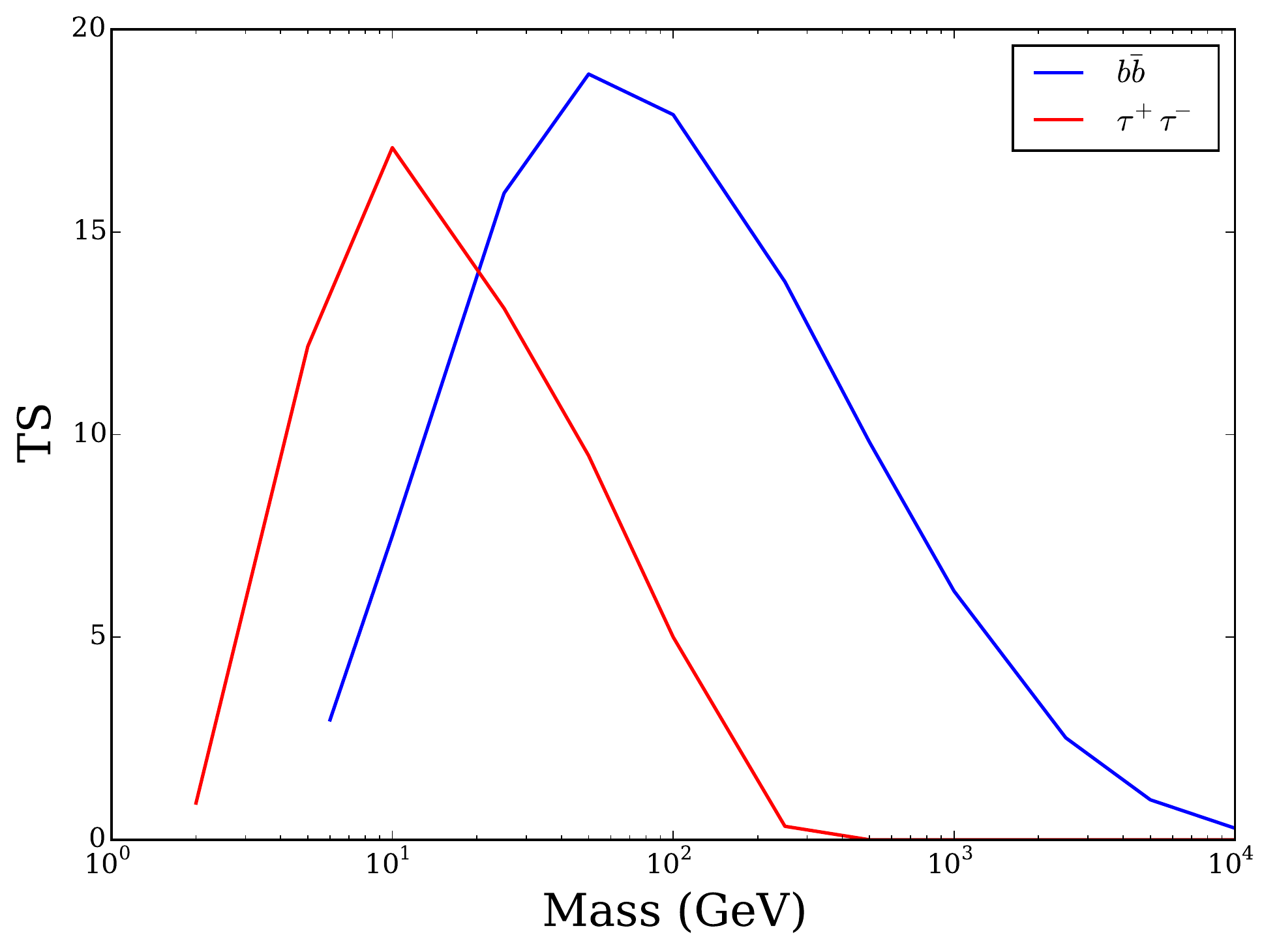}
\caption{TS of dark matter annihilation signal into either the $b\bar{b}$ or $\tau^+\tau^-$ channels over the background-only model for the {\tt sim-mean} profile at the center location that maximizes the TS ($\ell = 279.6,b = -33.1$), as a function of dark matter mass $m_\chi$. \label{fig:excess_TS}}
\end{figure}

That said, it is exactly this range of dark matter annihilation channels, masses, and cross sections that are of interest to explain the observed excess in the Galactic Center~\cite{Abazajian:2014fta,Daylan:2014rsa,Gordon:2013vta,Calore:2014zzz}. As we expect that these weaker-than-expected bounds in the LMC may attract some interest in regards to the Galactic Center excess, we wish to provide some extra context. In Figure~\ref{fig:excess_TS}, we show the TS for dark matter annihilating into $b\bar{b}$ or $\tau^+\tau^-$ in the {\tt sim-mean} profile at the location that has the maximum significance for the excess ($\ell = 279\fdg 6,b = -33\fdg 1$), $0\fdg 7$ from the {\tt HI} center and $1\fdg 1$ from the {\tt stellar}. Furthermore, the Galactic Center excess (when interpreted as dark matter annihilation) lies in the range of channels and masses that have an unfortunate degeneracy with the spectrum of baryons injected by astrophysical sources (see, {\it e.g.}, Figure~\ref{fig:fsyst} or discussion in Refs.~\cite{Calore:2014zzz,Abazajian:2010zy,Wharton:2011dv,Hooper:2013nhl,Carlson:2014cwa,Petrovic:2014uda}).

\begin{figure}[ht]
\includegraphics[width=0.5\columnwidth]{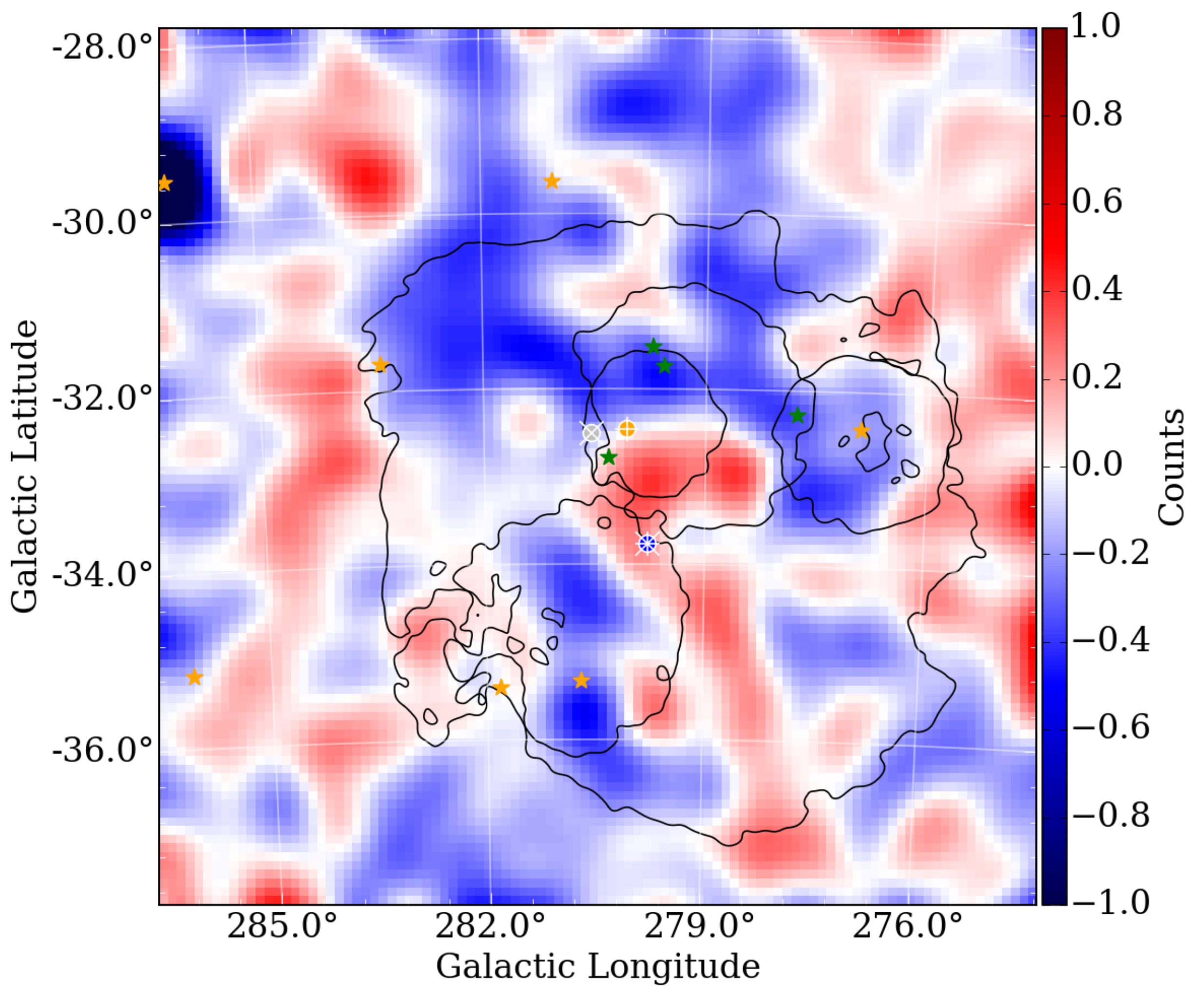}\includegraphics[width=0.5\columnwidth]{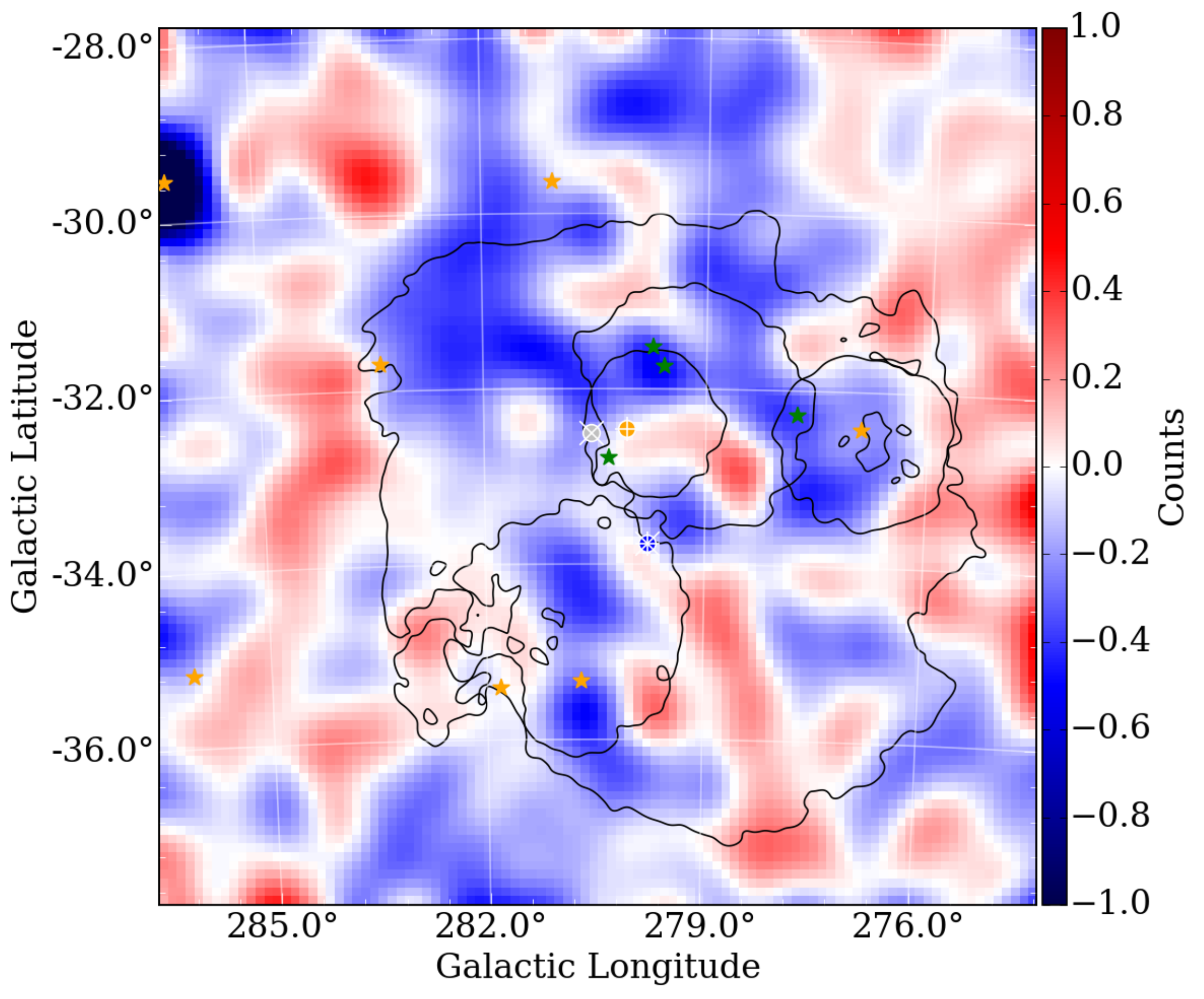}

\caption{Left: Residual map comparing the {\it Fermi} LAT data to the background-only model after fitting to data.  Black contours are extended components, orange stars are point-sources. 
Right: Residual map of data compared to the fit of the background model plus 50~GeV dark matter annihilating to $b\bar{b}$ in the {\tt sim-mean} profile, with the dark matter located at the center with maximum TS value ($\ell = 279\fdg 6,b = -33\fdg 1$).   Both maps are binned in $0\fdg 1 \times 0\fdg 1$ pixels and smoothed with a $\sigma=0\fdg 3$ Gaussian kernel. \label{fig:residual}}
\end{figure}

\begin{figure}[ht]
\includegraphics[width=0.6\columnwidth]{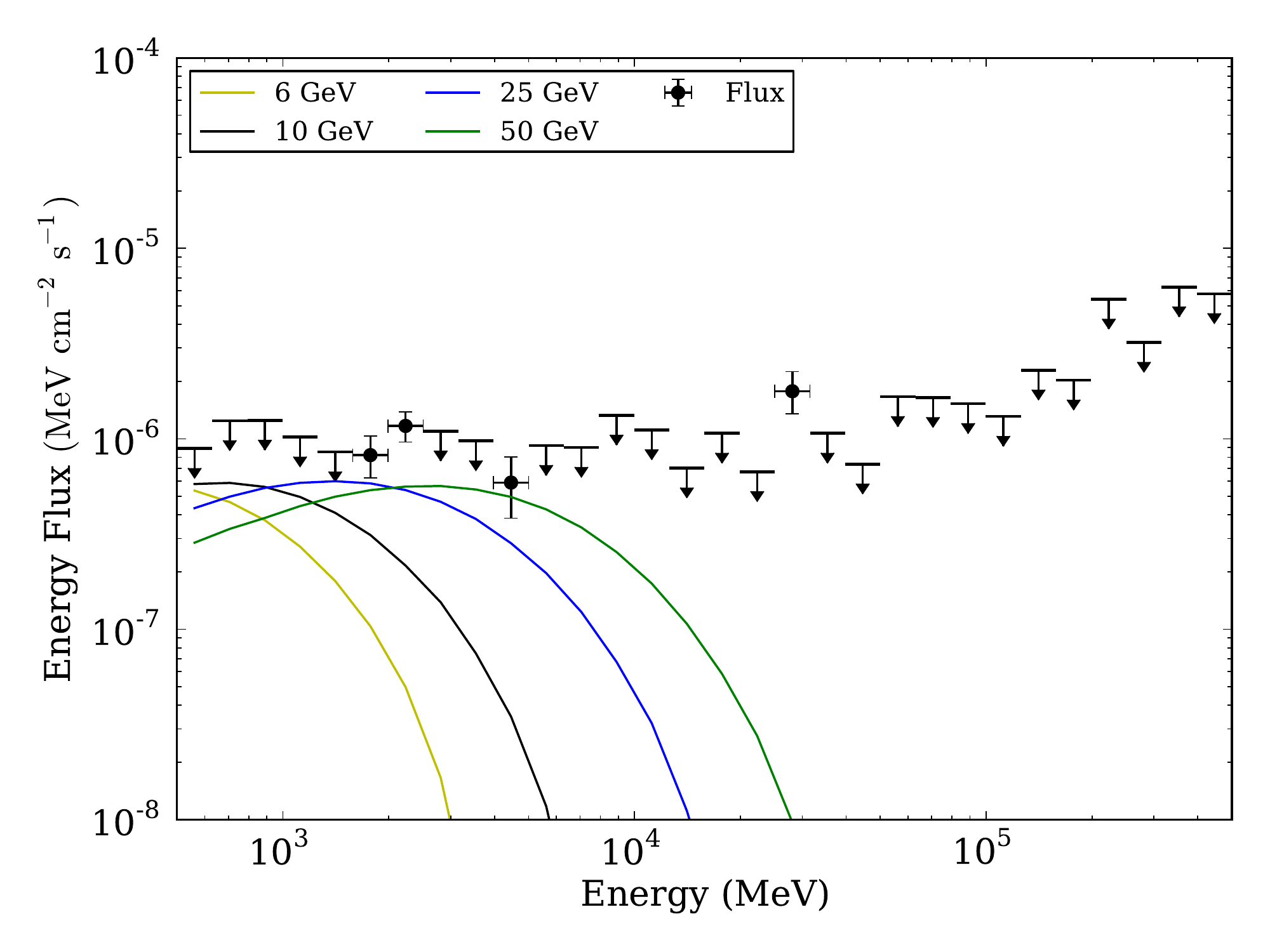}

\caption{Energy bin-by-bin fit to the gamma-ray flux after fitting to the {\tt sim-mean} profile located at the center with maximum TS value ($\ell = 279\fdg 6,b = -33\fdg 1$), compared to flux spectra of dark matter annihilation to $b\bar{b}$ states for representative values of $m_{\chi}$.   Upper limits (95\% CL) are given for all energy bins with ${\rm TS} < 4$.\label{fig:residual_sed}}
\end{figure}

We show in Figure~\ref{fig:residual} the smoothed residual gamma-ray map of the LMC after subtracting the best-fitting background-only model of the LMC. It is clear from this model that there is a region extending from between the {\tt stellar} and {\tt HI} center past the {\tt HI} center where the best-fit to the background model under-predicts the observed gamma rays. While large parts of the remainder of the LMC have gamma-ray counts that are somewhat over-predicted by the background-only model, other regions further from the prospective dark matter centers also have similar excesses. We show as well the residual map comparing the data with the background components after fitting to background plus dark matter in the {\tt sim-mean} profile located at ($\ell = 279.6,b = -33.1$). The improvement is noticeable, consistent with a $4-5\sigma$ {\it statistical-only} improvement to the fit. Recall that after systematic uncertainties are accounted for, this excess is only $1-2\sigma$, because the gamma rays driving the statistical fit are in the range in which baryonic backgrounds in our control region contribute greatly.
 The spectrum for this excess is not well defined by the spectral fit (Figure~\ref{fig:residual_sed}), which is unsurprising given the relatively low statistics associated with the excess. Only a few energy bins have best-fit fluxes that are not upper limits. Again, this is not unexpected from the level of statistics available.

Overall, we find that this excess, though located near our expected dark matter center and providing a significant {\it statistical} improvement to the fit for a number of likely dark matter profiles assuming similar dark matter spectra as claimed in the case of the Galactic Center excess, is {\it compatible} with our background-only assumption at the level of systematic uncertainty we expected to find in the LMC. While this may provide additional motivation to improve our understanding of baryonic backgrounds and the dark matter profile in the LMC, we caution the reader to not over-interpret the statistical significance of these results. We note that additional center locations exist with similar TS values for a dark matter signal component, but they lie outside the central region of the LMC, and thus cannot be dark matter-related. The existence of these locations lends weight to the conclusion that the observed excess is background-related.

\section{Conclusions}\label{sec:conclusion}

The LMC is the largest satellite of the Milky Way, as well as one of the closest. Though it is actively forming massive stars and therefore significant backgrounds are present, it is nonetheless an attractive target for indirect detection of dark matter annihilation. Using stellar and gas rotation curves, assuming an isothermal profile, and making conservative choices in the data analysis, the LMC still has an annihilation $J$-factor as large as the best (most constraining) dwarf spheroidal galaxies currently known. Simulations of galaxies similar to the LMC in both mass and stellar luminosity suggest a more cuspy profile, in which case the annihilation rate of dark matter would be at least an order of magnitude larger than in any of the dwarf spheroidals. In addition, as the LMC is spatially extended, the dark matter annihilation signal would have a characteristic morphology, which could be used in conjunction with the spectrum to distinguish it from backgrounds.

Given these advantageous properties, and the interest in the potential indirect detection of dark matter from the Galactic Center, it is an opportune time to analyze the {\it Fermi} LAT gamma-ray observations of the LMC for signals of dark matter. We have used five years of {\it Fermi} LAT data over a $10^\circ \times 10^\circ$ ROI. To understand the gamma rays originating from baryonic backgrounds, we used data-driven modeling of the gamma-ray backgrounds in the LMC, convolving Gaussian CR injection sources with the measured column density of gas. As the LMC is a unique environment in the gamma-ray sky, no ideal control regions exist.  As a result, we account for possible unmeasured backgrounds in our statistical fits by estimating the systematic errors from regions in the LMC outside of $3^\circ$ from the center, where the signal cannot be dark matter-dominated. This inclusion of systematic errors does weaken our potential limits in the energy range characteristic of the baryonic background.

\begin{figure}[th]
\includegraphics[width=0.7\columnwidth]{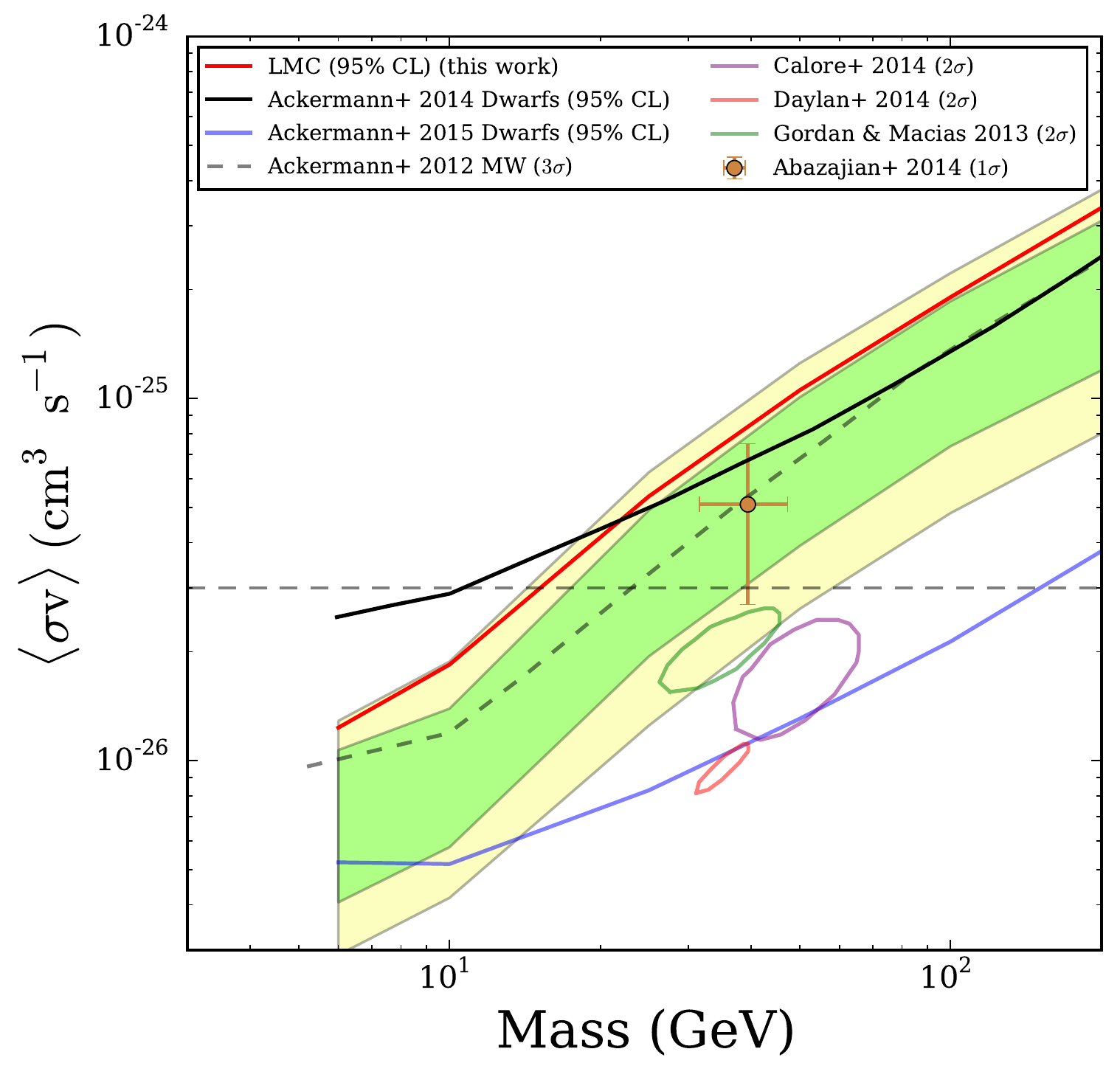}
\caption{Comparison between the 95\% CL upper limits from the LMC analysis (red solid line, predicted 84\%  and 95\% containment bands in green and yellow) and the upper limits set by the {\it Fermi} LAT analysis of the dwarf spheroidal galaxies using Pass 7 data (black solid). Also shown are the {\it Fermi} LAT upper limits from the Milky Way Galactic halo (dashed gray line) and the upper limits set by the dwarf spheroidals from the {\it Fermi} LAT analysis of Pass 8 data (solid blue lines)~\cite{2015arXiv150302641F}.  Confidence regions for cross section and mass determined by independent analyses of the Galactic Center excess are shown (brown~\cite{Abazajian:2014fta}, purple~\cite{Calore:2014zzz}, green~\cite{Gordon:2013vta} and red circles~\cite{Daylan:2014rsa}).  The horizontal dashed line shows the canonical thermal relic cross section.  The LMC upper limits are based on the {\tt sim-mean} profile at the {\tt HI} center. \label{fig:compare}}
\end{figure}

Our most conservative limits from this analysis are weaker than the existing limit from a joint analysis of dwarf spheroidal galaxies. However, assuming a profile of dark matter more representative of the results from simulation, we place bounds very competitive with those derived from the dwarf analysis. In Figure~\ref{fig:compare}, we show a direct comparison between the bounds we set in the $b\bar{b}$ channel (using the {\tt sim-mean} profile at the {\tt HI} center) and the existing dwarf analysis performed by the {\it Fermi} LAT Collaboration, which also used Pass 7 data. As can be seen, the limits set by our analysis of the LMC are stronger than those of the dwarf analysis in the low-mass region. We expected to find stronger bounds across the entire mass range, but did not due to an upward fluctuation compared to our statistical expectations. 

We also compare these limits with the values preferred by independent analyses of the Galactic Center excess~\cite{Abazajian:2014fta,Calore:2014zzz,Gordon:2013vta,Daylan:2014rsa}. We should note that the regions identified as good fits to the Galactic Center anomaly in Figure~\ref{fig:compare} are extrapolations of the results of Refs.~\cite{Abazajian:2014fta,Calore:2014zzz,Gordon:2013vta,Daylan:2014rsa} based on figures in those works
and do not include all sources of uncertainty described in the works.   In particular, the ellipses are given for specific dark matter spatial profiles, and do
not include uncertainties on $\langle\sigma v\rangle$ due to the uncertainty of the spatial profile.    Nevertheless, we include them in order to provide the reader with a sense of the parameter space of interest.

As this work was being prepared, preliminary results using the Pass 8 event analysis~\cite{2015arXiv150302641F} were presented by the {\it Fermi}-LAT Collaboration.  A reanalysis of the dwarf spheroidals using the Pass 8 data places significantly better upper limits on dark matter annihilation, compared to the previous dwarf spheroidal bounds, or our LMC analysis. Given the improvement seen from the use of Pass 8 data in the dwarf spheroidal analysis, it would be useful to re-examine the LMC using Pass 8 data and eventually a longer time interval.  

In parallel to this, additional efforts should be made to reduce the uncertainties in our dark matter models.  The most significant uncertainty in our modeling remains the dark matter profile and center location. Advanced simulations (including the effects of baryonic feedback) \cite{diCintio2014b,Governato2012} indicate that the LMC is unlikely to possess a cored dark matter profile (such as our isothermal profiles), in which case the resulting bounds on the dark matter annihilation cross section are expected to be as strong or stronger than those set by other dark matter dominated objects. While the results from simulation are compelling, future analyses of gamma rays from the LMC would greatly benefit from additional information from observations in other wavelengths that might resolve the core or cusp issue. 

The LMC is not a likely target for direct weak-lensing measurements of its dark matter profile; we have estimated that the critical surface mass density for a gravitational lens located at the LMC is some four or more orders of magnitude below the surface density of the LMC itself. Thus our knowledge of the dark matter profile of the LMC can come only from measurements of the motions of stars and gas. Presently, the leading uncertainty in the rotation curve is the inclination angle at which we observe the galaxy; it is unclear how much this uncertainty can be reduced.  However, the {\it GAIA} satellite will measure the six-dimensional position and velocity phase space of a billion stars, including targets in the LMC \cite{Perryman:2001sp}. While the measurement errors grow linearly with distance, when combined with {\it Spitzer} distance measurements, high-precision results may be obtained out to 60~kpc \cite{Price-Whelan:2013kpa}, beyond the LMC.  The ability of {\it GAIA} to constrain the LMC dark matter profile has not been investigated in detail.  

Given the appeal of the LMC as a target for the indirect detection of dark matter annihilation, efforts to reduce uncertainties in the dark matter profile and background characteristics are well motivated.
Additionally, comparison of the LMC dark matter profile extrapolated from observation to that of simulation could provide useful feedback on the accuracy of the simulations themselves.   

This work provides the first-ever constraints on the annihilation of dark matter into Standard Model particles from observations of the Large Magellanic Cloud. Using five years of {\it Fermi} LAT data, we place upper limits on the velocity-averaged cross section that reach the benchmark canonical thermal freeze-out value for low mass dark matter. Due to higher-than-expected flux, our limits were weaker than expected.  No signal was found with any statistical significance when systematic uncertainties were incorporated in the analysis.  Our results required construction of dark matter profiles for the LMC, which were derived using both observational results and state-of-the-art galaxy simulations, making conservative assumptions throughout. 

The main sources of uncertainties are systematic: in our dark matter profile, in the location of the center of the LMC, and in the background modeling of the LMC gamma rays originating from baryonic processes. It may be possible to reduce some of these uncertainties by application of results from other observations of the LMC.  We also note that estimates of the Small Magellanic Cloud's $J$-factor suggest it may be a promising target as well. Though not background-free, the Small Magellanic Cloud has a lower star-formation rate than the LMC, which results in a lower background flux.   Combined with the improvements possible with the {\it Fermi}-LAT Pass 8 event selection,  the Magellanic Clouds can continue to be important targets for indirect dark matter searches.

\begin{acknowledgments}
The \textit{Fermi} LAT Collaboration acknowledges generous ongoing support
from a number of agencies and institutes that have supported both the
development and the operation of the LAT as well as scientific data analysis.
These include the National Aeronautics and Space Administration and the
Department of Energy in the United States, the Commissariat \`a l'Energie Atomique
and the Centre National de la Recherche Scientifique / Institut National de Physique
Nucl\'eaire et de Physique des Particules in France, the Agenzia Spaziale Italiana
and the Istituto Nazionale di Fisica Nucleare in Italy, the Ministry of Education,
Culture, Sports, Science and Technology (MEXT), High Energy Accelerator Research
Organization (KEK) and Japan Aerospace Exploration Agency (JAXA) in Japan, and
the K.~A.~Wallenberg Foundation, the Swedish Research Council and the
Swedish National Space Board in Sweden.

Additional support for science analysis during the operations phase is gratefully
acknowledged from the Istituto Nazionale di Astrofisica in Italy and the Centre National d'\'Etudes Spatiales in France.

We thank Knut Olsen for his help in analyzing the LMC H\,{\sc i} rotational velocity data, and Charles Keeton for his help in the calculation of the magnitude of the gravitational lensing caused by the LMC.
Finally, we thank Johann Cohen-Tanugi, Gabrijela Zaharijas, Rouven Essig and Neelima Sehgal for help
with quantifying the effects of electroweak corrections to the dark matter spectra.

\end{acknowledgments}


\bibliography{LMC_DM}

\end{document}